\documentclass{article}
\usepackage{amsmath,amssymb,amsthm}
\usepackage{longtable}
\usepackage{cite}
\usepackage[shortlabels]{enumitem}
\usepackage{float}

\newcommand{\bnf}{binancefut\_BTC/USDT }

\newcommand{\bnfcm}{binancecmfut\_BTC/USD }

\newcommand{\hbcq}{hbdm\_BTC\_CQ }

\newcommand{\okcq}{okex\_BTC-USD-210326 }

\usepackage[utf8]{inputenc} 
\usepackage[T1]{fontenc}

\usepackage{url}            
\usepackage{booktabs}       
\usepackage{amsfonts}       
\usepackage{nicefrac}       
\usepackage{microtype}      
\usepackage{lipsum}
\usepackage{graphicx}

\usepackage{bbm}
\usepackage[ruled,vlined]{algorithm2e}
\usepackage{float}
\usepackage{epstopdf}

\usepackage{amsthm}

\usepackage[english]{babel}

\usepackage[colorlinks=true,linkcolor=blue,filecolor=green,citecolor=blue]{hyperref}

\usepackage[utf8]{inputenc}
\usepackage{titletoc}
    \titlecontents{chapter}
    [0em] %
    {\medskip\bfseries}
    {\thecontentslabel\quad}
    {}
    {\hfill\contentspage}

    \titlecontents{section}[1.72em]{\smallskip}%
    {\thecontentslabel.\enspace}
    {}
    {\titlerule*[1.2pc]{.}\contentspage}%

\usepackage[dvipsnames]{xcolor} 

\newcounter{noteMCctr} 

\newcounter{noteZZctr} 

\newcounter{noteXXctr} 

\newcommand{\bb}{\hspace{-1mm} $\bullet$ }

    \usepackage[top=20mm, bottom=20mm, left=20mm, right=20mm]{geometry}

\usepackage[%
    font={small,sf},
    labelfont=bf,
    format=hang,    
    format=plain,
    margin=0pt,
    width=0.8\textwidth,
]{caption}

\usepackage[list=true]{subcaption}

\usepackage{tcolorbox}

\begin{document}

\title{Fragmentation, Price Formation, and Cross-Impact in Bitcoin Markets}

\author{
 \textbf{Jakob Albers}\thanks{
 Denotes first author; the rest of the authors are listed in alphabetical order.
 } \\
 Department of Statistics, University of Oxford\\
 Oxford, UK \\
 \texttt{jakob.albers@merton.ox.ac.uk} \\
\\
 \textbf{Mihai Cucuringu} \\
 Department of Statistics and Mathematical Institute, University of Oxford, Oxford, UK\\
 The Alan Turing Institute, London, UK\\
 \texttt{mihai.cucuringu@stats.ox.ac.uk} \\
\\
 \textbf{Sam Howison} \\
 Mathematical Institute, University of Oxford\\
 Oxford, UK\\
 \texttt{howison@maths.ox.ac.uk} \\
\\
 \textbf{Alexander Y. Shestopaloff} \\
 School of Mathematical Sciences, Queen Mary University of London\\
 London, UK\\
 \texttt{a.shestopaloff@qmul.ac.uk} \\
}

\date{August 21, 2021}

\maketitle

\begin{abstract}
In light of micro-scale inefficiencies induced by the high degree of fragmentation of the Bitcoin trading landscape, we utilize a granular data set comprised of orderbook and trades data from the most liquid Bitcoin markets, in order to understand the price formation process at sub-1 second time scales. To achieve this goal, we construct a set of features that encapsulate relevant microstructural information over short lookback windows. These features are subsequently leveraged first to generate a leader-lagger network that quantifies how markets impact one another, and then to train linear models capable of explaining between 10\% and 37\% of total variation in $500$ms future returns (depending on which market is the prediction target). The results are then compared with those of various PnL calculations that take trading realities, such as transaction costs, into account. The PnL calculations are based on natural \emph{taker}  strategies (meaning they employ market orders) that we associate to each model. Our findings emphasize the role of a market's fee regime in determining its propensity to being a leader or a lagger, as well as the profitability of our taker strategy.  Taking our analysis further, we also derive a natural \emph{maker} strategy (i.e., one that uses only passive limit orders), which, due to the difficulties associated with backtesting maker strategies, we test in a real-world live trading experiment, in which we turned over 1.5 million USD in notional volume. Lending additional confidence to our models, and by extension to the features they are based on, the results indicate a significant improvement over a naive benchmark strategy, which we also deploy in a live trading environment with real capital, for the sake of comparison.
\end{abstract}

\newpage

\tableofcontents

\section{Introduction}

Cryptocurrency markets have seen an explosion of trade volumes over the past year, as the price of Bitcoin soared to an all time high of over 60,000 USD in April 2021, sparking a great deal of interest both from the broad public and academics alike.
The trading landscape for Bitcoin in particular has matured considerably over recent years,  with an ever-greater proportion of trade volume occurring in complex derivatives rather than spot (fiat for Bitcoin) markets.
With the rise of derivatives volumes came an accompanying flurry of academic works that investigate their role in the price formation process.
For instance,~\cite{ALEXANDER2020100776,Alexander_2019} examine the price impact of a set of popular unregulated derivatives exchanges,~\cite{HUNG2021107} performs a similar analysis, and~\cite{Aleti_2020} surveys market microstructural differences between spot and futures markets.
Further work such as~\cite{Soska_2021} provides insights on the way in which the proliferation of derivatives has exacerbated price jumps and thus contributed to greater volatility.

One of the main distinguishing elements of Bitcoin markets compared to traditional markets for stocks or bonds, aside from the much greater volatility for the former, is the considerable degree of \textit{fragmentation} inherent in the Bitcoin trading landscape.
This fragmentation stems from the existence of 5-10 highly liquid exchanges where the majority of trade volume takes place. Within each of these exchanges, the volume is typically further distributed across a number of different liquid instruments with slightly different specifications and properties. Exchanges operate independently from one another.
They are subject to different regulatory environments depending on where they base themselves out of. They often use different sever locations, resulting in cross-sectional arbitrages that can persist for at least as long as the time of information transfer between two distinct server locations. There is no cross-collateralization across exchanges, unlike in traditional markets where prime brokers play this role. 
That is, if a trader enters a long position on one exchange and an equally large short position on another exchange, he is market neutral but must still ensure that both positions are sufficiently collateralized in order to avoid a liquidation event on either exchange.


\begin{tcolorbox}[top=1mm, bottom=1mm, left=1mm, right=1mm]
\paragraph{Summary of main contributions. }  
The above factors, and others discussed in subsequent sections, altogether contribute to the aforementioned fragmentation of liquidity across exchanges. This type of environment raises a number of interesting questions one can pose. The main contributions of our work is to identify possible answers to the following questions.
\begin{enumerate}
\item Where does new price information tend to arrive first? In other words, which venue is most often the originator of price transmission?
\item What are the predominant directions in information flow?
\item In what way exactly is price on one venue affected by the arrival of new information on another market?
\item To what extent do time discrepancies in price impact enable price prediction on each venue? That is, can we leverage the cross-section of information from all markets to make reliable price predictions?
\item Can these predictions be leveraged to produce trading strategies in a natural way,  which give large PnL values? Do we obtain predictive power sufficient to produce ``alpha" in excess of transaction costs? 
\end{enumerate}

Our paper represents an addition to the rapidly growing body of literature on Bitcoin price formation in at least the following two ways. First, it is the first of its kind to comprehensively examine the consequences of the aforementioned fragmentation of the Bitcoin trading ecosystem. Our study encompasses the largest markets in terms of trade volume while, to the best of our knowledge  all other studies focus on the impact of either a single market or a limited subset of markets. Often such subsets are made up of illiquid spot markets or for example CME Bitcoin futures markets, which have little bearing on the price due to their lack in trade volume. Second, unlike other studies, we use extremely granular limit orderbook data which allows us to produce predictive models rather than merely explaining contemporaneous returns.

\end{tcolorbox}

\paragraph{Paper outline.} To give the reader a preview of how this is achieved, we first lay out the structure of this paper.  

\bb To begin with, Section~\ref{sec:2} provides necessary background knowledge pertaining to the Bitcoin trading ecosystem. This involves descriptions of the predominant markets and their trade volumes, along with some relevant information on their historical development. We will also describe relevant facts on execution cost,  before giving the reader an overview of the data sets that underlie the analyses presented in this paper. This includes a description of our method for its procurement and explanations of the processing steps we employed to render the data amenable to  subsequent analyses. 

\bb Next, in Section~\ref{sec:3} we define a set of microstructural features based on orderbook and trades information that build on notions of order flow imbalance~\cite{cont2014price} and price differences. The features are designed to quantify and make precise the vague notion of ``informational flow" mentioned in the above questions. We additionally undertake a number of preprocessing steps in order to render the features more suitable as inputs to the linear models that will be trained in later sections.

\bb In Section~\ref{sec:4} we begin to reap the rewards of our prior work. The previously defined feature set will be employed to answer the first three questions listed above. That is, we will investigate cross-impact between markets and derive various networks that illuminate the lead-lag relationships present between pairs of markets. We will then examine how a market's leadingness or laggingness translates to trading realities by defining natural trading strategies and performing PnL calculations. This will involve a foray into the crucial role played by transaction cost in determining not only PnL but also a market's proneness to being a leader or a lagger. 

\bb Section~\ref{sec:5} is devoted to the last two questions from our above list.
We will compare three different methodologies for developing powerful linear predictive models that make use of all available information.
These methodologies will be compared along the two axes of explanatory power and PnL.

\bb In Section~\ref{sec:6} we will address a follow-up question that arises from the results of the previous section.
The strategies we initially define for Sections~\ref{sec:4} and~\ref{sec:5} employ only taker orders (market orders or limit orders that lead to immediate execution) and we will find that they yield particularly low PnL values on markets where the taker fee is large.
But such markets typically also give large rebates to maker orders (passive limit orders).
This motivates the question of whether we can devise a natural maker strategy on the basis of our linear models and whether these strategies yield positive PnLs on markets with large maker rebates.
We will define such a strategy and test its performance on one market by means of a live trading experiment since, as it turns out, testing market making strategies on historical data is virtually impossible.

\bb Finally, in Section~\ref{sec:7} we conclude by summarizing our findings and discussing future research direction.

\section{Background}
\label{sec:2}

This section provides background information on various aspects pertaining to the structure of crypto markets. It is necessary to have a full contextual understanding,
in order to be able to appropriately describe results in later sections.

\subsection{Historical Developments and Market Specifications}

Crypto markets have matured considerably in recent years.
This can be seen by the increasing dominance of derivatives relative to spot volumes.
Nowadays, average daily volumes on derivatives markets eclipse those on spot markets by a factor of greater than five, whereas up until (and including) the year 2018 spot markets had greater average trade volumes~\cite{cmu_bitmex_study,cryptocompare_volume_report}.
On a spot market, traders transact the asset itself, while on derivatives markets traders exchange contracts whose value depends in some pre-specified way on the price of an underlying asset.
For instance, on a BTC/USD spot market buyers and sellers trade ``physical" Bitcoin for USD.
On the other hand, on derivative markets whose underlying is BTC/USD, traders can buy or sell contracts whose value derives (hence the name ``derivative") from that of the BTC/USD spot price. Derivatives provide two main advantages for traders. 
First, they greatly enhance the expressability of traders' opinions on the future price of, for instance, Bitcoin.
Traders that believe the price of Bitcoin will decline in the future can use derivative contracts to profit from the anticipated decline by purchasing a contract whose value rises as the price of Bitcoin declines.
The second advantage is that derivative contracts give traders the ability to use leverage, meaning that they can \emph{leverage} a small amount of initial capital to increase their buying power.
With the use of leverage comes the additional risk of forced liquidation.
Simply put, with leverage, traders can take extra risk for the chance of extra reward, which substantially boosts capital efficiency.
For more details on the mechanics of leveraged trading in crypto markets, we refer to~\cite{binance_leverage_trading}.
It is generally to be expected that the aforementioned advantages of derivatives, namely improved capital efficiency and greater expressability of views on price development, give rise to a more efficient marketplace.

The easy access to leverage and its accompanying increase in capital efficiency can, however, also bear some disadvantages.
This is particularly true for crypto markets, where the amount of leverage that many of the unregulated exchanges allow traders to employ is almost comically large when compared to traditional and regulated financial markets. It is not uncommon for exchanges to allow 100x leverage, meaning that traders of derivative contracts can increase their buying power by a factor of 100~\cite{bitmex_perpetual_guide,binance_leverage_trading}.
By comparison, regulated marketplaces of traditional assets such as stocks, bonds, and ETFs typically do not allow more than 5-10x leverage, with some prime brokers allowing leverage up to 20x on some exchanges (as in London and Tokyo Stock Exchanges). 
With such large leverage, however, comes considerable risk of ruin.
For example, a trader who holds a 100x leveraged position will lose his entire initial capital if he experiences an adverse price move of just 1\%.
A common theme in crypto markets is that of \emph{cascading liquidations}.
To illustrate what this means, suppose for instance that a large number of traders hold highly leveraged long positions.
This means that a relatively small price move down can cause their positions to be forcibly liquidated by the exchange, resulting in a great deal of unnatural selling.
This artificial (and unintentional) selling pressure drives the price down further which might result in more sell liquidations, and so on.
Extremely high leverage is akin to gunpowder in the powder keg, and a small initial price move acts like the match to light the keg, triggering a cascade of liquidations which greatly amplify the initial small price drop to an unnaturally large one.
The historical price development of Bitcoin is littered with interesting case studies of this dynamic playing out in real life. For more details, we refer the reader to~\cite{Soska_2021}.

\paragraph{Our selection of markets.} When we use the term \textit{``market"}, we are referring to a pair of \textit{exchange} and \textit{symbol}, where a symbol on a given exchange denotes either a spot market (for instance BTC/USD) or a derivative contract (for example a quarterly futures contract).

\begin{figure}[!ht]
\hspace*{-2cm}
\vspace{-3mm}
\centering
    \includegraphics[scale=0.48]{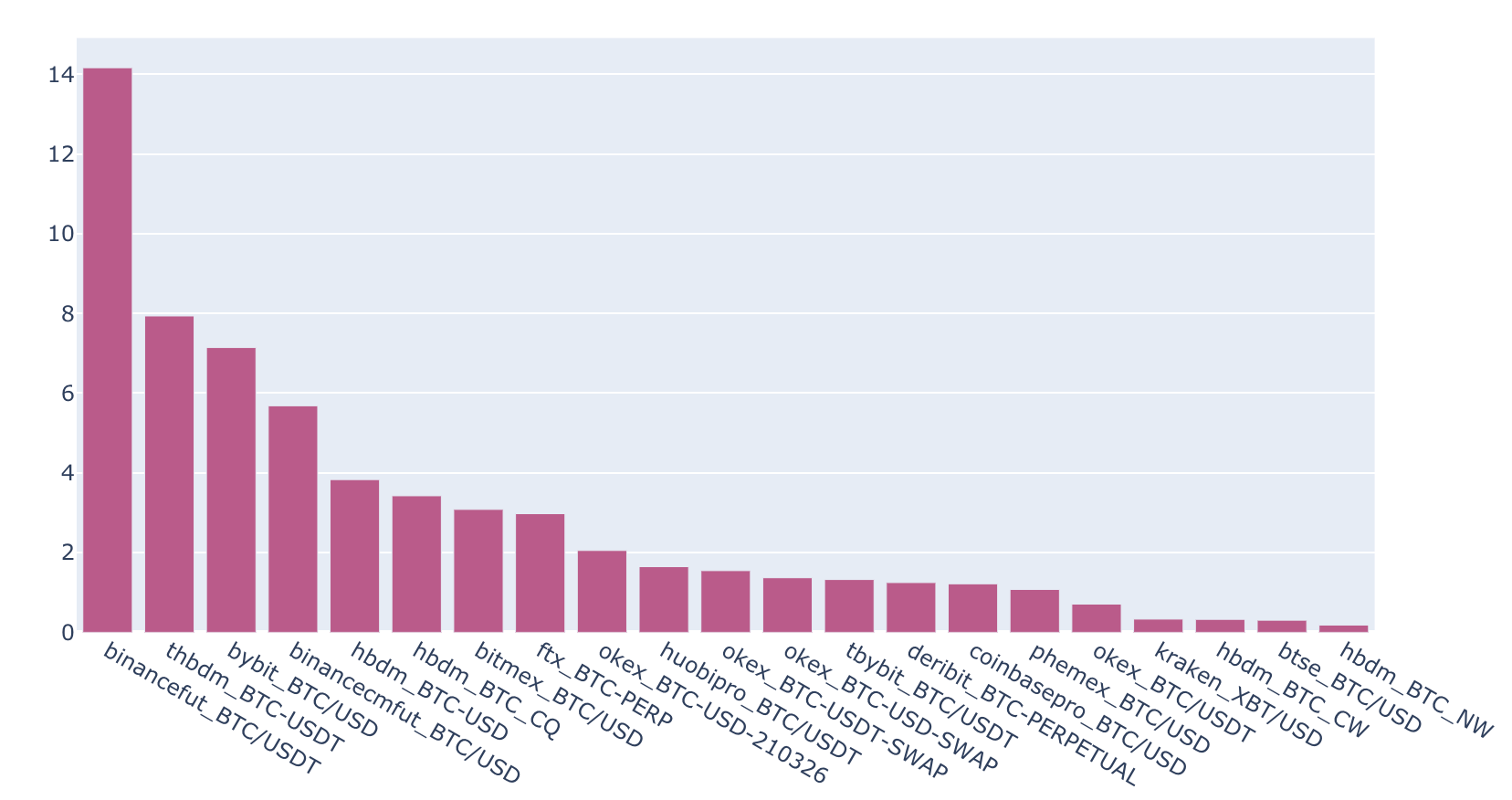}
    \caption{Average daily trading volume in billion USD per market in February 2021}
    \label{fig:feb_24h_volumes}
\hspace{-0.9\textwidth} 
\vspace{-3mm}
\end{figure}

New information marks its arrival on the markets as a whole in the form of traded volume.
That is, when a trader possesses knowledge which she/he believes is not yet reflected in the current price of an asset, they will make a trade that contributes towards closing the gap between the asset's current price and its estimated value. 
By this logic, if we compare two Bitcoin markets, the one with the larger average trade volume will have a greater importance to the Bitcoin trading ecosystem, since it is more often the venue on which new information initially arrives. Guided by this rationale, we choose as the basis of this paper those Bitcoin markets which have among the highest trading volumes. In Figure~\ref{fig:feb_24h_volumes}, we show a ranking of major Bitcoin markets by their average daily volumes over the course of February 2021.
In the remainder of this paper, we will chiefly concern ourselves with the top 14 markets from this bar plot. Next, we provide further context on each of these markets along with their specifications.

\paragraph{Market Specifications.} Each of the markets ranked in Figure~\ref{fig:feb_24h_volumes} can be classified as either a \textit{spot} market, a \textit{futures} contract market, or a so-called \textit{perpetual contract} market (sometimes also called a perpetual swap, or simply a perpetual). The latter market type was invented by the crypto exchange Bitmex in 2014, and, to the best of our knowledge, does not exist outside of the crypto trading world at the time of this writing (July 2021). From the previously mentioned three types of markets, the perpetual swap is the most popular among crypto traders and it has by far the largest trade volumes. Since its inception, the success and broad popularity of the perpetual contract has compelled many other exchanges to follow the blueprint originally laid out by Bitmex. Competing exchanges such as Binance, Huobi, Okex, FTX, Deribit, Bybit, and Kraken have either directly copied this derivative contract or launched a nearly identical product with some minor modifications. The combined trade volume across these perpetual contracts routinely exceeds 50 billion USD per day~\cite{ftx_volume_monitor}.

Given the large role this derivative product plays in the crypto markets, it is worthwhile giving some background on its mechanics. 
Loosely speaking, the perpetual swap can be described as a ``perpetually expiring futures contract" (hence its name). It allows traders to enter leveraged long or short positions without having to pay much attention to pricing a futures premium, since the perpetual swap is designed to closely track the price of a spot market (or a basket of spot markets). The mechanism by which the close link to the spot price is achieved is the so-called ``funding rate". This is a payment exchanged every eight hours (in case of the Bitmex and most other perpetual contracts) between the longs and the shorts. 
The magnitude and sign of this payment depends on the deviation of the perpetual's price from the spot price over the preceding eight hour time window. 
For instance, if the price of the perpetual over such an eight hour time window was, on average, greater than the spot price, then the longs must make a funding payment to the shorts. 
The size of this payment depends on what the average deviation of the perpetual from the spot market was. 
The more the perpetual strayed above the spot price, the larger is the cost incurred by the longs.  
The introduction of this additional cost to the longs (and, in this case, rebate to the shorts) incentivizes traders to short the perpetual contract and it disincentivizes traders to open long positions. 
These twin incentives, when acted upon, will cause the price gap between the perpetual and spot markets to close.
The reverse case where the price of the perpetual over an eight hour time window strays \emph{below} that of the spot market is of course completely analogous.
In this case, the longs will receive a rebate while the shorts incur a cost.
More details on how exactly the perpetual swap functions can be found in~\cite{bitmex_perpetual_guide}.

An important property of Bitcoin derivative contracts is the type of initial capital (also known as \textit{margin} or \textit{collateral}) with which the derivative can be purchased.
This is typically either Bitcoin, USDT (also known as Tether), or another so-called \emph{stablecoin}.
A stablecoin, of which USDT is the most prominent example, is a cryptocurrency whose value is (meant to be) pegged one-to-one to the value of a US dollar.
At the time of this writing in July 2021, the market capitalization of Tether is around 61 billion USD.
This provides an indication of the value that crypto traders place on being able to transact and store wealth in a quasi-USD asset which is void of many of the inconveniences associated with transacting actual USD, such as having to deal with slow bank transfers, stringnent identity verification procedures, and so on.
For more details on stablecoins in general and Tether in particular, we refer the reader to~\cite{GRIFFIN_2020}.
The FTX perpetual is unique among our set of 14 markets in that it allows positions to be collateralized with various stablecoins, actual USD, Bitcoin, Ethereum and many other tokens.

The underlying index of the derivative contract is typically computed as a (weighted) average value of a basket of BTC/USD spot markets or BTC/USDT spot markets.
It is worth pointing out that the BTC/USD price can deviate substantially from the BTC/USDT price under certain circumstances. 
For example, this can arise when investors lose confidence in the supposed peg of USDT to the US dollar, in which case the value of a USDT will decline. 
On the other hand, when there is particularly high demand for USDT (due to the conveniences associated with transacting quasi USD on a blockchain), investors might be willing to pay a premium for USDT. 
The exact composition of the baskets making up the index can vary from exchange to exchange.
The BTC/USD indices typically include at least the Coinbase, Bitstamp and Kraken spot markets, while the BTC/USDT indices typically include spot markets from Binance, Huobi, and Okex. 
The precise mechanism via which the index price is computed from its constituents is generally also different between exchanges. For instance, some exchanges exclude the markets with the minimal and maximal price in the calculation, in order to achieve greater robustness of the index price and resistance to price  manipulations~\cite{deribit_doc}. 
The fact that trade volumes on derivative markets far exceed those on the spot markets which make up the derivatives' underlying indices can create opportunities for market manipulations.
Colloquially speaking, the much larger volumes on derivative markets relative to spot markets is sometimes referred to as the ``tail wagging the dog" phenomenon. 
These realities necessitate a certain robustness in the calculation of the index price in order to have resistance to market manipulations. 
To give a hypothetical example of such a manipulation, suppose the index of a futures contract was made up of just one illiquid spot market, where a relatively small trade can cause a big price move while the futures contract is extremely liquid. 
A trader might then get the idea to, for instance, open a massively leveraged long position in the futures contract (which can be done cheaply due to its liquidity) and manipulate the price of the underlying index upward shortly before the expiration of the futures contract. The trader could then profit handsomely from their leveraged long futures position when it expires at the artificially inflated price index price.
These profits could be far in excess of the cost for performing this manipulation,  which consists mainly of the execution cost associated with the ``sloppy" manipulative buy orders on the illiquid spot market (which can be expected to be subject to high slippage cost). While this example was a hypothetical one, actual manipulations that resemble the above have actually occurred in reality.
An interesting case study can be found in~\cite{coinmetrics_market_manipulation}.

In Table~\ref{tbl:market_specifications}, we list some key properties of the markets  considered in the remainder of this paper.
We see that the 14 most liquid markets from Figure~\ref{fig:feb_24h_volumes} include two futures contracts, 11 perpetuals, and only one spot market.
The most liquid futures markets tend to be the ones whose expiry happens at the end of a quarter.
Huobi and Okex, the exchanges whose quarterly futures appear in our set of markets, both list a weekly expiration, a bi-weekly one, and a quarterly one.
Okex additionally offers a bi-quarterly futures contract. 
\setlength\LTleft{-1cm}
\begin{table}
\small  
\centering
\caption{Properties of markets under consideration in this paper}
\label{tbl:market_specifications}
\begin{tabular}{llllllll}
\toprule
{} &       Type & Exchange &    Symbol & Max. Leverage & Margin &      Expiration & Underlying \\
\midrule
hbdm\_BTC\_CQ           &    Futures &    Huobi &         - &          125x &             BTC &  Quarterly &    BTC/USD \\
okex\_BTC-USD-210326   &    Futures &    Huobi &         - &          125x &             BTC &  Quarterly &    BTC/USD \\
hbdm\_BTC-USD          &  Perpetual &    Huobi &         - &          125x &             BTC &               - &    BTC/USD \\
bitmex\_BTC/USD        &  Perpetual &   Bitmex &         - &          100x &             BTC &               - &    BTC/USD \\
deribit\_BTC-PERPETUAL &  Perpetual &  Deribit &         - &          100x &             BTC &               - &    BTC/USD \\
bybit\_BTC/USD         &  Perpetual &    Bybit &         - &          100x &             BTC &               - &    BTC/USD \\
okex\_BTC-USD-SWAP     &  Perpetual &     Okex &         - &          125x &             BTC &               - &    BTC/USD \\
ftx\_BTC-PERP          &  Perpetual &      FTX &         - &          101x &             Cross &               - &    BTC/USD \\
binancecmfut\_BTC/USD  &  Perpetual &  Binance &         - &          125x &             BTC &               - &    BTC/USD \\
binancefut\_BTC/USDT   &  Perpetual &  Binance &         - &          125x &            USDT &               - &   BTC/USDT \\
okex\_BTC-USDT-SWAP    &  Perpetual &     Okex &         - &          125x &            USDT &               - &   BTC/USDT \\
tbybit\_BTC/USDT       &  Perpetual &    Bybit &         - &          100x &            USDT &               - &   BTC/USDT \\
thbdm\_BTC-USDT        &  Perpetual &    Huobi &         - &          125x &            USDT &               - &   BTC/USDT \\
huobipro\_BTC/USDT     &       Spot &    Bybit &  BTC/USDT &             - &               - &               - &          - \\
\bottomrule
\end{tabular}
\end{table}
In the remainder of this paper, we shall refer to \bnf  as the ``Binance USDT-margined perpetual contract" or simply the ``Binance USDT perpetual". 
Similarly, \bnfcm shall be referred to as the ``Binance BTC-margined perpetual contract" or for short the ``Binance BTC perpetual". Other markets will be described in the same fashion  using their type, exchange, and when necessary, their margin currency.

\paragraph{Fake volume in crypto markets.} For the purposes of this study, we aimed to identify the set of the most liquid Bitcoin markets which account for the majority of Bitcoin trading volume. However, the task of determining what exactly these markets are can prove more challenging than one might initially suspect. 
The reason for this is the prevalence of fake volume.
The study~\cite{bitwise_fake_volume} by Bitwise from 2019 asserts that 95\% of reported volume in crypto markets is ``fake". The precise meaning of the word ``fake" in this context is somewhat ambiguous. It can for example simply mean that an exchange is reporting a trade volume value that does not correspond to the sum of volumes of actual trades processed in the matching engine. Volume can also be ``fake" in the sense that an exchange is printing wash trades where the exchange itself is engaging in self-trading. In other cases, exchanges have been known for partnering with market makers and incentivizing them to self-trade with certain type of rebates and bonus payment. 

The reasons for exchanges to fake their volume can be manifold. 
A new exchange might use it as a method of bootstrapping a user base.
Traders tend to trade on exchanges that are used by many other people and which therefore have large trade volume. If an incorrectly reported volume figure convinces a user to start trading on that exchange, then the exchange has successfully gained a new customer.
From the exchange's perspective, the cost of such practices is a loss of credibility which can negatively impact its long term success.

In most cases, it is straightforward to spot fake volume because exchanges tend not to be very sophisticated about the methods they employ to achive this. 
Some methods border on the comical, such as printing a trade at mid-price periodically. 
Another method used by some exchanges in the past was to report a 24th trade volume figure which is different from the one which can be obtained by summing up the volume of all trades printed on the exchange. For a more exhaustive list of the various ways employed by exchanges to fake volume, we refer the reader to~\cite{ftx_volume_report} that presents a detailed study of the subject. 

Seasoned traders generally know from experience where the real volume resides and are therefore able to steer clear of the many pitfalls of fake volume.
An excellent resource in this regard is~\cite{ftx_volume_monitor} which provides adjusted volume figures where reported volumes were purged of likely fake volume when applicable.

Our choice of markets for this study was made with the aforementioned caveats in mind, and on the basis of our prior and ongoing trading experience. We are confident to a high degree that the 14 markets which form the foundation of this study are not among those which are subject to fake volume.
Moreover, any volume figures reported in this study are based on our own calculations using data on individual trades published by the exchange's real time data feeds. We do not rely on values reported by the exchange, although we do find agreement between these figures and the  ones we independently calculated.

\subsection{Data and Preprocessing}

Our data set is comprised of orderbook and trades data. Both of these sources of data were mined by subscribing to the exchange's websocket API orderbook and trades endpoints for each of the markets under consideration in this paper. Detailed information on what exactly a websocket API is can be found in~\cite{wiki_websocket}, though we will give a cursory description below. The data we obtained in this way was compressed and placed into file storage for later processing. 

Websocket API endpoints generally provide the most granular crypto market data that one can procure. These endpoints push new data to the subscriber in real time. For trades data, this means that one receives information about all executed trades on the market to which one is subscribed. In the case of orderbook data, exchanges typically publish orderbook snapshots every 5-200ms depending on market activity. Note that we therefore do not have order-by-order information (typically referred to as Level 3 data), since exchanges batch changes over a short previous time window and then publish these updates as part of a new orderbook snapshot. The frequency at which new orderbooks arrive not only depends on market activity but it also varies from exchange to exchange. In some cases, there is even variation in the update frequency within an exchange when a market has multiple websocket API endpoints with slightly different properties. For instance, the Bybit BTC-margined perpetual offers a feed that publishes a new snapshot containing 25 orderbook levels every 20ms, while at the same time offering another slower feed which publishes updates every 100ms but provides information for the top 200 orderbook levels~\cite{bybit_websocket_doc}. A common minimum update frequency used by exchanges (even when they do not document this) is around 20ms. That is, orderbook updates arrive at a frequency not higher than one per 20ms, even when market activity is extreme. Exceptions are Deribit and Bitmex where update frequencies are sometimes in the single digit milliseconds when market volatility is high.   

There is little uniformity across exchanges in the properties of their data feeds.
That is, the number of orderbook levels varies from exchange to exchange (and even from market to market), the update frequencies are different, and the data comes in a different file format for each exchange. For each market in the set of 14 markets we consider in this paper, we receive between 25 and 75 orderbook levels. In our choice of the websocket API feed, we generally preferred speed over orderbook depth. A single day of orderbook data for all markets combined is over 30 gigabytes in size. Creating uniformity and comparability within our large and disorganized data set posed a major challenge. 

Furthermore, since the data is gathered asynchronously, timestamps of orderbooks generally differ. In order to more easily compare orderbooks across markets, we resample the data to a 50ms frequency using the last seen observation~\cite{pandas_resample_last}.
Missing values can appear when a market does not publish an orderbook update for, say, 100ms.
In this case, we fill the missing value by propagating the last valid observation forward.
For the sake of consistency, we also resample trades data to the same 50ms frequency. 
After applying this processing step, the buy (or sell) amount in a given 50ms window will be the sum of all buy (sell) trade amounts that occurred in this time window. The trade amounts can have different denominations on different markets. For example, for certain perpetual contracts, the value of one contract is 1 USD while on others it might be 100 USD or 0.001 Bitcoin. In order to have an ``apples-to-apples" comparison between trade volumes across markets, we normalize all trade amounts to a USD amount, using the last seen Bitcoin price to convert amounts in Bitcoin to USD amounts. 

Since we are working at a 50ms granularity (our resampling frequency), we shall define the trade price over a 50ms period as the volume-weighted average price computed over all trades (irrespective of side) which took place during this 50ms window. If no trade occurs during a time window, we fill the missing price value, as before, with the previous valid price observation.

With these preprocessing steps, successive samples in our data set are equally spaced in time (50ms apart), making cross-market comparisons easier. For each observation, we have one orderbook per market (potentially with differing numbers of levels), and information (buy amount, sell amount, average price) about a ``meta trade" on each market. This is our data set in its final form, which will be the foundation for any further analysis.

\subsection{Remarks on Execution Cost}

The cost of execution of a taker order (one which leads to immediate execution) is composed of two different components.
The first component is the taker fee charged by the exchange.
Fees exhibit great variation across exchanges and even across markets within a single exchange. For instance, the taker fee for a derivatives contract tends to be, by and large, substantially lower than that of a spot market. A number of exchanges additionally offer  tiered fee structures, where certain groups of traders receive lower fees than others. 
These fee rebates can be very substantial. Traders can ascend to a better fee tier by having high transaction volume. The volume cutoffs dictating a trader's fee tier are different from market to market. Being in the top tier typically requires on the scale of 500 million to 1 billion USD of monthly notional volume. Certain exchanges (Binance, FTX, and Huobi) have issued their own cryptocurrency and offer additional fee discounts for traders that hold this cryptocurrency in their exchange wallet.
The magnitude of this additional fee discount is usually proportional to the holdings in the exchange's native cryptocurrency. An alternative (often undocumented and under-the-table) way of obtaining a preferred fee status is through the formation of strategic partnerships with exchanges. Particularly newly launched exchanges with little liquidity are typically interested in attracting liquidity, which they can achieve, for example, by partnering with market making firms whom they incentivize to provide the desired liquidity by offering various fee rebates. Table~\ref{tbl:taker_fee} displays the lowest possible (documented) taker fee that a trader in the top fee tier (who also owns a sufficient amount of the exchange's native cryptocurrency, when applicable) receives. 

\begin{table}
\centering
\caption{Lowest possible and default taker fees per market in basis points}
\label{tbl:taker_fee}
\begin{tabular}{lll}
\toprule
{} & Lowest Possible Fee & Ordinary Fee \\
\midrule
ftx\_BTC-PERP          &                 1.5 &            7 \\
binancefut\_BTC/USDT   &                1.53 &            4 \\
binancecmfut\_BTC/USD  &                 1.8 &            5 \\
huobipro\_BTC/USDT     &                1.93 &         4.75 \\
hbdm\_BTC\_CQ           &                   2 &            4 \\
okex\_BTC-USD-210326   &                 2.5 &            5 \\
thbdm\_BTC-USDT        &                 2.7 &            4 \\
okex\_BTC-USD-SWAP     &                   3 &            5 \\
okex\_BTC-USDT-SWAP    &                   3 &            5 \\
hbdm\_BTC-USD          &                 3.7 &            5 \\
deribit\_BTC-PERPETUAL &                   5 &            5 \\
bitmex\_BTC/USD        &                 7.5 &          7.5 \\
bybit\_BTC/USD         &                 7.5 &          7.5 \\
tbybit\_BTC/USDT       &                 7.5 &          7.5 \\
\bottomrule
\end{tabular}
\end{table}
Notice that these heavy fee rebates for certain ``VIP" traders lead to a wealth concentration whereby VIP traders, who presumably have some measure of wealth by virtue of the requirements they must have met to be a member of the best fee tier, are capable of turning greater profits (or trading profitably at all) in their trading since they incur far less fee cost than, for instance, a novice trader would.

\begin{figure}[!ht]
\vspace{-3mm}
\centering
\includegraphics[scale=0.48]{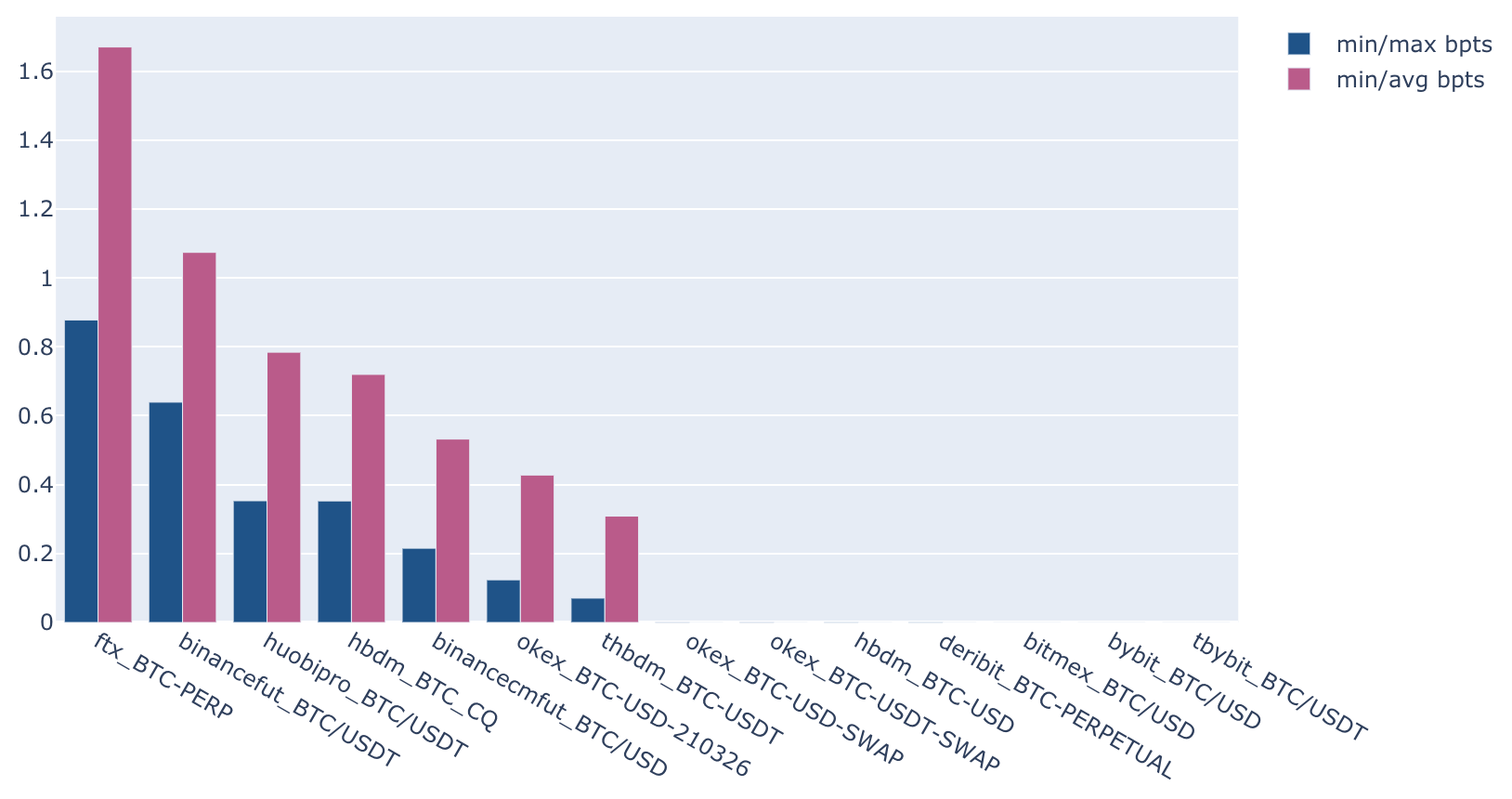}
\caption{Price impact of taker orders of size $>100$k USD.}
\label{fig:trade_impact}
\hspace{-0.9\textwidth}
\vspace{-3mm}
\end{figure}

The second component of the execution cost can be broadly described by the term ``spread".
Since the aggressor at least crosses the bid-ask spread, the minimum spread cost is at least the bid-ask spread at the time of execution. If the size of their order is greater than the liquidity available at the first orderbook level which is aggressed against, the trader additionally ``walks the book". That is, part of her/his execution occurs at one or multiple prices which are worse than the top quote. In Figure~\ref{fig:trade_impact}, we provide an illustration of the size of the spread cost incurred by an aggressor. 
In this plot, the set of blue bars represents the median difference in bpts between the minimum and the maximum price of a single order, while the red set of bars represents the median difference in bpts between the best price (minimum for buy orders, maximum for sell orders) and the average price of a single order.
The sample consists of all orders of size $>100$k USD between February 15--27, 2021.
It is interesting to note that there seems to be an inverse relationship between a market's VIP taker fee and the price impact (in other words, spread cost) of moderately large to large orders (greater than 100k USD). 
Writing 
\begin{equation}
  \text{execution cost} = \text{taker fee} + \text{spread},
\end{equation}
it appears that the two terms in this equation balance one another out, in a sense.
If a market has a high taker fee, the orderbook appears to be more liquid and hence the spread cost is small. On the other hand, a market with a low taker fee has little liquidity at or near the top of the book, and hence the spread cost is large. We will later see variants on this theme reappear, when we perform PnL calculations that take fees into account.

\section{Features and Preprocessing}
\label{sec:3}

\subsection{Base Features}

The fundamental microstructural quantity that determines price movements in a market that employs a limit orderbook is, broadly speaking, \textit{order flow} which comes in two forms.
The first one is aggressive, where traders submit market orders or limit orders that lead to immediate execution.
The second one is passive, where market participants submit limit orders that do not lead to immediate execution or where such orders are canceled. 
Our objective in this subsection is to develop a set of base features that build on order flow, and which achieve high efficacy for certain tasks tackled later in this paper, such as predicting future returns or characterizing cancellation behaviour of HFT market makers.
We shall discuss our base features one by one, and provide a motivation and justification for their inclusion. In later subsections, we discuss ways of transforming and optimizing the base features for the tasks that lay ahead.

\subsubsection{Orderbook Imbalances}

It is well established in the literature that the shape of the orderbook has significant impact on the distribution of future returns~\cite{Bouchaud_2002, Alfonsi_2009,cont2014price,Bouchaud2008HowMS}. The intuitive reason for this is straightforward: if the liquidity on the bid side far outweighs that of the ask side, the volume required for a price move up is much smaller than the volume required for a price move down.
Thus, if we assume that the arrival of a sell order is not much more likely than the arrival of a buy order, the probability of a future price move up is larger than that of a price move down.
Clearly, similar reasoning applies to the dual case where the orderbook shape takes the opposite form.

The empirical evidence cited above provides ample justification for the inclusion of an orderbook imbalance feature in our set of features. However, the question remains of the precise notion of orderbook imbalance that is most appropriate to our setting.

The classical notion of orderbook imbalance is given by 
\begin{equation}
\frac{b-a}{b+a} \in [-1, 1], 
\label{eqn:ob_imb}
\end{equation}
where $a$ and $b$ represent the top of the ask and bid liquidity, respectively.
However, this quantity suffers from an obvious drawback, namely that it only takes into account the liquidity at the two tops of the book. In other words, information from deeper orderbook levels is discarded. This is particularly problematic for markets where the liquidity at the top of the book tends to be very small and instead concentrates more strongly around deeper orderbook levels. 

Extensions of the classical orderbook imbalance~\cite{Xu_2019} typically try to account for liquidity on deeper levels by means of a weighting scheme which replaces the quantity $a$ in Eqn.~\eqref{eqn:ob_imb} by $\sum_{j} w_j a_j$, where $a_j$ denotes the volume on orderbook level $j$ and $w_j\in \mathbb R_{>0}$ is a weight.   The quantity $b$ is of course replaced in an analogous manner.
The weight $w_j$ is usually constructed in a way which reflects the probability of execution of an order on level $j$.
It can, however, be challenging to get a handle of execution probability at each orderbook level.
Conventional models of orderbook dynamics imply that the probability of execution is an exponentially decaying function of the distance to the midprice.
Yet, in practice, it is not only the distance to midprice that matters, but also the fee structure (and possibly other factors). 
We saw evidence of this in Figure~\eqref{fig:trade_impact}, where we noted that a low taker fee seems to imply more liberal ``walking of the book" (that is, greater trade impact and hence a larger execution probability on deeper levels compared to markets where the taker fee is large).
This makes it difficult in practice to choose such weights, since they must be chosen individually in a bespoke manner for each market. Further complicating things, different markets often have different tick sizes.
For instance, the tick size on FTX is 1 USD, while Binance uses a tick size of $0.1$ USD.

Another concern with the classical orderbook imbalance (and most of its extensions) is that it does not adequately reflect uncertainty of future returns in times of high volatility when (effective) spreads are large.
\begin{figure}[!ht]
\vspace{-3mm}
\centering
\subcaptionbox[]{  
Bitmex orderbook during non-volatile times. Sizes are in USD.
\label{fig:non_volatile_orderbook}
}[ 0.42\textwidth ]
{ \includegraphics[width=0.35\textwidth, trim=0cm 0cm 0cm 0.0cm,clip] {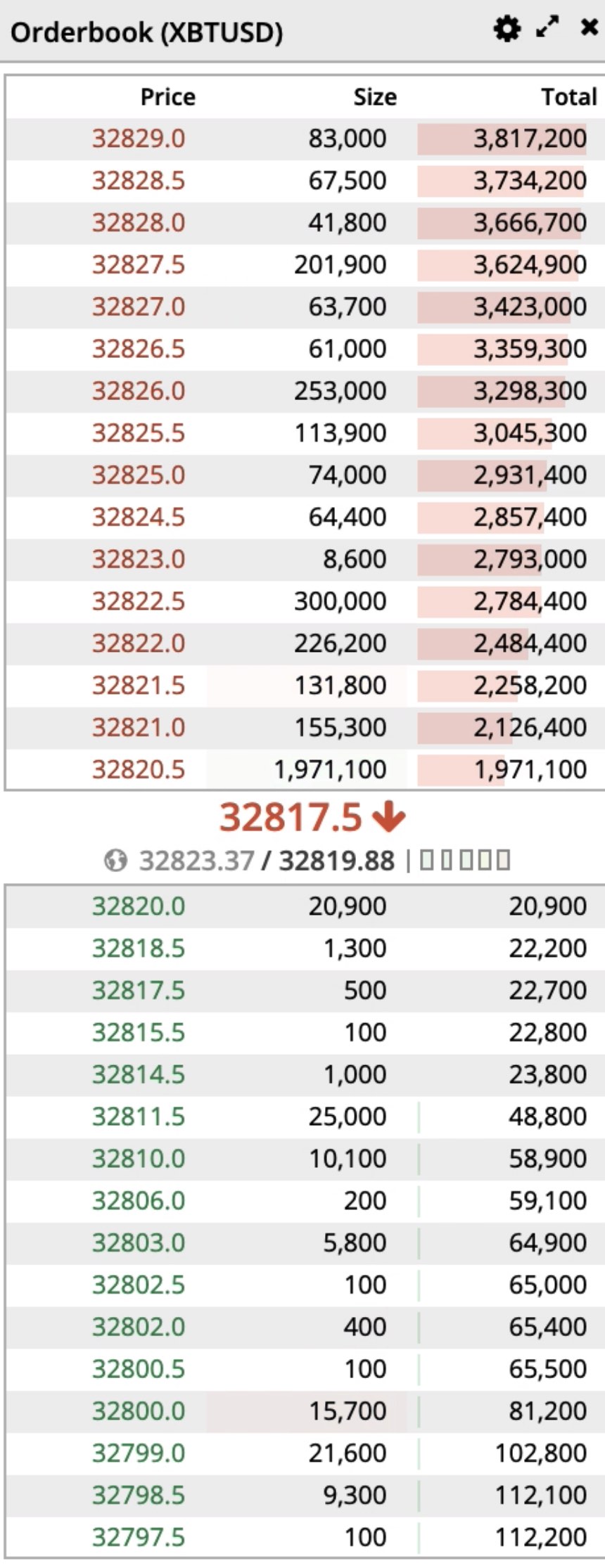} }
\hspace{0.04\textwidth} 
\subcaptionbox[]{ 
Bitmex orderbook during volatile times. Sizes are in USD
\label{fig:volatile_orderbook}
}[ 0.42\textwidth ]
{\includegraphics[width=0.35\textwidth, trim=0cm 0cm 0cm 0.0cm,clip] {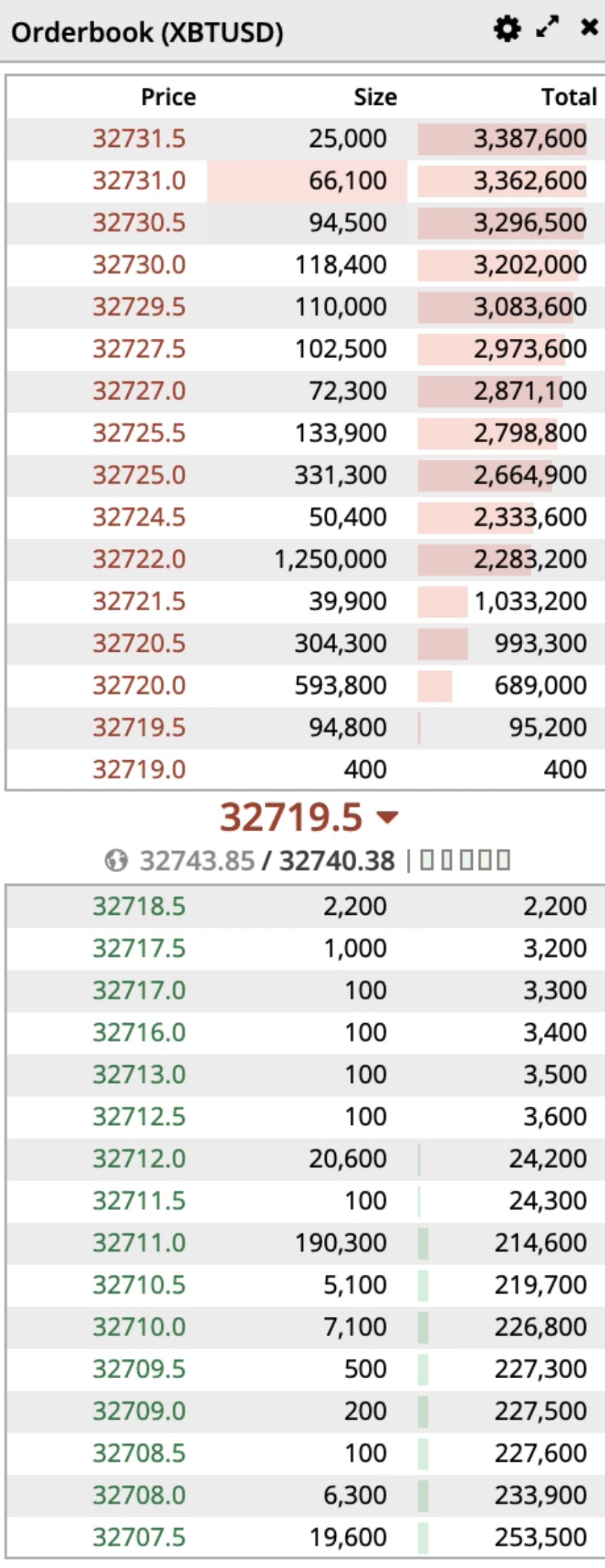} }
\vspace{-2mm}
\captionsetup{width=0.98\linewidth}
\label{fig:bothxyz}
\vspace{-3mm}
\end{figure}
To illustrate this point, consider an example of an orderbook during non-volatile times~\ref{fig:non_volatile_orderbook} and one shortly after a volatility spike~\ref{fig:volatile_orderbook}.
If we compute the classical orderbook imbalance for the first orderbook, we obtain the value $-0.98$ which reflects well what one would intuit from looking at the orderbook: the ask liquidity is large and overwhelming compared to the bid liquidity, hence we would expect, all else being equal, a price move down to be much more likely than a price move up.
The second orderbook gives a classical imbalance of $0.69$ which one would interpret as a relatively strong predictive price signal of a price move up.
However, a glance at deeper levels quickly reveals that also in this orderbook a price move down should be far more likely than a price move up.
However, a glance at deeper levels quickly reveals that the liquidity distribution in this orderbook implies a far greater probability of a price drop than a price rise.
What causes the discrepancy between the computed orderbook imbalance value and our intuition are the minuscule amounts posted at the two tops of the book, which ultimately have little bearing on the distribution of future returns due to their small sizes.
Such small top-of-the-book amounts are a common occurrence during volatility breakouts and they often persist for several hundred milliseconds.
For example, a submission of a mere 5,000 USD to the top ask would result in a vastly different orderbook imbalance of $-0.42$. 
This phenomenon makes the classical orderbook imbalance values extremely noisy, rendering it unsuitable for practical purposes.

We propose an alternative pair of orderbook-based features (one for each of the ask and bid sides) which are more robust during volatility breakouts, and which do not depend on a complicated choice of weights. Let us define the following quantities for each market $i=1,\ldots, 14$ and for fixed quantities $N_1,\ldots,N_{14} \in \mathbb R_{\geq 1}$
\begin{eqnarray}
    IMB^{a,i}_t &:= \left(\frac{p_{a, t}^i(N_i)}{{p_{a, t}^i(1)}} - 1 \right)\cdot 10000,  \\
    IMB^{b, i}_t &:= \left(\frac{p_{b, t}^i(1)}{p_{b, t}^i(N_i)} - 1 \right)\cdot 10000, 
    \label{eqn:imb}
\end{eqnarray}
where $p_{a, t}^i(x)$ denotes the average price one would pay at time $t$ for a market \emph{buy} order of size $x$ on market $i$.
Note that $p_{a, t}^i(1)$ is simply the top ask price.
Similarly $p_{b, t}^i(x)$ is the average price for a market \emph{sell} order of size $x\in \mathbb{R}_{\geq 1}$, so that $p_{a, t}^i(1)$ is just the top bid price.
The quantity $IMB^{a,i}$ can be described as the difference in basis points between the top ask price on market $i$ and the average price of a market order of size $N_i$, and analogously for the bid version $IMB^{b,i}$.

What are sensible choices of $N_i\in [1, \infty)$ for $i=1,\ldots, 14$?
Note that $N_i = 1$ always yields $IMB^{a,i}_t = 0$, while letting $N_i\to \infty$ we have $IMB^{a,i}_t \to \infty$; thus the two extreme ends of the spectrum are void of any signal. We set $N_i$ to be the median liquidity within the top five basis points of the top of the book on market $i$. More precisely, we compute this median value for the ask side and the bid side separately, and then take the average of the two resulting values. Preliminary experiments indicate this to be a sensible choice, in the sense that the resulting features~\eqref{eqn:imb} appear to contain much signal and little noise.
We leave it to future work to calibrate the choice of $N_i$ in a more rigorous fashion.  

Let us revisit the two orderbooks considered above.
The orderbook from Figure~\ref{fig:non_volatile_orderbook} has $(IMB^{a,i}_t, IMB^{b,i}_t) \approx (0, 7)$.
The small $IMB^{a,i}_t$ value means that the ask side is heavily populated, while the large $IMB^{b,i}_t$ points at the sparsity of the bid side. 
From the comparison of these two values, we would unequivocally conclude that a price drop is far more likely than an increase. 
For the orderbook from Figure~\ref{fig:volatile_orderbook} we have $(IMB^{a,i}_t, IMB^{b,i}_t) \approx (1, 7.4)$.
As in the previous case, the far larger $IMB^{b,i}_t$ value suggests a much greater probability of a price decrease, which agrees with our intuition from observing the liquidity distribution in the orderbook, and which is in contrast to the classical orderbook imbalance value that pointed to a greater likelihood of a price increase. 

Moreover, our pair of features retains its high fidelity during highly volatile times when liquidities on both sides of the book are sparse In such cases, the greater uncertainty in the distribution of futures returns is represented by values $IMB^{a,i}_t$ and $IMB^{b,i}_t$ which are both large.

As further empirical evidence of the suitability of the features defined in~\eqref{eqn:imb}, we compare their explanatory power with that of the classical orderbook imbalance.
Specifically, for each market $i=1,\ldots,14$, we fit the following pair of univariate linear models 
\begin{equation}
fret_t = \alpha + \beta \left(IMB^{a, i}_t - IMB^{b, i}_t\right)  + \epsilon_t,
\end{equation}
\begin{equation}
fret_t = \alpha' + \beta' \left( \frac{b_t^i - a_t^i}{b_t^i + a_t^i} \right)  + \epsilon'_t,  
\end{equation} 
where $a_t^i$ and $b_t^i$ are the top ask and bid on market $i$, respectively.
The prediction target $fret_t$ used here consists of 500ms future returns on the Bybit BTC perpetual.
We fixed this target market for the sake of concreteness; other markets give similar results.   
The above linear models are fitted using OLS on training data spanning February 22nd until February 27th.
See Table~\ref{tbl:r2_comparison_classic_vs_new} for a comparison of the corresponding $R^2$ values.
\begin{table}
\centering
\captionsetup{width=0.85\linewidth} 
\caption{Comparison of $R^2$ values of classical orderbook imbalance and our proposed imbalance}
\label{tbl:r2_comparison_classic_vs_new}
\begin{tabular}{lll}
\toprule
{} &     $R^2_{\text{classical}}$ &     $R^2_{\text{new}}$ \\
\midrule
bybit\_BTC/USD         &  0.1176 &  0.1646 \\
hbdm\_BTC-USD          &  0.0893 &    0.13 \\
thbdm\_BTC-USDT        &  0.0957 &  0.1059 \\
tbybit\_BTC/USDT       &  0.0724 &  0.0818 \\
bitmex\_BTC/USD        &  0.0547 &  0.0786 \\
binancecmfut\_BTC/USD  &  0.0228 &  0.0645 \\
hbdm\_BTC\_CQ           &  0.0179 &  0.0425 \\
okex\_BTC-USD-SWAP     &  0.0315 &  0.0358 \\
okex\_BTC-USDT-SWAP    &  0.0103 &  0.0275 \\
ftx\_BTC-PERP          &  0.0157 &  0.0273 \\
huobipro\_BTC/USDT     &  0.0135 &  0.0204 \\
okex\_BTC-USD-210326   &  0.0035 &  0.0159 \\
deribit\_BTC-PERPETUAL &  0.0115 &  0.0155 \\
binancefut\_BTC/USDT   &     0.0 &  0.0061 \\
\bottomrule
\end{tabular}
\end{table}
With the new imbalance measure, we find an average increase in $R^2$ of $0.019$ compared to the classical one. The $R^2$ value corresponding to the classical imbalance is only approximately 58\% that of the new imbalance which, additionally, is higher in every single case.

\subsubsection{Trade Imbalances}

Let us define the so-called \textit{trade flow imbalance} indicators, for each market $i=1,\ldots,14$, computed over time horizons $\delta \in \left\{100\text{ms}, 250\text{ms}, 500\text{ms}, 1000\text{ms}, 2000\text{ms}\right\}$
\begin{equation}
    TFI^{i, \delta}_{t} := B_{[t-\delta, t]}^i - S_{[t-\delta, t]}^i, 
    \label{eqn:tfi}
\end{equation}
where $B_{[t-\delta, t]}^i$ represents the total volume of buy orders on market $i$ in the time interval $[t-\delta, t]$.
Similarly, the quantity $S_{[t-\delta, t]}^i$ represents the total volume of sell orders on market $i$ over the same time interval.

This feature can be construed as the ``aggressive" component (relating to taker flow rather than submissions or cancellations of passive limit orders) of the order flow imbalance indicator defined by Cont et al. in~\cite{cont2014price}
\begin{align}
ofi_t & = \left[ \text{buy flow} - \text{sell flow} \right]_{[t-\delta, t]} \\
& = [ (\underbrace{\text{buy trades - sell trades})}_{\text{``aggressive" component}} + (\text{bid submissions - ask submissions}) + (\text{ask cancels - bid cancels}) ]_{[t-\delta, t]}.
    \label{eqn:ofi}
\end{align}   
The order flow imbalance indicator was demonstrated in~\cite{cont2014price} to have significant explanatory power over \emph{contemporaneous} returns. In ongoing work by a subset of the authors, similar results are demonstrated for future returns, when employing a variation of the order flow imbalance from~\cite{cont2014price}, showcasing the usefulness of this family of indicators.

The fragmented nature of the Bitcoin trading universe means that volumes are scattered across several venues.
This implies an asynchronous information arrival: trade flow typically arrives first on one venue and then ``trickles down" to other venues in the form of cancellations or taker orders. Following this logic, we would expect to also have explanatory power over future returns based on the cross-section of trade flows.
This motivates our inclusion of the trade flow imbalance indicators~\eqref{eqn:tfi} into our set of base features.
Additional remarks are in order.
\begin{itemize}
\item We believe that the aggressive component in~\eqref{eqn:ofi} holds a fundamentally greater importance (more signal) than the remaining two ``passive" terms.
The underlying rationale is that passive flow (submissions or cancellations) are almost exclusively the work of HFT market makers who are largely only reacting to the cross-section of aggressive flow and orderbook imbalances. That is, passive flow tends to be \emph{reactive} in nature rather than contain genuine new information. A rigorous investigation of this topic is left for future work. 

\item Notice that future returns are a dimensionless quantity measured in basis points,  while the unit of trade flow imbalance is USD. Based on this discrepancy of units, one might predict that future returns vary non-linearly on trade flow imbalance. We will see later that this is indeed the case, and will devise a methodology for addressing this concern.

\item One might get the idea to consider a relative measure of the form $\frac{B-S}{B+S}$, where $B$ and $S$ denote buy and sell volume (respectively) instead of the trade flow imbalance as we defined it. However, this relative indicator suffers from some of the same drawbacks as does the classical orderbook imbalance, namely a far noisier predictive signal than its non-relative USD denominated analogue. For instance, if the sell volume is zero, a trivial buy volume of size $1$ USD gives rise to the same imbalance value as a significant buy volume of size, say, 10 million USD.

\item The nature of our orderbook data (snapshots data) makes it impossible to precisely compute the remaining two ``passive" terms that make up the order flow quantity~\eqref{eqn:ofi}. A precise computation requires order-by-order feeds, and is left as future work, for the case of equity data, for which such feeds are available\footnote{To this end, intraday LOBSTER data could be employed, which provides detailed tick-by-tick limit order book data for NASDAQ stocks. 
LOBSTER contains all limit order submissions, cancellations and executions on each trading day. All events are time-stamped with millisecond precision 
\cite{Bibinger2017}, and are reconstructed from NASDAQ's historical TotalView-ITCH data \cite{HuangPolak2011}.}.  
\end{itemize}

\subsubsection{Past returns}

Following a similar logic as above for the trade flow imbalance, it is to be expected that a drastic price change on one market should cross-impact short-term future returns on another market.
It is therefore natural to include past returns on each market in our feature set.
We use the same time horizons here as for the trade flow imbalances.
Formally, for each market $i=1,\ldots 14$ and time horizon $\delta \in \left\{100\text{ms}, 250\text{ms}, 500\text{ms}, 1000\text{ms}, 2000\text{ms}\right\}$, we define the following indicator at each time $t$
\begin{equation}
PRET^{i, \delta}_{t} := \left( \frac{p_t^i}{p_{t-\delta}^i} - 1 \right)\cdot 10000, 
\label{eqn:pret}
\end{equation}
where $p_t^i$ denotes the price on market $i$ at time $t$.
This quantity in~\eqref{eqn:pret} computes the price change in basis points on market $i$ over the time interval $[t-\delta, t]$.
It needs to be clarified precisely what notion of price we are using in $p_t^i$, since there are many possible choices (mid price, last trade price, and so on). 
Our initial choice of last trade price gave rise to ambiguities in the case when a number of trades are executed simultaneously at multiple different price levels. 
This happens for example for ``sweeping" orders which consume multiple orderbook levels. To circumvent this concern, we compute the average price in these cases. 
More precisely, we will use the average price over a short lookback window of 50ms (which is the minimum time step in the preprocessing or our data). 
That is, we define $p_t^i$ as  
\begin{equation}
    p_t^i := \frac{1}{A} \sum_{(a, p)\in T_t^i} a\cdot p,
\end{equation}
where $T_t^i$ is the set of all trades, i.e.\ pairs $(a, p)$ of amount and price, executed on market $i$ in the time period $[t-50\text{ms}, t]$ and $A = \sum_{(a, p)\in T_t^i} a$.
Note that we include both buy and sell trades in $T_t^i$.

\subsubsection{Mean Divergence}

The instruments traded on the markets under consideration in this paper are closely related to one another, although not identical. The similarity stems from the fact that they are all either perpetuals or futures contracts on a Bitcoin underlying
(a BTC/USDT or BTC/USDT index), or a BTC/USDT spot market.
We would, of course, expect price differences to be mean reverting processes since if this were not the case there would supposedly be arbitrage opportunities.
This furnishes the motivation for the inclusion in our set of base features of the price differences between pairs of markets, as indicators for future returns. One would expect, for instance, that if a particular market is cheap relative to some (or all) other markets (e.g. due to a price move up on the other markets) there should be a short-term price increase on the cheap market, as arbitrageurs and market makers rush to profit from the arbitrage. However, there are several assumptions underlying this hypothesized market behaviour which do not hold in practice, and which therefore make the actual dynamics more complicated. We list some of these confounding factors below. 

\begin{enumerate}
    \item The presence of trading fees restricts the set of arbitrages. That is, price differences that would represent arbitrages in a world without trading fees are in fact not arbitrage opportunities. The consequence of this is that the price difference between a pair of markets can fluctuate within a ``no arbitrage band" whose size is given by the sum of the taker fees of the two markets. Only when the price difference is greater than the sum of the two taker fees does there exist an arbitrage that can be immediately profited from with a pair of taker orders.

    \item The futures premium of, say, quarterly futures contracts is (mostly) nonzero, so that the price difference between a futures contract and another market is, a-priori, not indicative of where the price will move in the near future. Rather, what matters is how far the price difference between the futures contract and the other market strays from its ``equilibrium" level. This equilibrium is difficult to get a handle of in practice, since it depends on future expected interest payments whose magnitude is uncertain. 

    \item The price difference between a BTC/USDT based market (e.g. a perpetual whose underlying is a basket of BTC/USDT spot markets) and a BTC/USD based market alone cannot be expected to be indicative of future returns since it depends on the USD/USDT rate, which is generally only \emph{approximately} 1, but can fluctuate by several basis points, and in rare cases even percentage points. As we discussed before, such fluctuations are usually related to varying investor confidence in the peg between USD and USDT which can decrease the value of USDT (see~\cite{GRIFFIN_2020}). On the other hand, the value of one USDT can temporarily exceed that of an actual USD during periods where investors are willing to pay a premium for the conveniences associated with transacting in USDT rather than USD (usually when sentiment is extremely bullish). Deviations of the USD/USDT rate from $1$ usually cannot be \emph{immediately} arbitraged since the conversion typically takes several business days and involves a bank transfer. 

    \item The dominant trader demographics differ from exchange to exchange. For instance, US citizens are completely prohibited from trading on Bitmex and must therefore resort to other exchanges. These types of restrictions can result in persistent price differences, say, between Bitmex and another exchange which is popular with US traders. One reason for this could be that prevailing investor sentiment and expectations of future returns differ between distinct demographics, for example due to divergent macroeconomic backdrops. Other reasons for persistently large price divergences include differences in capital controls, financial regulation, ``know-your-customer" (KYC) requirements, etc. across different countries and  exchanges. A prominent example is the premium of the Bitcoin price on Korean exchanges relative to US exchanges, dubbed as the \textit{Kimchi premium}, which at various points in the past was notoriously large for extended periods of time. More details can be found in~\cite{kimchi_premium}. 
\end{enumerate}

\begin{figure}[!ht]
\vspace{-3mm}
\centering
\subcaptionbox[]{Price difference between the Binance USDT perpetual and the Huobi BTC perpetual 
\label{fig:price_diff_time_series}
}[ 0.49\textwidth ]
{ \includegraphics[width=0.49\textwidth, trim=0cm 0cm 0cm 0.0cm,clip] {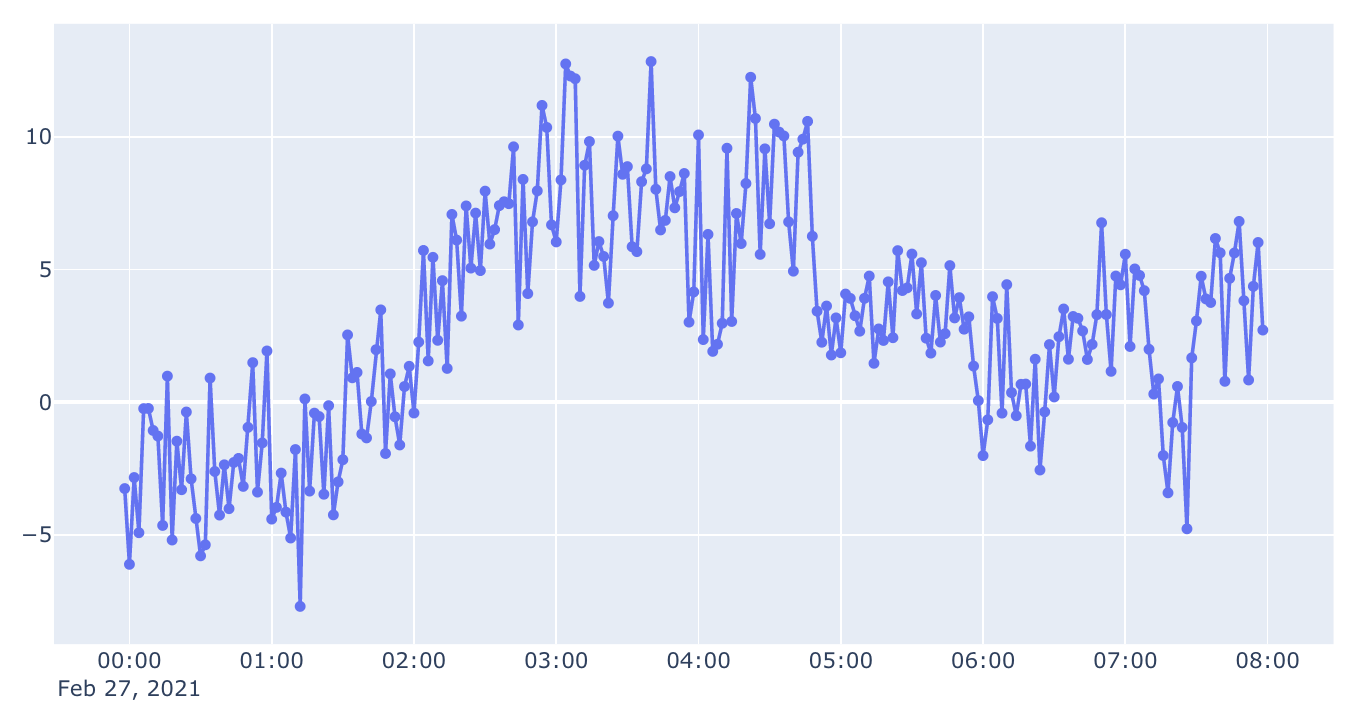} }
\hspace{0.0\textwidth} 
\subcaptionbox[]{Mean divergence feature for the Binance USDT perpetual and the Huobi BTC perpetual 
\label{fig:div_time_series}
}[ 0.49\textwidth ]
{\includegraphics[width=0.49\textwidth, trim=0cm 0cm 0cm 0.0cm,clip]{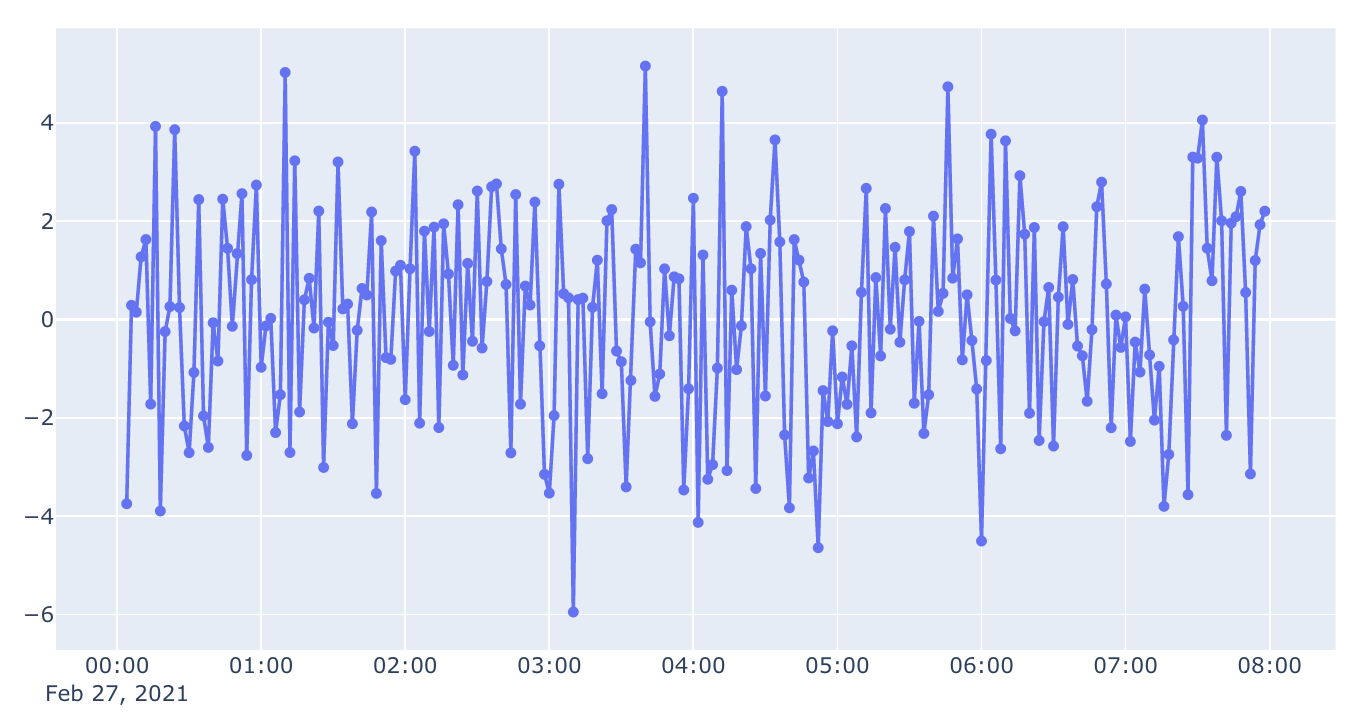} }
\vspace{-2mm} 
\captionsetup{width=0.98\linewidth}
\caption[Short Caption]{To avoid an excessively large number of samples (and hence to facilitate legibility), the data was resampled to a 2 minute frequency.}
\label{fig:bothxyzzzzzzzzzzz}
\vspace{-3mm}
\end{figure}

Altogether, these factors diminish the predictive signal of price differences alone.
For visual evidence of this, consider~\ref{fig:price_diff_time_series} where we plot the price difference over an eight hour period between a pair of perpetuals where one uses BTC/USD as its underlying, while the other uses a BTC/USDT index.

We propose to circumvent the concerns listed above by correcting for average price differences over a certain lookback window. That is, instead of computing ``vanilla" price differences, we compute the deviation of the price difference from the mean price difference over a past time period ranging from several seconds to several minutes.
The rationale for this is that persistent large price differences arising for some of the reasons mentioned above (nonzero futures premium, USD/USDT rate different from $1$, or differing investor outlook between distinct groups of traders) are usually unchanged over such short time horizons. For instance, barring the occurrence of an extreme outlier event, the repricing of a futures premium most often happens over the course of several hours or days as investors gradually adjust expectations of future interest payments.
With our proposed new indicator, we are therefore, in a sense, ``purging" the price difference of such confounding longer time scale effects. In  Figure~\ref{fig:div_time_series} we show the same price difference depicted   in Figure~\ref{fig:price_diff_time_series} purged by its five minute rolling mean.
Note the seemingly much stronger tendency for mean reversion.

This altogether motivates the definition of the following features for pairs of markets $(i, j) \in \{1,\ldots, 14\}^2$ and time horizons $\Delta \in \{5\text{s}, 9\text{s}, 19\text{s}, 38\text{s}, 75\text{s}, 150\text{s}, 300\text{s}, 600\text{s}\}$
\begin{equation}
DIV_t^{i, j, \Delta} = d(p_t^{i}, p_t^{j}) - \operatorname{rolling}^\Delta\left( d(p_t^{i}, p_t^{j}) \right), 
\end{equation}
where $d(p, q) = \left( \frac{p}{q} - 1 \right) \cdot 10000$ is the difference in basis points between $p, q \in (0, \infty)$, and the $\operatorname{rolling}^\Delta(\cdot)$ function returns the rolling mean of samples from its input over the past $\Delta$ seconds.
We shall call $DIV_t^{i, j, \Delta}$ the \textit{mean divergence} feature between markets $i$ and $j$.


As further evidence of the superiority of this indicator compared to the vanilla price difference, we display in Table~\ref{tbl:r2_comparison_diff_vs_div} a comparison of $R^2$ values corresponding to the univariate regression models
\begin{equation}
fret_t = \alpha + \beta \left( d(p_t^{i_0}, p_t^{j}) \right)  + \epsilon_t, 
\end{equation}
and
\begin{equation}
fret_t = \alpha' + \beta' \left( DIV_t^{i_0, j, 150\text{ms}} \right)  + \epsilon'_t, 
\end{equation} 
for each $j=1,\ldots, 14$. 
For sake of concreteness, we fix a reference market $i_0$, namely the Bybit BTC perpetual, whose future returns are used in the above models.
That is, we are comparing the explanatory power over Bybit's future returns of, one the one hand, the ``plain" price difference between Bybit and other markets, and, on the other hand, the mean divergence feature of Bybit with other markets.
\begin{table}
\centering
\captionsetup{width=0.98\linewidth}
\caption{Comparison of $R^2$ values of vanilla price difference (left column) with divergence from mean (right column)}
\label{tbl:r2_comparison_diff_vs_div}
\begin{tabular}{lll}
\toprule
{} &     $R^2_{\text{diff}}$ &     $R^2_{\text{div}}$ \\
\midrule
binancecmfut\_BTC/USD  &   0.057 &  0.1426 \\
binancefut\_BTC/USDT   &  0.0252 &  0.1331 \\
hbdm\_BTC\_CQ           &  0.0017 &  0.1224 \\
okex\_BTC-USD-210326   &  0.0015 &   0.119 \\
okex\_BTC-USD-SWAP     &  0.0337 &  0.1169 \\
okex\_BTC-USDT-SWAP    &  0.0175 &  0.1113 \\
thbdm\_BTC-USDT        &  0.0099 &  0.0961 \\
hbdm\_BTC-USD          &  0.0175 &  0.0778 \\
ftx\_BTC-PERP          &   0.031 &  0.0754 \\
deribit\_BTC-PERPETUAL &  0.0149 &  0.0744 \\
huobipro\_BTC/USDT     &  0.0118 &  0.0661 \\
tbybit\_BTC/USDT       &  0.0032 &  0.0393 \\
bitmex\_BTC/USD        &  0.0038 &   0.016 \\
\bottomrule
\end{tabular}
\end{table}
On average, the explanatory power of our new indicator is larger than that of the previous one by $0.074$. This represents an average \emph{15-fold} increase. Note that the improvement is particularly drastic for the two quarterly futures contracts {\hbcq} and {\okcq} where we see a more than 70-fold increase.

\subsection{Eliminating Nonlinearities}

Our overarching objective in this paper is to train powerful \emph{linear} price prediction models using the features defined above.
The performance of a linear model is, however, limited by the extent to which the future returns vary \emph{linearly} as a function of the feature variables.
That is, if the relationship between variates and covariates is highly nonlinear, we cannot possibly expect to achieve good generalizability of our model. 

In this subsection, we will demonstrate the nonlinear nature of the dependence between future returns and some of our indicators. We will then devise a methodology that addresses these concerns. Our approach involves the construction of bespoke nonlinear transformations which will subsequently be applied to our features prior to their usage in linear models. 

\begin{figure}[!ht]
\vspace{-3mm}
\centering
\subcaptionbox[]{(Original) trade flow imbalance
\label{fig:fret_vs_tv_by_bucket}
}[ 0.49\textwidth ]
{ \includegraphics[width=0.49\textwidth, trim=0cm 0cm 0cm 0.0cm,clip] {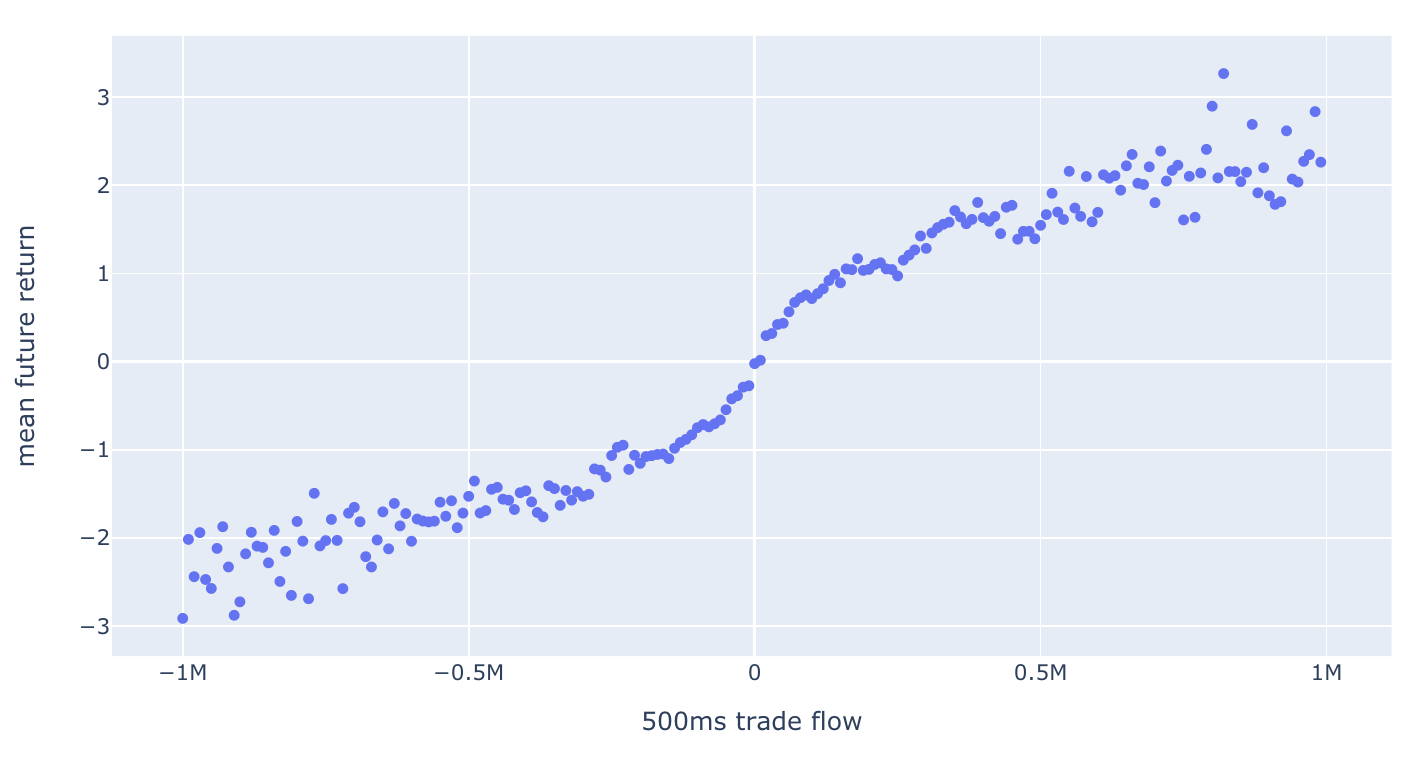} }
\hspace{0.0\textwidth} 
\subcaptionbox[]{Transformed trade flow imbalance
\label{fig:fret_vs_transformed_tfi_scatter}
}[ 0.49\textwidth ]
{\includegraphics[width=0.49\textwidth, trim=0cm 0cm 0cm 0.0cm,clip]{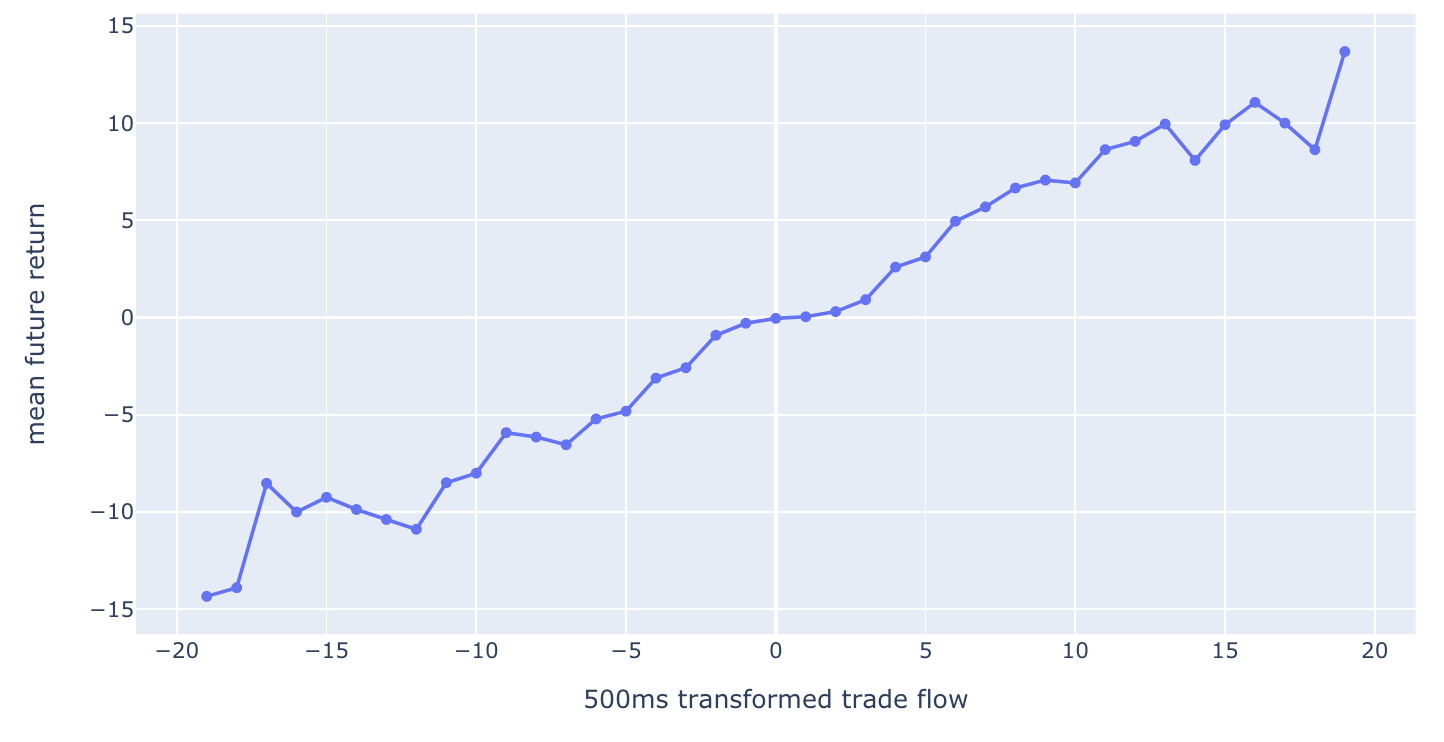} }
\vspace{-2mm}
\captionsetup{width=0.98\linewidth}
\caption[Short Caption]{500ms future returns of the Huobi BTC perpetual as a function of Binance USDT perpetual 500ms original and transformed trade flow imbalance.}
\label{fig:bothxyzzzzzzzzzzz222}
\vspace{-3mm}
\end{figure}

To illustrate the presence of some of the aforementioned nonlinearities consider Figure~\ref{fig:fret_vs_tv_by_bucket}.
This scatter plot exhibits the relationship between 500ms trade flow imbalance of the Binance USDT perpetual and 500ms future returns of the Huobi BTC perpetual.
Each $x$-value in this plot corresponds to a trade flow imbalance bucket 
$$B_n := \left[ 10000\text{ USD}\cdot (n-1), 10000\text{ USD}\cdot n \right]$$
for $n \in \{-100, 100\}$.
The $y$-value associated to the bucket $B_n$ was obtained by computing the mean 500ms future returns on the Huobi perpetual over all samples where the Binance trade flow falls into the interval $B_n$. The functional form we observe in  Figure~\ref{fig:fret_vs_tv_by_bucket} clearly resembles a sigmoid or hyperbolic tangent function more than it does a linear one (although samples for buckets $B_n$ with large $|n|$ are sparser, hence the data noisier). The sigmoid-like relationship between trade flow imbalance and future returns agrees well with prior work, which noted the same empirical observation~\cite{POTTERS2003133,Plerou_2002}.

Why does the price impact decrease as trade flow imbalance grows? One possible economic interpretation is that extreme trade flow values (either very large or very small) could correlate negatively with trader sophistication and correlate positively with the  ``sloppiness" of execution.
That is, an informed trader wishing to enter a large position will rarely do so by using, for example, one large market order or a large ``meta order" executed over a time window as small as 500ms. It is perhaps more likely that extreme trade flow values originate from uninformed traders, or, in the extreme case, are the result of forced liquidations where traders involuntarily liquidate their entire positions as a consequence of a margin call. 
If our hypothesis is correct, the relative lack of sophistication reflected by extreme trade flow imbalances could be correctly recognized by other (more astute) market participants who would then only minimally adjust their estimation of the ``fair price",  and hence contribute towards a more limited price impact.

Further below, we describe our methodology for mapping our (original) features to nonlinearly transformed ones. Before we proceed with this description, it is interesting to visually compare Figure~\ref{fig:fret_vs_tv_by_bucket}, which we already considered earlier, with Figure~\ref{fig:fret_vs_transformed_tfi_scatter} where we plot the functional dependence of Huobi returns on the \emph{transformed} Binance trade flow imbalance feature. Note the stronger degree of linearity in the dependence of future returns on the transformed feature.

\textbf{Methodology for feature transformation.} The intuition behind our methodology is to define, for each feature, a sequence of transformations that starts with the identity map and ends with a function which yields a significant boost in explanatory power when it is applied to the feature. This sequence is defined through an iterative procedure where each step of the iteration achieves an $R^2$ improvement.

Before specifying our general algorithm, it is instructive to see what happens in case of the trade flow imbalance feature.
To this end, let us preliminarily define, for a finite set of real numbers $T$, a mapping $f_T:\mathbb{R} \rightarrow \mathbb{R}$ as follows 
\begin{equation}
f_T(x) := \operatorname{sgn}(x) \sum_{t\in T} \mathbf{1}{\left\{|x| \geq t\right\}}, 
\end{equation}
where $\operatorname{sgn}(\cdot)$ is the sign function, and $x\mapsto \mathbf{1}{\left\{|x| \geq t\right\}}$ denotes the indicator function which takes the value $1$ when $|x|\geq t$ and $0$ else.
Now suppose $X = \{x_t\}_t \subset \mathbb Z$ is the set of observations of the trade flow imbalance on some fixed market.
Let us then set $m:=\min(X), M:=\max(X)$ and $T_0:=\left\{ m, m+1, \ldots, M-1, M \right\}$.
Note that we can then write the identity function in a more complicated way as follows
\begin{equation}
\operatorname{id}_X = \left(x\mapsto \operatorname{sgn}(x) \sum_{t\in T_0} \mathbf{1}{\left\{|x| \geq t\right\}} = f_{T_0}(x) \right).
\end{equation}
This stems from the fact that, in the sum, we have precisely $|x|$ many nonzero indicator functions. 
The next step of our procedure involves the elimination of one element $i^*$ from the set $T_0$, giving rise to an updated set $T_1 = T_0 \setminus \{ i^* \}$, which in turn defines a new mapping
\begin{equation}
\left(x\mapsto \operatorname{sgn}(x) \sum_{t\in T_1} \mathbf{1}{\left\{|x| \geq t\right\}} = f_{T_1}(x) \right), 
\end{equation}
representing a slight alteration of the identity mapping. How do we decide which element  $i^*$ to prune from the original set $T_0$? We do so by trying all possibilities, and then ultimately choosing the element which provided the largest improvement in average explanatory power of future returns. 

Specifically, we fit the following univariate linear models using OLS regression for every market $j = 1,\ldots 14$ and every element $i\in T_0$
\begin{equation}
fret_t^{{500\text{ms}}, j} = \alpha + \beta f_{T_0\setminus \{ i \}}(x_t)  + \epsilon_{t}.
    \label{eqn:model_t0_ij}
\end{equation}
We denote the coefficient of determination of the model specified in Eqn.~\eqref{eqn:model_t0_ij} by $R^2_{ij}$.
That is, $R^2_{ij}$ is the explanatory power of the linear model that predicts future returns on market $j$ using the trade flow imbalance feature of the fixed market which has been transformed by the function $f_{T_0\setminus \{ i \}}$.
We then form an average across market $j$, and select the transformation that resulted in the greatest average coefficient of determination. That is, we set
\begin{equation}
\overline{R^2}_{i} := \frac{1}{14} \sum_{j=1}^{14} R^2_{ij}, 
\end{equation}
and finally define $i^* := \operatorname{argmax}_{i\in T_0} \overline{R^2}_{i}$.
This yields the set $T_1 := T_0 \setminus \{ i^* \}$ with corresponding transformation $f_{T_1}$. 
We then prune another element in the same manner as above, in order to obtain a next set $T_2$ which defines a new mapping $f_{T_2}$, and so on. 
This process of successive elimination is terminated when none of the eliminations yields an improvement in average explanatory power. That is, we halt the process when the average explanatory power under a transformation $f_{T_{k+1}}$ is not greater than that  corresponding to the previous transformation $f_{T_{k}}$. At this point, we have obtained our final set $T^* := T_k \subset T$ which defines the (nonlinear) feature transformation $f_{T^*}$ for the trade flow imbalance feature we fixed in the beginning. 
In practice, it is computationally infeasible to begin with the initial set $T_0$ defined above, whose cardinality is $|M-m+1|$, which, for the trade flow imbalance feature, is on the scale of tens of millions. We instead choose an initial set $T_0$ of cardinality $100$ representing an evenly spaced partition of the interval $[m, M]$. Moreover, to mitigate the impact of outliers, we do not use $m=\min(X)$ and $M=\max(X)$, but instead define the two quantities $m$ and $M$ as a very small quantile and a very large quantile, respectively, of the set $X$ of feature observations. 
The data used to calibrate the feature transformation spans February 21st 2021, the day prior to our one week training and test period.

We now specify the general algorithm for obtaining the final ``optimal" feature  transformation for a fixed feature with observations $X= \{x_t\}_t$. 
The initialization is as follows. First, we define $\overline{R^2}_{0}$ to be the average of the $R^2$ values of the linear models
\begin{equation*}
fret_t^{{500\text{ms}}, j} = \alpha + \beta x_t  + \epsilon_{t}, 
\end{equation*}
for $j=1,\ldots, 14$. Now let $m$ and $M$ denote the $q$-th and $(1-q)$-th quantiles of $X$, respectively, and initialize $T_0$ as an evenly-spaced partition with $100$ elements of the interval $[m, M]$.
In these calculations, we employed $q=0.0001$ which was revealed as a sensible choice in preliminary experiments.

The iterative step for obtaining $T_{k+1}$ from $T_k = \{ t_1, \ldots, t_n \}$ proceeds in the following manner.
\begin{enumerate}
    \item Fit the linear model $fret_t^{{500\text{ms}}, j} = \alpha + \beta f_{T_k\setminus \{ t_i \}}(x_t)  + \epsilon_{t}$ for each market $j=1,\ldots, 14$ and every $i=1,\ldots, n$ and denote the corresponding coefficient of determination by $R^2_{j, i, k}$.
    \item With $\overline{R^2}_{i, k} := \frac{1}{14} \sum_{j=1}^{14} R^2_{j, i, k}$ we make the definitions $i^* := \operatorname{argmax}_{i=1,\ldots, n} \overline{R^2}_{i, k}$ and $\overline{R^2}_{k+1} := \overline{R^2}_{i^*, k}$
    \item If $\overline{R^2}_{k+1} > \overline{R^2}_{k}$ we define $S:=\{ t_{i^*} \}$, else we let $S:=\emptyset$.
    \item Finally, we set $T_{k+1} := T_k\setminus S$. 
\end{enumerate}
Clearly, this iteration terminates eventually since either $S$ becomes $\emptyset$ or the set $T_k$ itself becomes empty. If the process has terminated after $K$ steps, we will further define $T^* := T_{K}$ and $\overline{R^2}_* := \overline{R^2}_{K}$.

\begin{figure}[!ht]
\hspace*{-0cm}
\vspace{-3mm}
\centering
\includegraphics[scale=0.45]{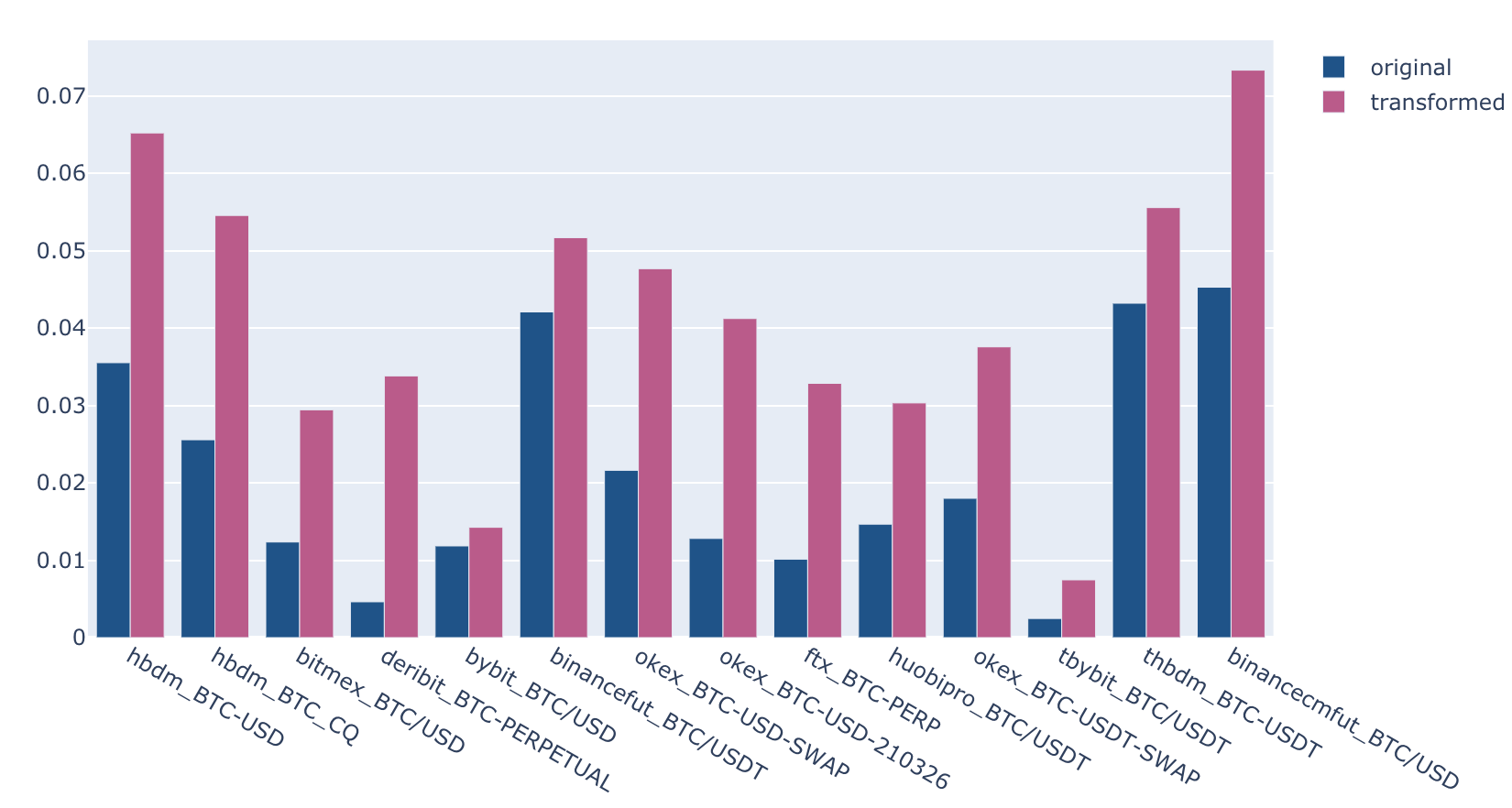}
\caption{Trade flow imbalance improvement of transformation versus original feature.}
\label{fig:transformed_vs_original_tfi}
\hspace{-0.9\textwidth} 
\vspace{-3mm}
\end{figure}

\textbf{Results.} For each feature, we compute its optimal transformed feature according to the above method. 
A natural question to ask is for which features does the above transformation procedure result in a significant improvement in explanatory power of future returns by a linear model? 
By far the biggest improvement was found for the trade flow imbalance feature. 
In Figure~\ref{fig:transformed_vs_original_tfi} we plot the average $R^2$ values obtained using the (untransformed) 500ms trade flow imbalance feature on the one hand, and its optimal transformed version on the other hand. 

More specifically, we compare the two quantities $\overline{R^2}_0$ and $\overline{R^2}_*$ (as defined above) which capture the average explanatory power of the original and the transformed feature, respectively. The results are shown in  Figure~\ref{fig:transformed_vs_original_tfi}. 
The transformation yielded a significant improvement for every market.
The average improvement is by a factor of $2.5$.
The improvement is largest on the Deribit perpetual contract, where we observe an increase by a factor of $7.3$. The smallest improvement occurs on Bybit's perpetual contract, where we nevertheless note an increase of 20\%. 

\begin{figure}[!ht]
\hspace*{-0cm}
\vspace{-3mm}
\centering
\includegraphics[scale=0.45]{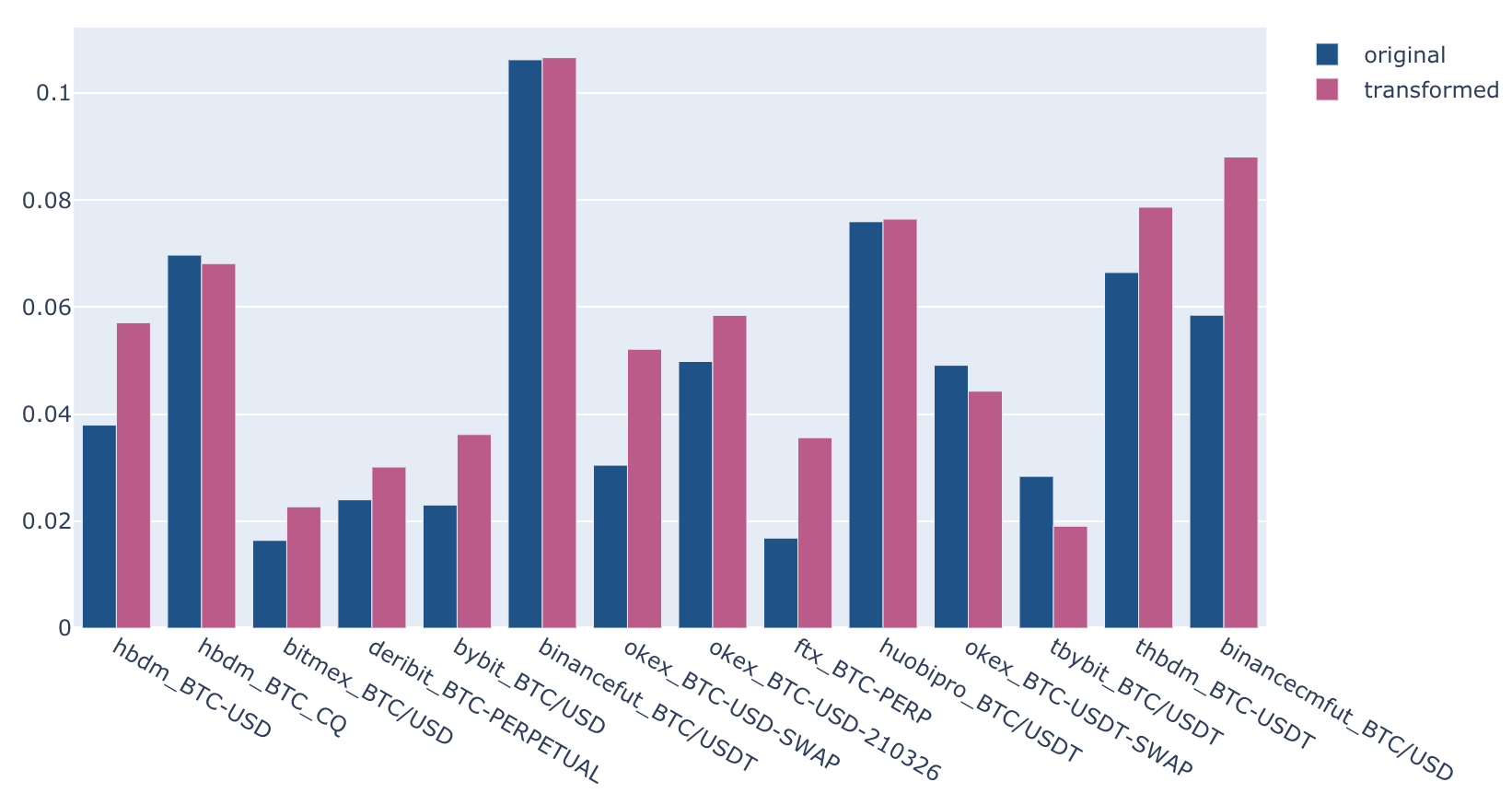}
\caption{Past returns improvement of transformation versus original feature.}
\label{fig:transformed_vs_original_pret}
\hspace{-0.9\textwidth} 
\vspace{-3mm}
\end{figure}

The only other feature where we found, on average, a significant improvement using our previously described transformation procedure is the past returns feature.
See Figure~\ref{fig:transformed_vs_original_pret} for a barplot of the results.
On average, we note an improvement of 28\%. This is far lower than the one attained for the trade flow imbalance feature, but it is still significant. On the FTX perpetual, the increase in average explanatory power was largest, with an increase by a factor of $2.11$.
The Bybit USDT-margined perpetual actually saw a decrease in explanatory power by around 33\% (although it was quite low to begin with). This was the largest decrease of any markets. 

The two remaining features (orderbook imbalances and mean divergence) did not show a significant average boost in explanatory power of the transformed feature over the original version. For the orderbook imbalance feature, we found an average increase of around 3\%, while the mean divergence feature with lookback window 150s produced an average decrease of around 5\% (with similar results holding true for other lookback windows). This lack of $R^2$ gain points towards an already linear relationship between the aforementioned features and future returns.

In the remainder of this paper, we will use the transformed trade flow imbalance feature instead of the original version. On markets where the transformation yielded an  improvement for the past returns feature, we shall use the transformed version.
Where a decrease was found, we will continue using the original version. 
For the orderbook imbalance and mean divergence features, where we did not find a significant improvement, we will simply continue using the untransformed features.

\subsection{Optimal Time Horizon Selection}

With the exception of the orderbook imbalance features, all of our base features are defined over multiple time horizons.
From a standpoint of complexity reduction, however, it is desirable to select only one time horizon per base feature.
This is because base features of a single feature category at different time horizons share a high degree of collinearity.  For instance, it is clear that 500ms past returns are highly correlated to 1s past returns (the values will often coincide).
In this subsection, we set out to devise a methodology by which we select a single time horizon for any feature defined over multiple time horizons. With an eye towards  eventually training large composite models that employ the full feature set, this additional feature selection step will be useful in ameliorating concerns of overfitting  (it is common knowledge that OLS regression is prone to overfitting when the number of covariates is large and the covariates exhibit high cross-correlations). 
Additionally, it is an interesting research question in and of itself to investigate which lookback windows for which features are most predictive of, say, 500ms future returns.

Our methodology for selecting the ``optimal" time horizon involves a simple process of computing an average predictability score based on $R^2$ values for each time horizon, and then choosing the one with the highest score.

\textbf{Methodology for selection of optimal time horizon.} Suppose we are trying to determine the optimal time horizon $\delta_{k^*}$ corresponding to some set of features $f^{\delta_1}, \ldots, f^{\delta_n}$.
We proceed in two steps. The first step consists of using OLS regression to fit the following model for every time horizon $k\in \{1,\ldots, n\}$ and target market $j=1,\ldots, 14$ 
\begin{equation}
fret_t^{500\text{ms}, j} = \alpha + \beta f^{\delta_k}_t  + \epsilon_{t}, 
\end{equation}
whose coefficient of determination we denote by $R^2_{k, j}$.
For the second step, we compute the following averages for each time horizon $k\in \{1, \ldots, n\}$
\begin{equation}
\overline{R^2}_k := \frac{1}{14} \sum_{j=1}^{14} R^2_{k, j}, 
\end{equation}
and then set $k^* := \operatorname{argmax}_{k} \bar{R^2_k}$. 
In this case, we shall call $\delta_{k^*}$ the \textit{optimal time horizon} for the feature $f$. In summary, a feature's optimal time horizon is the one whose average explanatory power over 500ms future returns is largest.

\textbf{Empirical results.} We now examine the empirical results of the procedure described above.

\begin{figure}[!ht]
\hspace*{-0cm}
\vspace{-3mm}
\centering
    \includegraphics[scale=0.45]{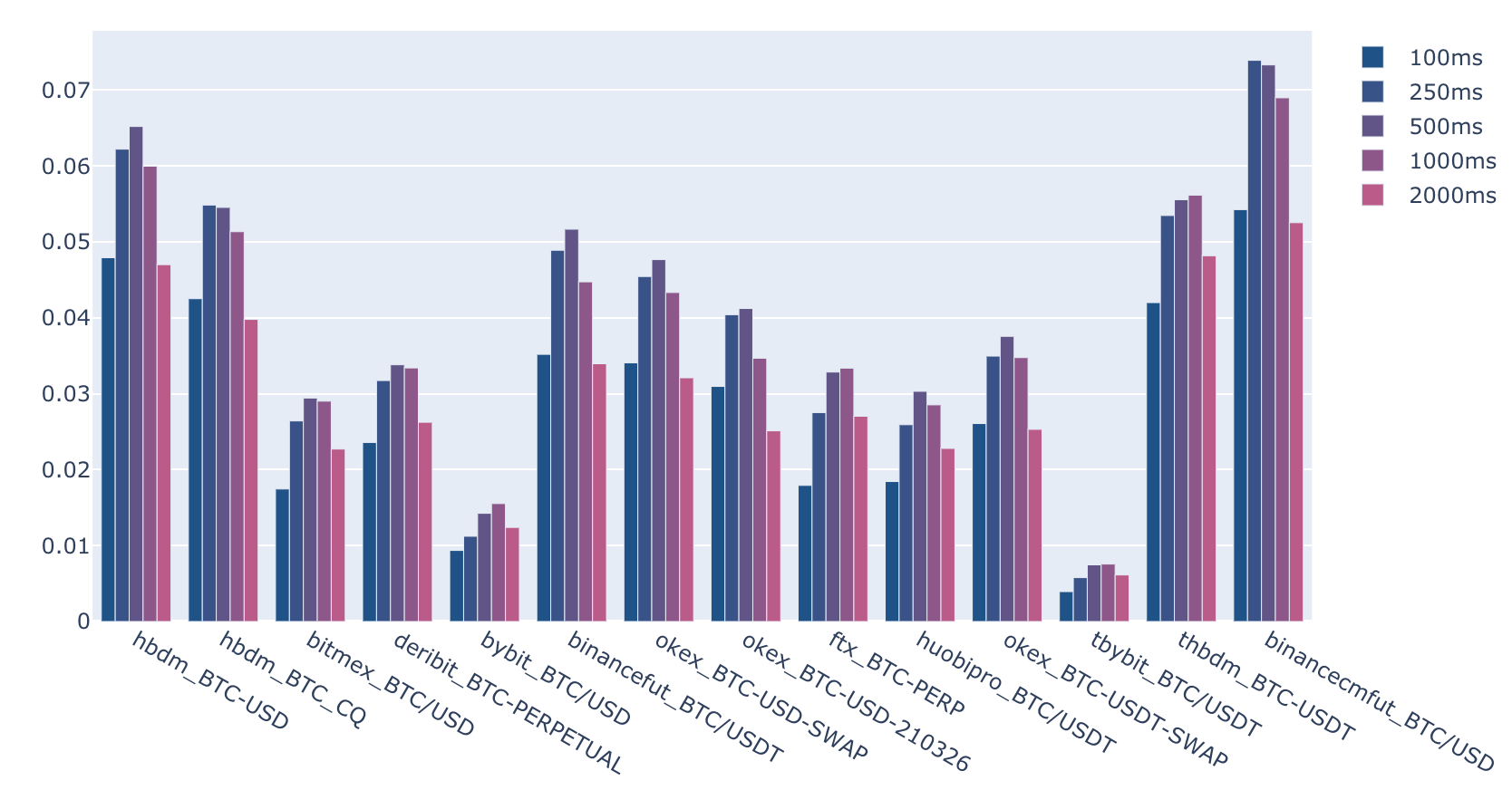}
    \caption{Average explanatory power of TFI features for each time horizon}
    \label{fig:horizon_selection_tfi}
\hspace{-0.9\textwidth} 
\vspace{-3mm}
\end{figure}

(1) Let us first consider the \textbf{trade flow imbalance (TFI) indicators}. 
Since we found that the \emph{transformed} TFI yielded a considerable improvement over the untransformed one, we will in fact use the transformed version, though we shall, for the sake of brevity, generally omit the qualifier ``transformed".
In Figure~\ref{fig:horizon_selection_tfi} we illustrate for each market's (transformed) TFI the quantities $\overline{R^2}_1, \ldots \overline{R^2}_4$ corresponding to the five time horizons.

We observe that the smallest and largest time horizons are not selected for any market, indicating a sensible choice of lower and upper bound for the size of the lookback windows with the optimal value falling somewhere in-between.
Next, we note that the most commonly preferred horizon (eight cases) is 500ms, while the 250ms horizon and the $1000$ms horizon are preferred in two and four cases, respectively.
Finally, it is noteworthy that the five markets whose trade flows have the most predictive power are all on either Binance or Huobi.
Specifically, this set of markets is comprised of the two perpetual contracts (one of them being margined with USDT, the other one with Bitcoin) on Binance and Huobi, as well as the Huobi quarterly futures contract. This furnishes our first indication of the leading behaviour of Binance and Huobi, which is a theme we that will repeatedly reappear later in this work.

\begin{figure}[!ht]
\hspace*{-0cm}
\vspace{-3mm}
\centering
\includegraphics[scale=0.45]{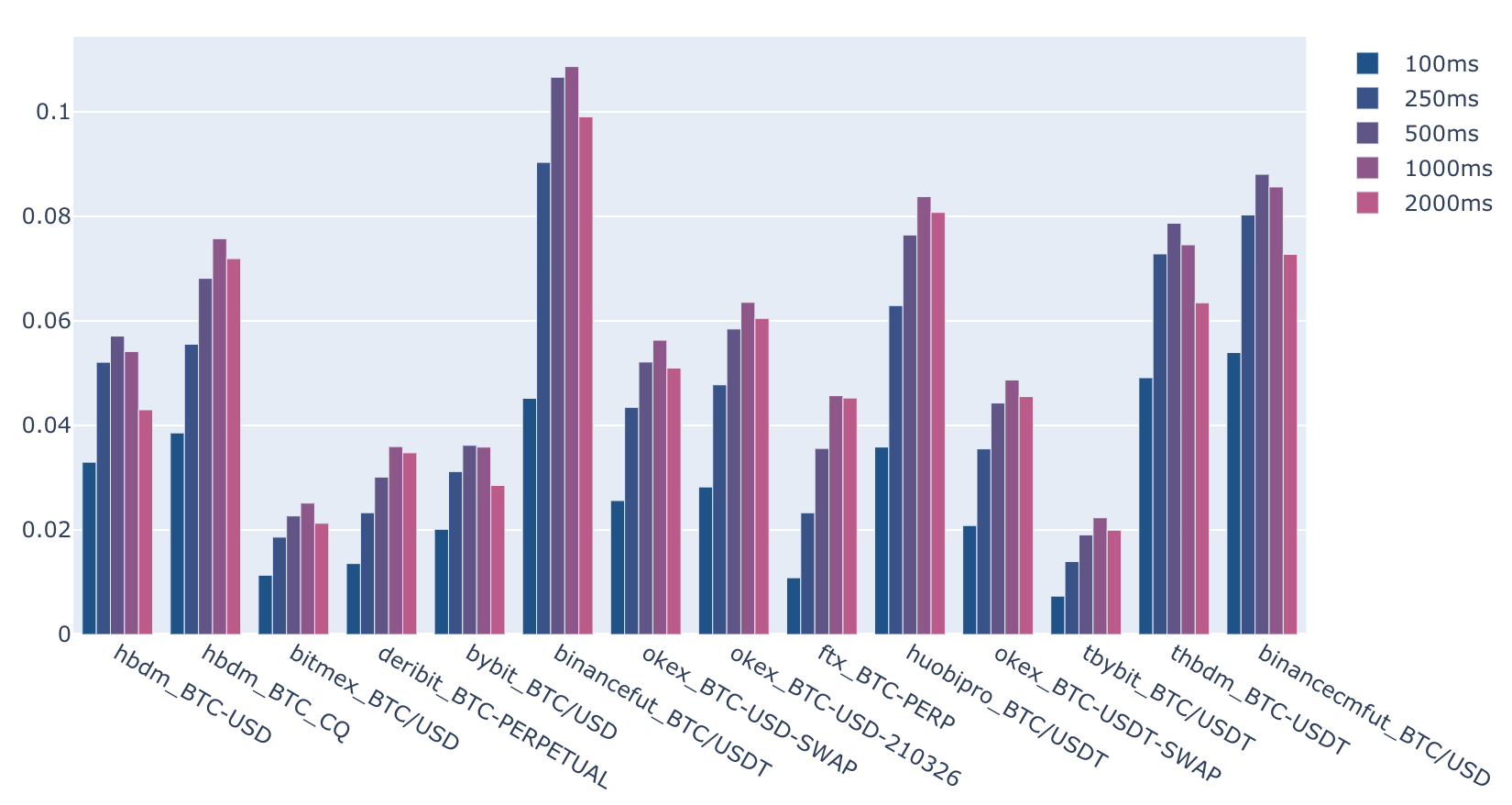}
\caption{Average explanatory power of past returns features for each time horizon}
\label{fig:horizon_selection_pret} 
\hspace{-0.9\textwidth} 
\vspace{-3mm}
\end{figure}

(2) We now examine the \textbf{past returns indicators}. For these features, we previously also found an improvement when using the transformed feature instead of the original one, so we shall again use the transformed version in the following analysis. 
In Figure~\ref{fig:horizon_selection_pret}, we show for each market the average explanatory powers $\overline{R^2}_1, \ldots \overline{R^2}_5$ corresponding to the five  time horizons.  
As was the case with the TFI indicators, we note that also for the past returns features, neither of the optimal time horizons is attained at the minimal or maximal time horizon, providing further validation for the choice of time windows.
In four cases, the $1000$ms time window is preferred, while the $250$ms horizon is selected in two cases. The most commonly selected optimal horizon is $500$ms for six markets. It is noteworthy that the two markets where the smaller time horizon ($250$ms) is preferred are also among the markets where the smaller time horizon is preferred for the TFI feature.
Furthermore, the markets which appeared to be leaders based on their comparatively large $R^2$ values for the TFI features, also have among the highest explanatory powers in terms of their past returns features. This lends additional   confidence to the previously conjectured leading behavior of the two Binance and Huobi perpetual contracts, as well as the Huobi quarterly futures market.

\begin{figure}[!ht]
\hspace*{-0cm}
\vspace{-3mm}
\centering
\includegraphics[scale=0.45]{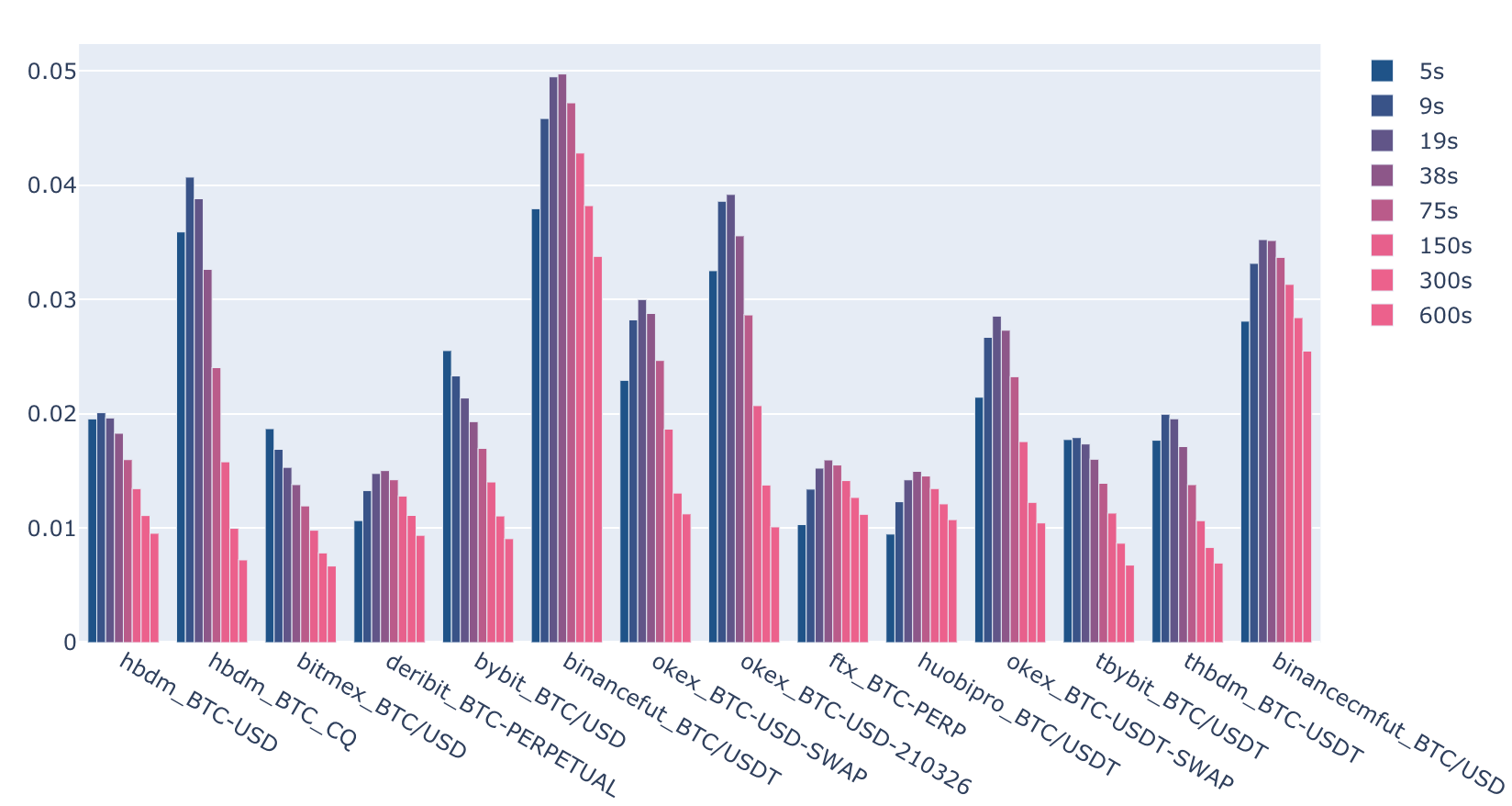}
\captionsetup{width=0.97\linewidth}
\caption{Average explanatory power of mean divergence features for each time horizon.}
\label{fig:horizon_selection_div}
\hspace{-0.9\textwidth} 
\vspace{-3mm}
\end{figure}

(3) The third and final feature whose definition relied on the choice of a time horizon is the \textbf{mean divergence feature}, which we shall examine now. 
Here, the feature transformation did not yield an improvement, so we will simply use the original feature. The corresponding illustration of each market's average $R^2$ values can be seen in Figure~\ref{fig:horizon_selection_div}.
Our smallest time horizon (5s) appears twice as the optimal choice. The time horizons 9s, 19s, and 38s were each selected as the optimal choice for exactly four markets. The previously noted pattern of leading markets persists in this feature class, as well with the Binance USDT-margined perpetual achieving the largest average $R^2$ value, followed by the Huobi futures contract and then the Binance BTC perpetual.

\vspace{-1mm}
\section{Network Effects in Bitcoin Markets}
\label{sec:4}
\vspace{-1mm}

With our feature set in its final form, we now turn to examining network effects amongst the markets considered in this work. We begin by investigating the presence and strength of \textit{lead-lag relationships} between pairs of markets. After performing this analysis, we will examine how it translates into the realities of trading.

\subsection{Leader-Lagger Network}
 
What does it mean for a market $j$ to lead another market $i$? 
Intuitively, one would say that a market $j$ \emph{leads} a (lagging) market $i$ if future returns on the lagging market $i$ can reliably be anticipated based on information from market $j$. In our case, this information consists of microstructural data encapsulated in the features we previously calculated.

Let us make the previous formulation more precise. A market $i$ is said lag another market $j$ if a large portion of the total variation in future returns of market $i$ can be explained by features of the leading market $j$. It is therefore natural to fit the following linear models for each pair of markets $(i, j) \in \{1, \ldots, 14\}$
\begin{equation} 
fret_{t}^{\delta, i} = \mu_{ij} + \beta_{ij,1} IMB_{j}^{a, j} + \beta_{ij,2} IMB_{t}^{b,j} + \beta_{ij,3} TFI_{t}^j + \beta_{ij, 4} PRET_{t}^j + \beta_{ij,5}  DIV_{t}^{ij} + \epsilon_{ij,t}, 
    \label{eqn:model_ij}
\end{equation}
and inspect the resulting coefficients of determination.
We fit these models using OLS regression using training data spanning the period from February 22-27, 2021. The data was normalized by subtracting the mean and dividing by the  standard deviation. For the sake of readability, we have suppressed in our notation the time horizons over which those features are calculated, 
with the understanding that the optimal time horizon has been selected. 
Furthermore, for those features where the nonlinear transformation yielded an increase in linear explanatory power, we will let it be implicit in the notation that the transformed version is used. The models will be fitted for both future returns lookahead windows $\delta \in \{500\text{ms}, 1000\text{ms}\}$, although the results turn out to be similar in both cases, and we will report results primarily based on the $500$ms horizon. 

Let us denote the coefficient of determination of the above fitted linear model by $R^2_{ij}$.
This is the total variation in future returns on market $i$ explained with features from market $j$, and it will be our first indicator of the lead-lag relationship between the two markets. 

\begin{figure}[!ht]
\vspace{-3mm}
\centering
\includegraphics[scale=0.4]{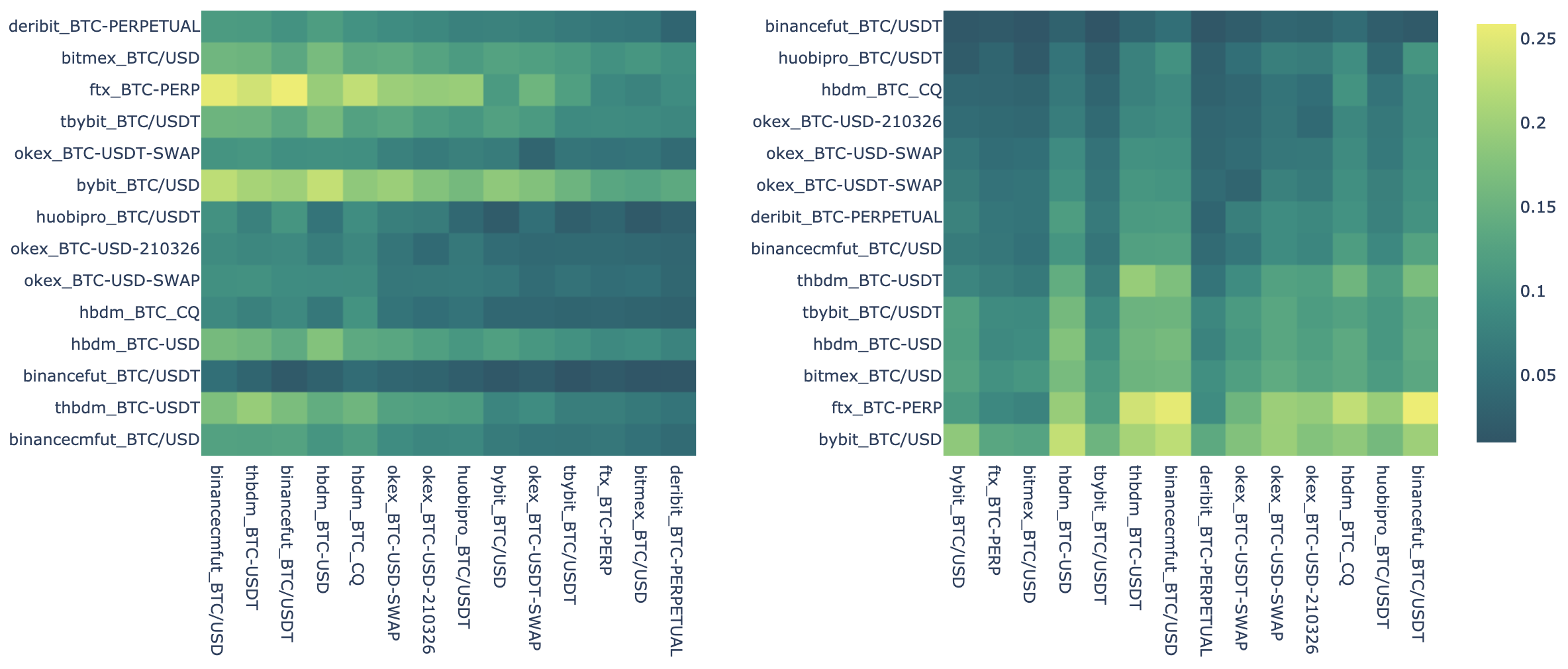}
\captionsetup{width=0.97\linewidth}
\caption{Entries in the matrix $R$ illustrating the lead-lag relationships between pairs of markets.}
\label{fig:r2heatmaps_500ms}
\vspace{-3mm}
\end{figure}

We visualize the matrix  $R := \left(R^2_{ij}\right)_{i,j=1,\ldots, 14}$ in  Figure~\ref{fig:r2heatmaps_500ms} as a pair of heatmaps, where the first one uses an ordering by column sum, and the second uses an ordering by row sum.

The largest $R^2$ value, of about $25.9$\%, is achieved when predicting FTX using features from the Binance USDT perpetual. The smallest $R^2$ value is about 1\% for predicting the Binance USDT perpetual using features of the Bybit USDT perpetual. Overall, we achieve an average $R^2$ of $9.6$\% and a median of 9\%. The results are similar when we use $\delta=1000$ms as our future returns time horizon, where we find average and median values of $10.4\%$ and $9.2\%$, respectively. 
This is quite remarkable considering the fact that only information of a single market was used in these regression models. In other words, even information from just a single market can be quite effective in explaining future returns. It demonstrates that it is possible and indeed quite feasible to anticipate future price moves in Bitcoin markets on sub-1s time scales, and it is a reassuring demonstration of the utility of our features.

\begin{figure}[!ht]
\hspace*{-1cm}
\vspace{-3mm}
\centering
\includegraphics[width=0.8\textwidth]{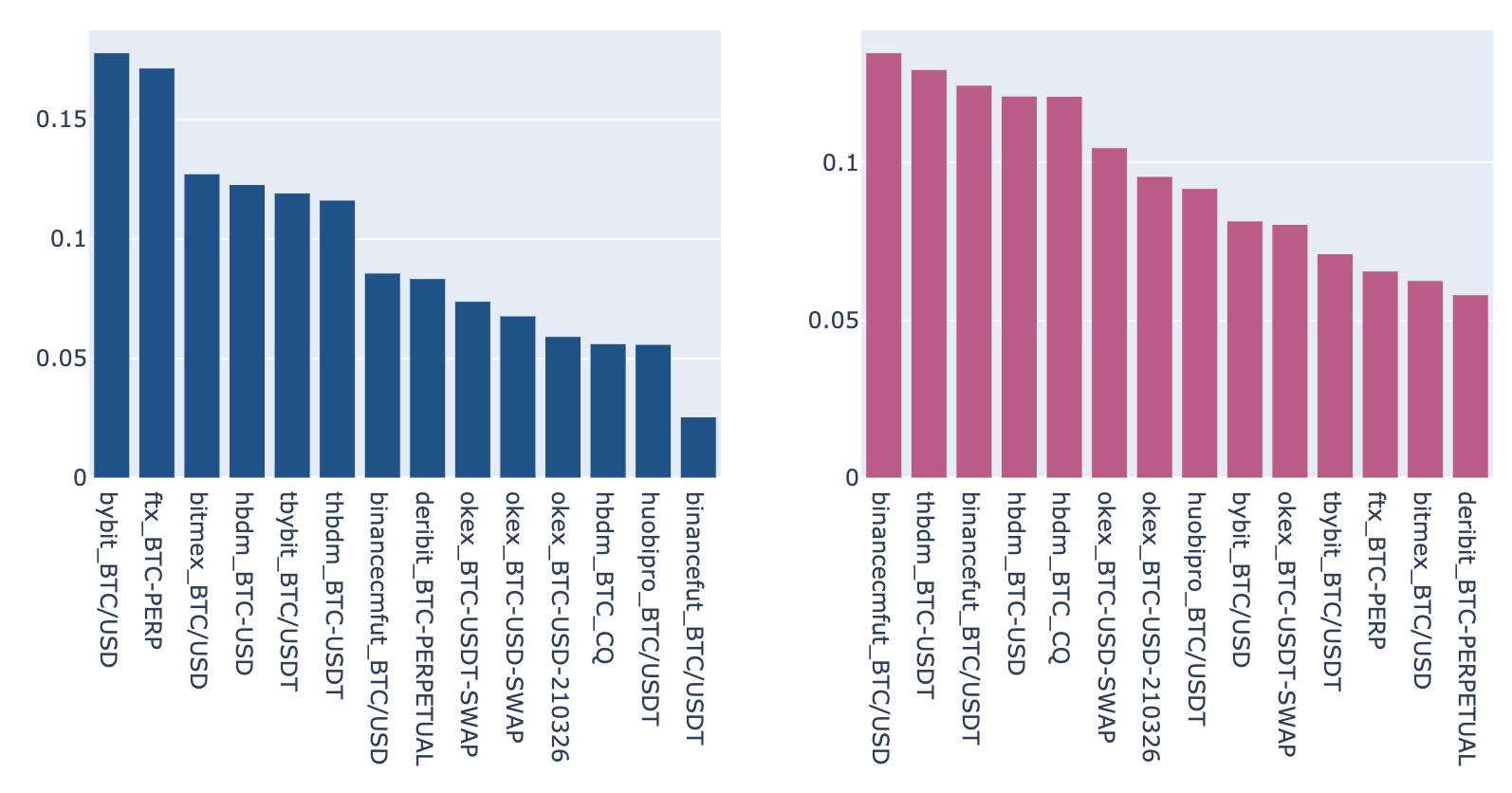}
\captionsetup{width=0.97\linewidth}
\caption{The left bar plot shows row averages of the matrix $R$, while the right bar plot shows column averages.}
\label{fig:r2barplots_500ms}
\hspace{0.0\textwidth} 
\vspace{-3mm}
\end{figure}

One can compare the $R^2$ values we obtain here with those reported in~\cite{cont2014price} where the authors achieve values of around 60\% using order flow imbalance to explain \emph{contemporaneous} returns.
Our next set of observations pertains to the hierarchy of markets implied by our results. See Figure~\ref{fig:r2barplots_500ms} for a visualization of the column averages and the row averages. A large column average shows that the market is easily predicted by others (hence a laggard), while a large row average shows that a market is particularly useful in predicting price action on other markets. Notably high are the column averages of the Bybit, FTX, and Bitmex perpetuals. Many of the markets that are especially useful in predicting Bybit and Bitmex are from the exchanges Binance and Huobi, as it can be seen in the above heatmaps when one searches, for instance, in the Bybit row, for the cells corresponding to the Binance USDT perpetual or the Huobi perpetual. Indeed, the top 5 markets in terms of row average are all from Binance and Huobi, providing strong evidence for the central role occupied by these exchanges in the price formation process. Note also that this agrees well with the ranking of markets in terms of volume: the Binance USDT perpetual has the largest average daily volume in the month of February by a sizable margin, as seen in Figure~\ref{fig:feb_24h_volumes}.

\subsection{Accounting for Trading Realities}
\label{sec:trading_realities}

So far we have assessed the goodness-of-fit of our models only by an in-sample measure, namely their $R^2$ values.
This measure suffers from two complications.
First, the $R^2$ of a model is an inherently dimensionless quantity, and second, it is not clear how an $R^2$ value translates to the realities of trading. 
Out-of-sample measures like RMSE or misclassification error can give a more interpretable quantity having a dimension, but they too suffer from the latter complication mentioned above. 
It is unclear how, for example, an accuracy score maps to PnL.
One can conceive of cases where a model has an extremely high accuracy score which maps to a large number of slightly profitable trades, but where a single misclassification results in a disastrous loss that wipes out any prior profits and more. 
To address these concerns, we propose an alternative out-of-sample way of evaluating goodness-of-fit which is more in line with trading realities.
Our measure will be defined as the PnL of a natural trading strategy associated with the model from Eqn.~\eqref{eqn:model_ij}. 
This PnL value will be computed from a synthetic walk-forward on an out-of-sample data set, consisting of the two days following our training period. The training period spanned February 22-27 2021, and the test period includes February 28th and March 1st.

\textbf{Mapping a linear model to a trading strategy.} We now describe how we map each model from Eqn.~\eqref{eqn:model_ij} to a trading strategy.
The basic idea is very simple: we buy when the prediction is large (and positive), and sell when the prediction is small (and negative). 
More precisely, the starting point is the strategy which generates predictions on out-of-sample observations and places a hypothetical buy order at the top ask price when the model predicts a value $>T$. 
When the prediction is $<-T$, the strategy places a hypothetical sell order. Clearly, this strategy involves the choice of a threshold $T$. 
However, there is a relatively canonical choice that we can make here: we set $T$ as a quantile value of the set of predictions that the model produced on the training period. Let us introduce some notation to make this more precise. We denote by $M_{ij}$ the model from Eqn.~\eqref{eqn:model_ij}, and let us write $\operatorname{preds}_{\text{in-sample}}(M_{ij})$ for the set of in-sample predictions produced by $M_{ij}$. Then we define
\begin{equation}
T_{ij} := \text{95-th percentile of the set } \operatorname{preds}_{\text{in-sample}}(M_{ij}).
\end{equation}

The strategy in its current state still suffers from (at least) two technical issues which we circumvent with a simple additional constraint on the strategy.
The first problem is that when we, for example, place a hypothetical buy order anytime the prediction is $>T_{ij}$, we often obtain bursts of many buy orders over a period of time while the prediction remains large.
However, of the trades belonging to such a burst of buy orders often only the first one is actionable since it will impact the market and usually change the top ask price. 
The second issue is that the PnL calculation becomes more complicated and noisy when we have frequent bursts of buy or sell orders (compared with our alternative approach outlined below). 
For these reasons, we impose the additional constraint on the strategy that the hypothetical trader's position shall never exceed ``one unit".
That is, in our sequence of hypothetical trades, a buy order may only be followed by a sell order (and a sell order only by a buy order). 

We can summarize the strategy associated to $M_{ij}$ by the following set specifications
\begin{enumerate}
    \item Initialize the trader's position as $p=0$.
    \item If $p\leq 0$ and the prediction of $M_{ij}$ on an unseen observation is $>T_{ij}$, we execute a hypothetical buy order at the top ask price and set $p=1$.
    \item If $p\geq 0$ and the prediction of $M_{ij}$ on an unseen observation is $<-T_{ij}$, we execute a hypothetical sell order at the top bid price and set $p=-1$.
\end{enumerate}
By chronologically scanning over the out-of-sample period (February 28th and March 1st) in this manner, we obtain a sequence of hypothetical trade prices of the following form
\begin{equation}
\mathcal S_{ij} = \left( \ldots, p^{ij, buy}_k, p^{ij, sell}_{k+1}, p^{ij, buy}_{k+2}, \ldots  \right).
    \label{eqn:pnl_ij}
\end{equation}
Next, we will compute various measures of PnL from the above sequence.

\textbf{Results without execution fees.} The first PnL we calculate from the sequence~\eqref{eqn:pnl_ij} is the simplest one: we compute the trader's total PnL over the test period without accounting for the exchange's execution fees yet. 

Note that we already implicitly incorporated the spread component of transaction cost since the trader buys at the top ask price and sells at the top bid price. However, it is also important to observe that oftentimes the top quote (ask or bid) contains only a fleetingly small amount of volume. Not infrequently, this volume is a sub 1000 USD amount and occasionally it is just pennies. This implies that the strategy in its current form is often not scalable.

With this caveat stated, we define our first PnL measure by the following equation
\begin{equation}
\operatorname{PnL}_{1, ij} := \sum_{k=1}^{|\mathcal S_{ij}|} \left( \frac{p^{ij, sell}_{k+1}}{p^{ij, buy}_k} - 1 \right)\cdot 10000, 
\label{eqn:pnl1}
\end{equation}
which can be regarded as the difference in basis points between the average buy price and the average sell price over the sequence $\mathcal S_{ij}$,  multiplied by the total number of trades.

\begin{figure}[!ht]
\hspace*{-2cm}
\vspace{-3mm}
\centering
\includegraphics[width=1\textwidth]{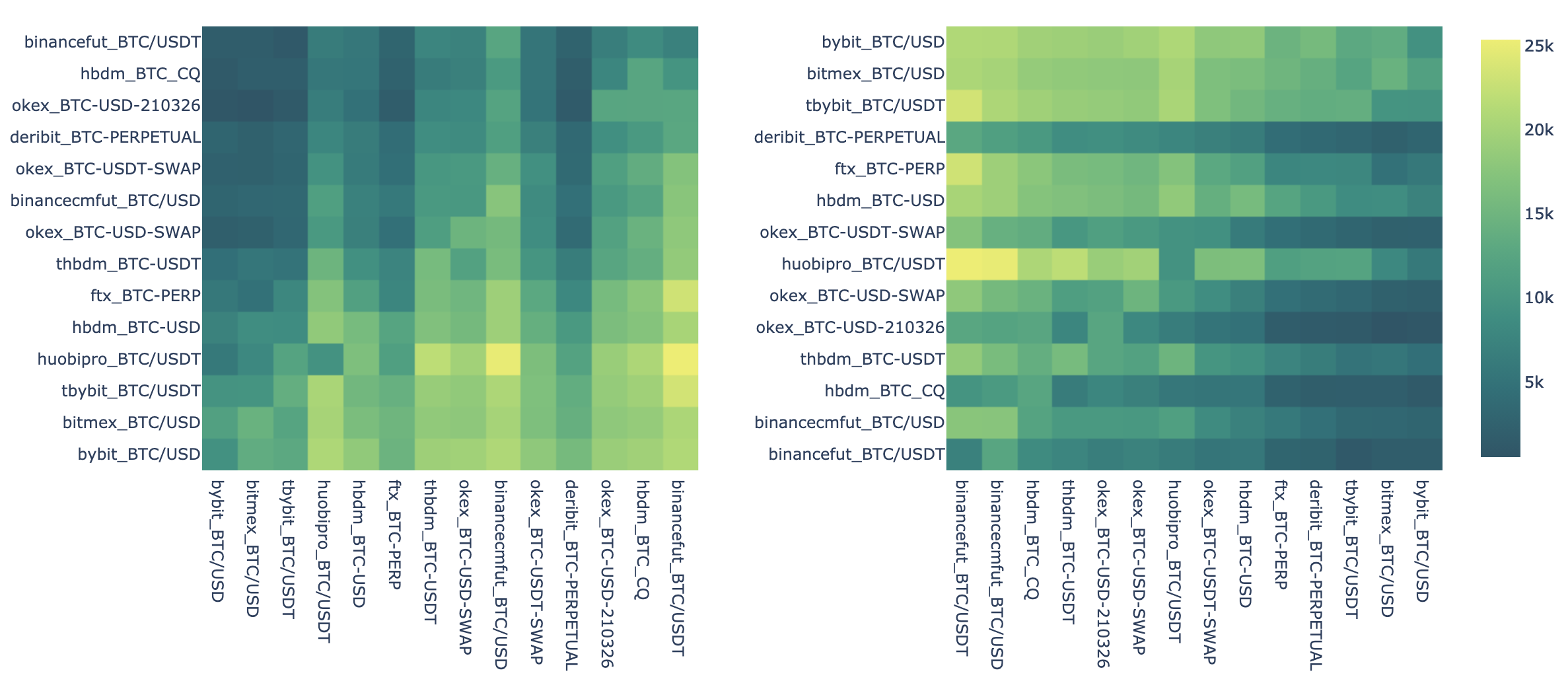}
\captionsetup{width=0.97\linewidth}    
\caption{Visualization of the matrix $\operatorname{PnL}_{1}$ with two orderings, one by row sum and the other by column sum.}
\label{fig:pnl1_heatmaps}
\hspace{-0.9\textwidth} 
\vspace{-3mm}
\end{figure}

The resulting matrix $\operatorname{PnL}_{1} = \left(\operatorname{PnL}_{1, ij} \right)_{i,j=1\ldots,14}$ is visualized as a pair of heatmaps in  Figure~\ref{fig:pnl1_heatmaps}, where the left heatmap is ordered by row sum and the right heatmap by column sum. 

\begin{figure}[!ht]
\hspace*{-1cm}
\vspace{-3mm}
\centering
\includegraphics[width=0.9\textwidth]{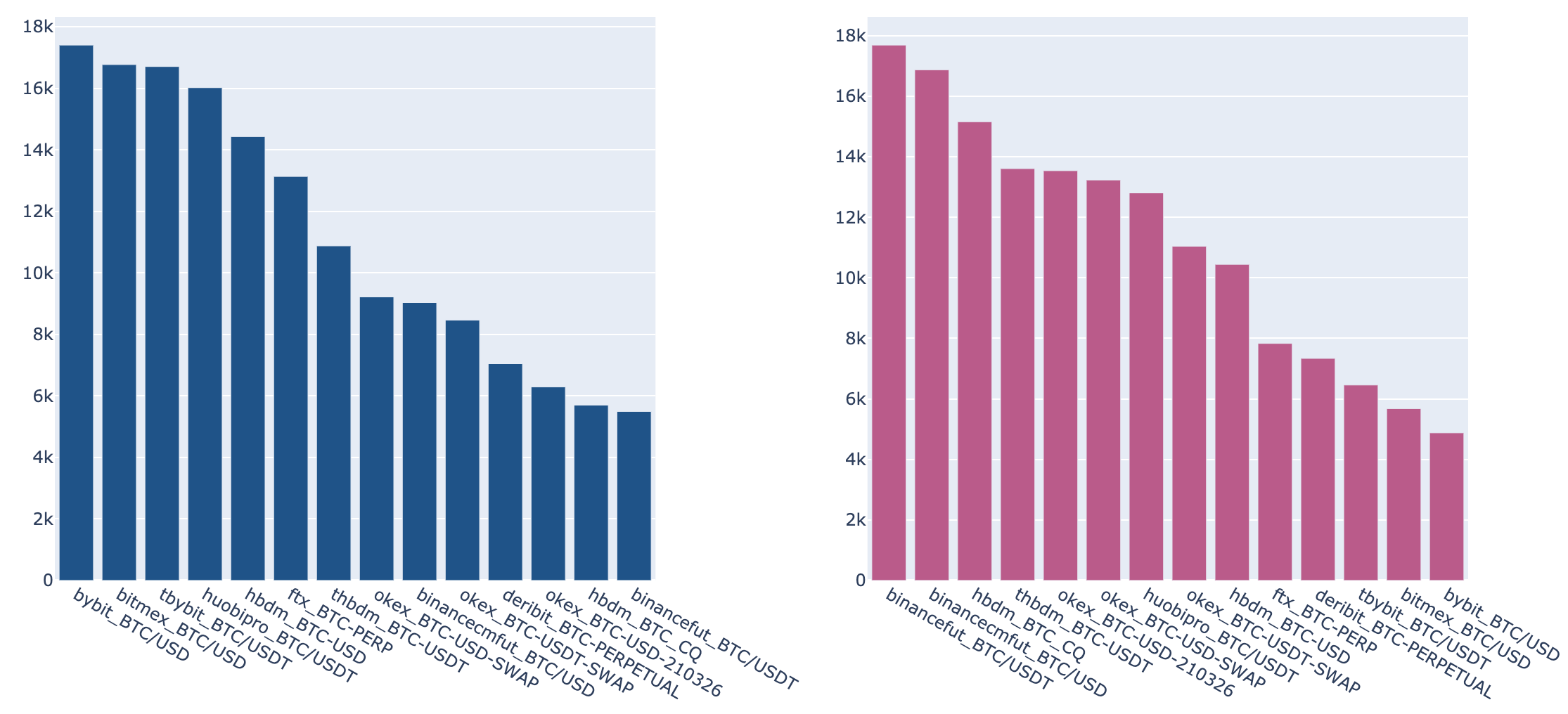}
\captionsetup{width=0.97\linewidth}
\caption{Row and column averages of the matrix $\operatorname{PnL}_{1}$.}
    \label{fig:pnl1_barplots}
\hspace{0.0\textwidth} 
\vspace{-3mm}
\end{figure}

Row and column averages are provided in Figure~\ref{fig:pnl1_barplots}.
Immediately we are struck by the agreement of the results here with those  found in the previous subsection, where our analysis was based on $R^2$ which, reassuringly, provides evidence of the consistency of our approaches. Markets that yield among the largest PnL values are the Bybit at Bitmex perpetuals, which were previously identified as some of the most predictable ones. The most useful markets in producing high PnL values on other markets are the two Binance perpetuals and the Huobi quarterly futures contract, as well as the Huobi USDT perpetual. 

The mean and median value of the matrix $\operatorname{PnL}_{1}$ are both approximately 11000.
Note that this number is the total number of basis points ``accumulated" by the strategy.
This basically says that, on average, we accumulate 11000 basis points over the course of the two day test period, or, in other words, we achieve roughly a doubling of the initial capital.
Of course, this is not realistic, at least not with any sizable amount of capital. How do these results change when we take execution fees into account? This is the question we will turn to next.

\textbf{Results with default execution fees.} As previously noted, exchanges typically have tiered fee structures whereby highly active traders often pay a far lower execution fee than what the default fee is that novice traders are subject to. 
Another way of obtaining fee discounts on some exchanges can be by holding a large amount of the exchange's native cryptocurrency.
This applies in particular to Binance and FTX where steep discounts can be obtained this way.
We refer the reader to Table~\ref{tbl:taker_fee} where, in our previous remarks on execution cost, we display the default and the lowest possible taker fee per market.

\begin{figure}[!ht]
\vspace{-3mm}
\centering
\includegraphics[width=1\textwidth]{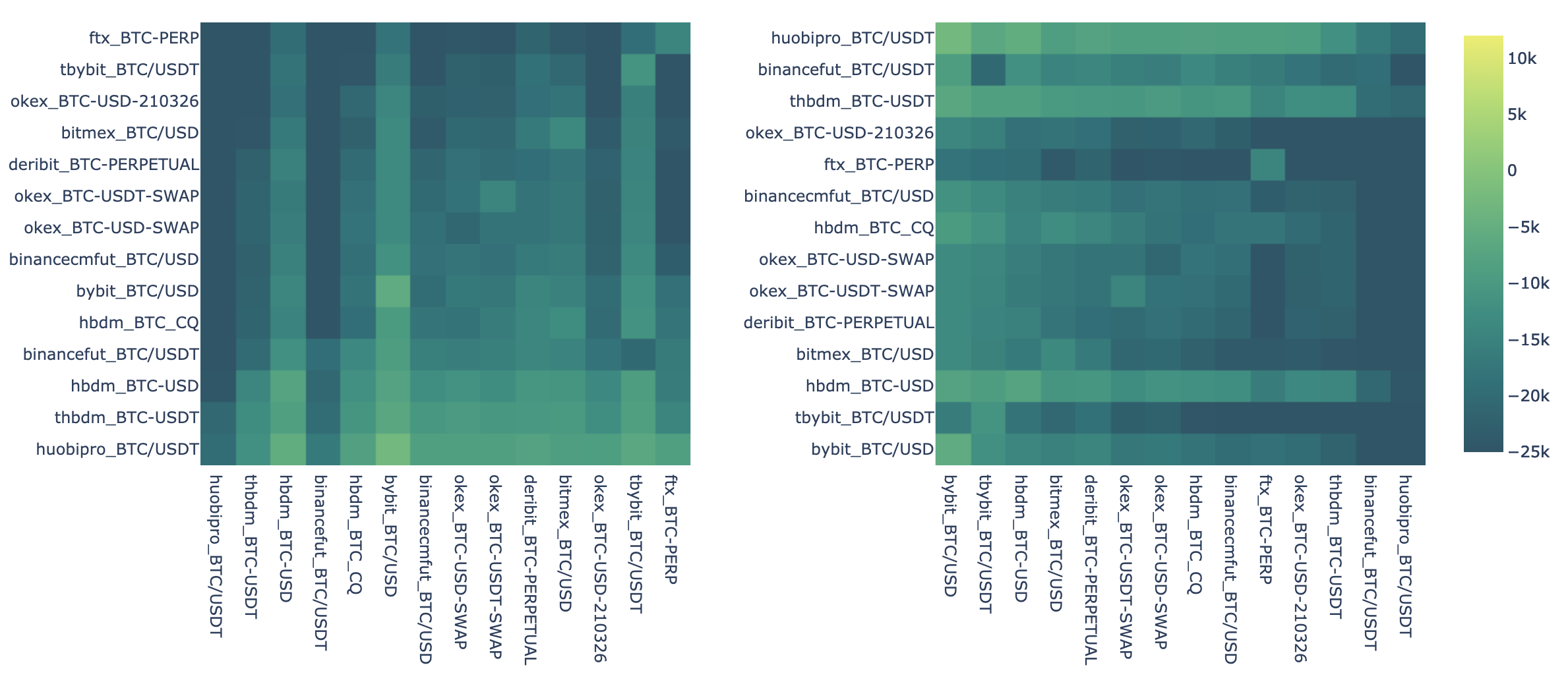}
\captionsetup{width=0.97\linewidth}    
\caption{Visualization of the matrix $\operatorname{PnL}_{2}$ with two orderings, one by row sum and the othre by column sum.}
\label{fig:pnl2_heatmaps}
\hspace{-0.9\textwidth} 
\vspace{-3mm}
\end{figure}

\begin{figure}[!ht]
\vspace{-3mm}
\centering
\includegraphics[width=0.9\textwidth]{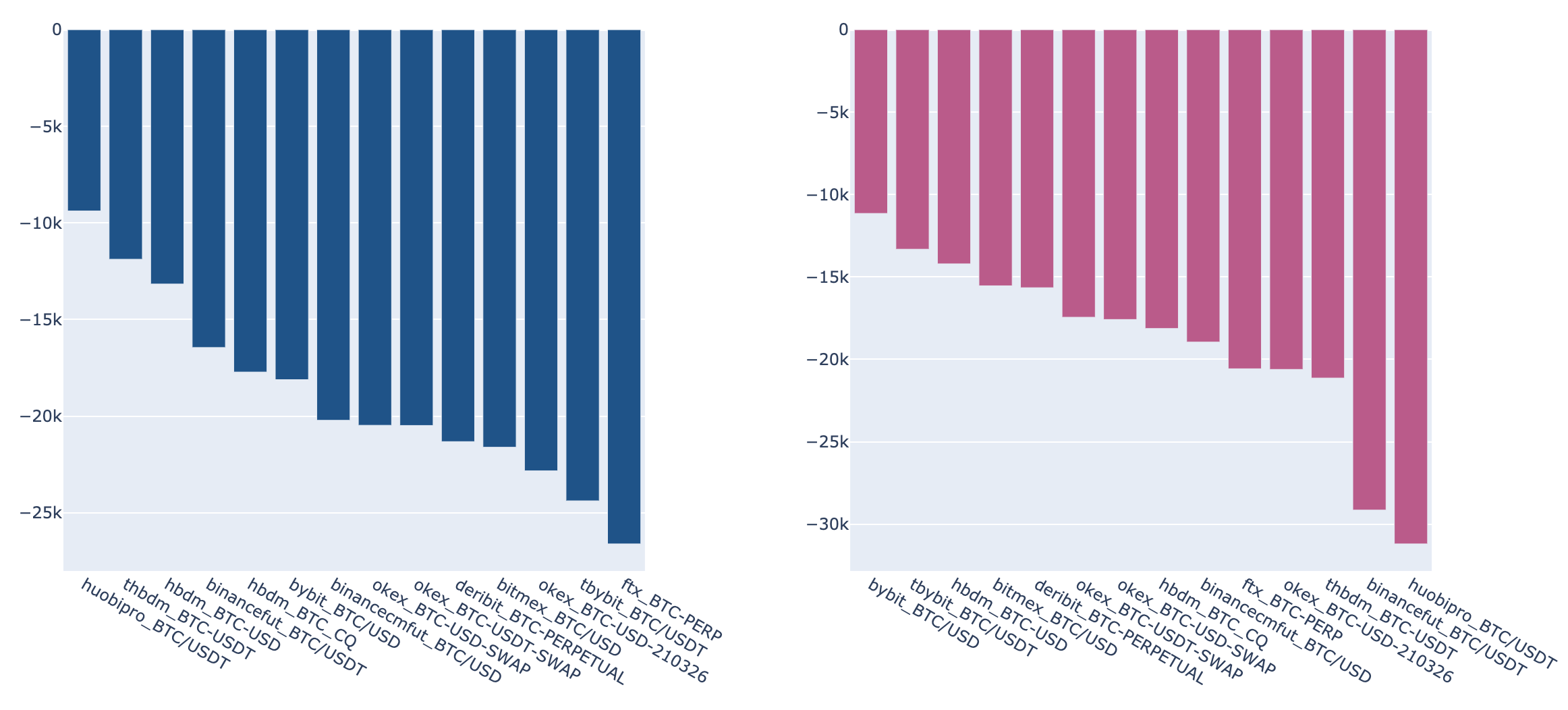}
\captionsetup{width=0.97\linewidth}
\caption{Row and column averages of the matrix $\operatorname{PnL}_{2}$.}
\label{fig:pnl2_barplots}
\hspace{0.0\textwidth} 
\vspace{-3mm}
\end{figure}

For our second PnL measure, we will assume that the default execution fee applies to the trader on each executed trade.
Specifically, we define
\begin{equation}
\operatorname{PnL}_{2, ij} := \operatorname{PnL}_{1, ij} - |\mathcal S_{ij}|f_{i}^{\text{default}}, 
\label{eqn:pnl2}
\end{equation}
which is the first PnL measure in \eqref{eqn:pnl1}, adjusted for the default execution fee. 
We are assuming each trade incurs a fee of $f_{i}^{\text{default}}$, which is our notation for the default taker fee in basis points on market $i$.
The corresponding matrix will be denoted by $\operatorname{PnL}_{2} = \left( \operatorname{PnL}_{2, ij} \right)_{i,j=1,\ldots, 14}$ and it is  illustrated in Figure~\ref{fig:pnl2_heatmaps} along with its column and row averages in Figure~\ref{fig:pnl2_barplots}.

The obtained results are more economically sensible. All PnL values are negative, as one would expect from such a simple model which incorporates information of only one market. The smallest loss is 2819 basis points,  while the overall average loss is 18899 basis points.

A striking observation is that markets which were formerly identified as the most predictive are now among those which yield the highest (although still negative) PnL values. The top five markets by row average are all either on Huobi or Binance, while the lowest row average is attained by the FTX perpetual which we formerly noted as one of the most predictable ones.

\begin{figure}[!ht]
\vspace{-3mm}
\centering
\includegraphics[width=1\textwidth]{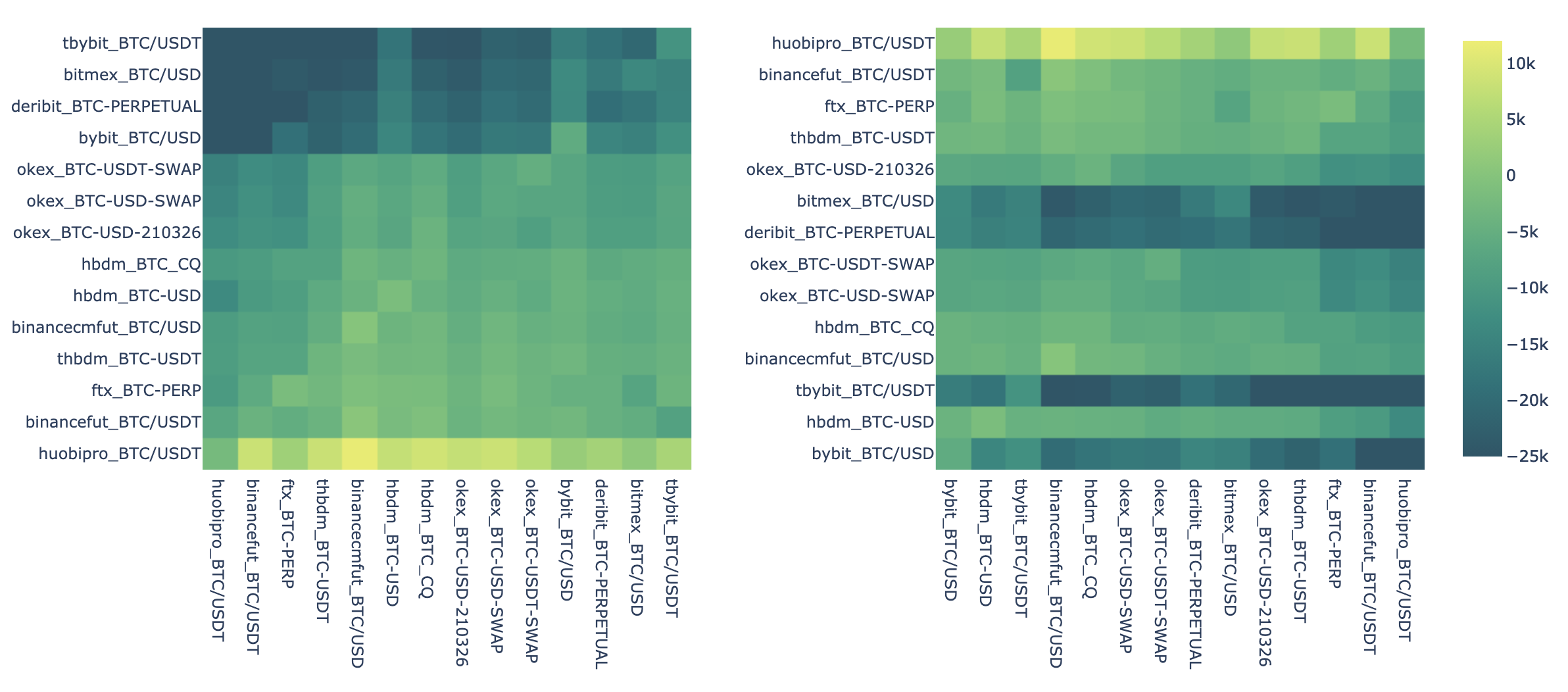}
\captionsetup{width=0.97\linewidth}
\caption{Visualization of the matrix $\operatorname{PnL}_{3}$ with two orderings, one by row sum and the other by column sum.}
\label{fig:pnl3_heatmaps}
\hspace{-0.9\textwidth} 
\vspace{-3mm}
\end{figure}

\begin{figure}[!ht]
\vspace{-3mm}
\centering
\includegraphics[width=0.9\textwidth]{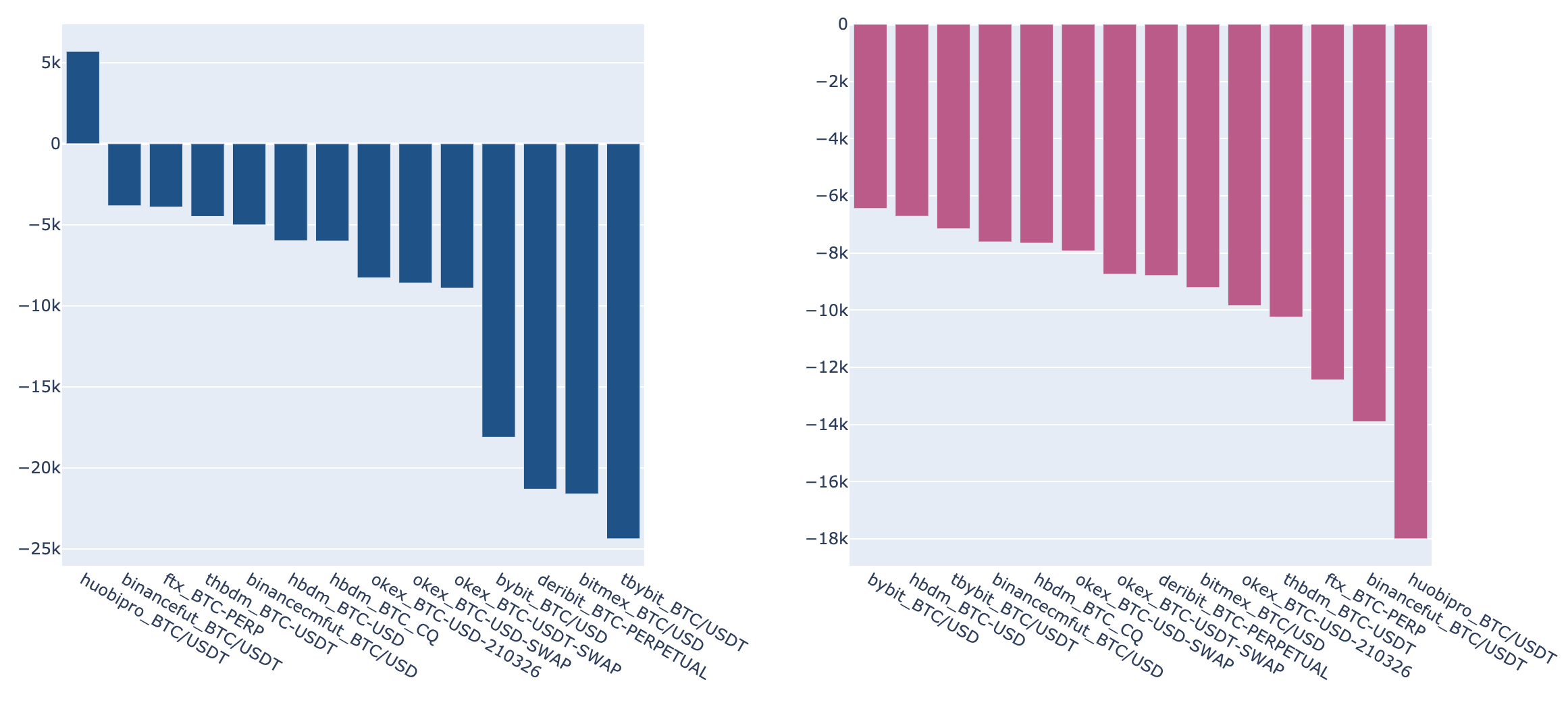}
\captionsetup{width=0.97\linewidth}
\caption{Row and column averages of the matrix $\operatorname{PnL}_{3}$.}
\captionsetup{width=0.97\linewidth}
\label{fig:pnl3_barplots}
\hspace{0.0\textwidth} 
\vspace{-3mm}
\end{figure}

\textbf{Results with VIP execution fees.} The third and final PnL measure is similar to the second one, with the only difference that we use the lowest possible execution fee instead of the default fee.
The lowest fee typically requires on the scale of around 500m USD monthly notional volume, and in some cases significant holdings of the exchange's native cryptocurrency in the trader's exchange wallet. For these reasons, there are only a small number of wealthy individual traders or significant trading firms that receive this fee. 
To this end, we define 
\begin{equation}
\operatorname{PnL}_{3, ij} := \operatorname{PnL}_{1, ij} - |\mathcal S_{ij}|f_{i}^{\text{VIP}}, 
\label{eqn:pnl3}
\end{equation}
which is the first PnL measure adjusted for the lowest possible execution fee $f_{i}^{\text{VIP}}$ on market $i$.
We denote the corresponding matrix by $\operatorname{PnL}_{3} = \left( \operatorname{PnL}_{3, ij} \right)_{i,j=1,\ldots, 14}$.
A visualization can be found in Figure~\ref{fig:pnl3_heatmaps}.
The row and column averages can be seen in the bar plots in Figure~\ref{fig:pnl3_barplots}.

The surprising implication of these results is that the most leading 
markets are the ones where the largest PnL can be achieved. 
For instance, the Binance USDT perpetual has the second largest row average. Conversely, many of the most lagging markets are the ones where the lowest PnL is achieved.
An example of this would be the Bitmex perpetual which has the second lowest row average.
The Huobi spot market seems to be an outlier, where even when averaged across markets, a positive PnL can be achieved. It should be noted that this market is one with often particularly small top of the book amounts which implies that the capital that can be deployed towards this strategy is particularly small (perhaps a single digit USD amount) since we are assuming execution at the top of the book price.
 
Overall, in the $\operatorname{PnL}_{3}$ matrix, we see an average loss of 9621 basis points. The highest value is observed for trading on Huobi spot using information from Binance BTC perpetual, where the PnL is 11426 basis points. The largest loss (of size -42435 basis points) is incurred by the model trading on the Bybit USDT perpetual using features of the Huobi spot market.

\textbf{Remarks on the Role of Exchange Fees.} Exchange fees have been used in studies of price formation, e.g. by~\cite{MALINOVA_2015}. We identified a set of leading markets and a set of lagging markets using an approach based on $R^2$ values of the models $M_{ij}$. In our first comparison with the PnL results where we ignored execution fees, we found strong agreement with the $R^2$ based lead-lag analysis. In line with basic intuition, we found that the most lagging markets yielded the largest PnL values. With execution fees accounted for, however, we obtained a very different picture. Some of the most leading markets attained the highest PnL values, while some of the most lagging markets attained among the lowest PnL values. What are some possible explanations of this phenomenon?
We will argue that a market's fee regime naturally predisposes it to either being a leader or a lagger (or something in between).
This natural tendency in and of itself does not correspond to trading ``alpha" (positive PnL).
Only when a market exhibits ``leadingness" in excess of its natural disposition to being a leader, can we extract alpha to produce positive PnLs on other markets.
By the same token, a market's ``laggardness" in and of itself only implies alpha (positive PnL) if the lag effect is greater than what is implied by the market's fee regime.

So how does the fee structure on a market correspond to a natural disposition to being a leader of laggard? 
To this end, let us consider the example of a pair of markets $(1, 2)$ with differing fee structures and reason about how these affect the exchange's lead-lag behaviour in each case.
Let us assume that the difference between taker fee and maker rebate is the same on both exchanges.
That is, there exists a scalar $c$ such that for $i \in \{1, 2\}$
\begin{equation}
\operatorname{fee}_i - \operatorname{rebate}_i = c, 
\label{eqn:taker_maker_diff}
\end{equation}
where $\operatorname{fee}_i$ denotes the taker fee on market $i$ and $\operatorname{rebate}_i$ denotes the maker rebate on market $i$.
Note that the taker fee is the execution cost in basis points paid for a market order (or a marketable limit order).
The maker rebate is the rebate, also in basis points, received by the market maker upon receiving a fill for a passive limit order.
The maker rebate can also be negative, in which case the market maker incurs a cost for getting a fill.
In practice, one finds a somewhat uniform difference
\begin{equation}
\operatorname{fee} - \operatorname{rebate} \approx 5, 
\end{equation} 
when one compares across exchanges (with VIP traders often enjoying a smaller difference, hence paying less net fees).
Eqn.~\eqref{eqn:taker_maker_diff} of course does not impose any restrictions on the values of $\operatorname{fee}_i$ or $\operatorname{rebate}_i$ themselves. 
Let us suppose that $c=5$ and that the first market has fee structure $(\operatorname{fee}_1, \operatorname{rebate}_1) = (10, 5)$ while the second one uses $(\operatorname{fee}_1, \operatorname{rebate}_1) = (5, 0)$. 

An informed market taker (a trader contemplating the placement of a taker order) will rationally place a buy order on market $i$ if and only if they expect an upward price move of at least $\operatorname{fee}_i$ basis points over the course of their expected holding period.
In other words, the trader sends a taker buy order if and only if the following condition is met
\begin{equation}
\mathbb E\left[ \left( \frac{p_{t+h, i}}{p_{t, i}} -1\right)\cdot 10000 \right] > \operatorname{fee}_i,
\label{eqn:taker_calculus}
\end{equation}
where $p_{t, i}$ is the price on market $i$ at time $t$, and $h$ is the trader's expected holding time. 
Since $\operatorname{fee}_1 > \operatorname{fee}_2$, this implies that the trader requires a stronger signal to send a taker order on market $1$ and is therefore disincentivized from sending taker orders.
From the perspective of the market maker, the larger rebate $\operatorname{rebate}_1$ on market $1$ implies a disincentive to cancel passive orders, since a larger amount of adverse selection can be compensated for by the maker rebate on market $1$ relative to market $2$.
The combination of these twin effects implies a natural tendency of market $1$ to be a laggard.
In the limiting case $\operatorname{fee}_1, \operatorname{rebate}_1 \to \infty$, the orderbook would be completely static with no transactions ever occurring as market makers can tolerate an arbitrary amount of adverse selection, and takers require an infinitely strong price signal to overcome the taker fee. The conclusion therefore is that market $2$ is naturally disposed toward lagging behaviour when compared to market $1$ (which is consequently a leader in this comparison).

This line of reasoning leads us to conclude that the extent to which a market is a leader or a lagger does not, in and of itself, have any implications regarding trading ``alpha" that can be achieved for trading on said market or for utilizing its information to trade on other markets.
The reality is more subtle due to potentially different fee regimes (even when the difference between taker fee and maker rebate is uniform across exchanges).
Only when a market exhibits lagging behaviour \emph{in excess} of the laggardness to which it is naturally predisposed by its fee structure do we obtain trading ``alpha" that can be exploited to generate positive PnL values on the market in question.
For instance, counter-intuitively, the Binance USDT perpetual, while generally a strong leading market, is nevertheless sufficiently predictable, owing to its fee structure, to give rise to positive PnL values.

\section{Cross-Sectionally Combining Information}
\label{sec:5}

Only by taking into account information from the full set of relevant markets, can one hope to reliably make accurate predictions on short term order flow, and hence price action on any given market.
Evidence in support of this assertion was provided in the previous subsection, where we observed that a single market's features are insufficient to produce ``alpha" in excess of execution cost on another market (with rare exceptions for VIP traders that receive heavy fee discounts).
The underlying explanation is that market makers, who are responsible for the bulk of liquidity in the orderbook, continuously adjust their quotes on the basis of \emph{all} available information sources  which consistently bear relevance for price moves.
For if this were not the case, they would be vulnerable to being picked off by quick arbitrageurs who are ``on the prowl" for resting limit orders which are no longer in line with the rest of the market.
For instance, suppose the price moved up substantially on all markets except one.
The market participant who continues to post a passive sell order on the market where a price move up has not yet occurred is apparently ignorant to the new information dispersed among other markets, and will no doubt shortly incur a loss by having his order consumed by an arbitrageur.
Hence the market makers that ``survive" (the profitable ones that do not consistently incur such losses), and therefore ultimately dictate the bulk of the liquidity in the orderbook and thus also the price, are precisely the ones that are capable of parsing all available relevant information so as to prevent being adversely selected.

Therefore, if we keep in mind our objective of predicting order flow and price change on sub-second time scales, we ought to search for ways to combine information across markets to achieve greater predictive power.
One natural starting point is to simply fit linear regression models (as before with OLS) using the full set of features from all markets instead of just a single market.
This will be the starting point and a baseline for comparison of our analysis in this subsection. One concern with this approach is that the model might be overfitted due to a significant number of  
irrelevant
features and/or a strong degree of collinearity between features.
We will therefore compare the baseline approach with the following two alternative approaches.

The first alternative approach is to deal with the large number of features by using $L^1$-regularization in the regression model.
That is, instead of fitting a simple OLS model on the full feature set, we will train a LASSO regression model which ends up selecting a subset of the features.

The second alternative approach involves linearly combining features across the markets to form a set of ``meta features" which are subsequently leveraged to generate parsimonious models for price prediction on each market.
The linear combinations will be formed using coefficients that are proportional to predictive power, as measured by $R^2$.

\subsection{Baseline Approach}

\textbf{Model specification.} Our baseline model consists of a linear regression trained on the full feature set.
Each market has 5 features and there are a total number of 14 markets, resulting in a total of 70 covariates. 
Formally, we use OLS regression to fit the following linear models for all $i\in \{1, \ldots, 14\}$ and $\delta \in \left\{ 500\text{ms}, 1000\text{ms} \right\}$
\begin{equation}
fret_{t}^{\delta, i} = \mu_{i} + \sum_{j=1}^{14} \left( \beta_{ij,1} IMB_{t}^{a, j} + \beta_{ij, 2} IMB_{t}^{b,j} + \beta_{ij, 3} TFI_{t}^j + \beta_{ij, 4} PRET_{t}^j + \beta_{ij, 5} DIV_{t}^{ij} \right) + \epsilon_{i, t}.
    \label{eqn:baseline_model_i}
\end{equation}
As usual, to keep notation simple, we let it be implicit that each feature is dependent on the choice of a time horizon, using the one that was previously deemed optimal for it.
It is also implicit in the notation that we use the transformed feature rather than its original version, whenever an improvement was found for the feature. 
Let us denote the model corresponding to Eqn.~\eqref{eqn:baseline_model_i} by $M_{i,\delta}^{\text{baseline}}$.

\textbf{Results.} We now examine the coefficients of determination of the models $M_{i,\delta}^{\text{baseline}}$ for $i=1,\ldots, 14$ and $\delta~\in~\left\{ 500\text{ms}, 1000\text{ms} \right\}$.
We compute this number both in-sample and out-of-sample.
In other words, we report the portion of total variation in future returns explained by each model on the training period, as well as on the two-day test period following the training period.
The results can be found in  Table~\ref{tbl:baseline_models_r2s}.
\begin{table}
\centering
\caption{In-sample and out-of-sample $R^2$ values for baseline models}
\label{tbl:baseline_models_r2s}
\begin{tabular}{lllll}
\toprule
{} & \multicolumn{2}{l}{$R^2$} & \multicolumn{2}{l}{$R^2_{oos}$} \\
{} &  500ms & $1000$ms &       500ms & $1000$ms \\
\midrule
ftx\_BTC-PERP          &  0.364 &  0.342 &       0.273 &  0.264 \\
bybit\_BTC/USD         &  0.294 &  0.331 &       0.289 &   0.32 \\
thbdm\_BTC-USDT        &  0.273 &  0.231 &        0.25 &  0.215 \\
hbdm\_BTC-USD          &  0.229 &  0.262 &       0.214 &  0.243 \\
huobipro\_BTC/USDT     &  0.219 &  0.136 &       0.191 &  0.126 \\
bitmex\_BTC/USD        &  0.217 &  0.292 &       0.193 &   0.27 \\
tbybit\_BTC/USDT       &  0.214 &  0.279 &       0.219 &   0.28 \\
binancecmfut\_BTC/USD  &  0.209 &  0.193 &       0.187 &  0.177 \\
deribit\_BTC-PERPETUAL &  0.187 &  0.212 &       0.178 &  0.214 \\
hbdm\_BTC\_CQ           &  0.186 &  0.152 &       0.171 &  0.144 \\
okex\_BTC-USDT-SWAP    &  0.173 &  0.196 &       0.156 &  0.184 \\
okex\_BTC-USD-210326   &   0.17 &  0.164 &       0.158 &  0.155 \\
okex\_BTC-USD-SWAP     &  0.164 &   0.18 &       0.151 &  0.167 \\
binancefut\_BTC/USDT   &  0.106 &  0.071 &       0.094 &  0.069 \\
\bottomrule
\end{tabular}
\end{table}
Significance values and p-values for all models are reported in Tables~\ref{tbl:baseline_models_tstats} and~\ref{tbl:baseline_models_pvalues} in the Appendix.

We observe that the difference between the two time horizons $500$ms and $1000$ms is generally rather small.
As one would expect, the 500ms time horizon is somewhat more easily predictable.
On average, the in-sample $R^2$ values for this horizon are $4.6$\% larger than their $1000$ms counterparts.
When we perform the same computation on the out-of-sample values, we find that the $R^2$ on the 500ms horizon is, on average, greater than that the $1000$ms $R^2$ by $1.4$\%.
There are, however, some noteworthy outliers.
Notably larger values on the 500ms horizon are found on the Binance USDT perpetual and the Huobi spot market, while the Bitmex and Bybit USDT perpetuals exhibit considerably larger $R^2$ values for the $1000$ms horizons.
What are some possible explanations of these phenomena?

When a market's $1000$ms $R^2$ is much greater than the $500$ms $R^2$ it means that the features at time $t$ contain information that is not yet reflected in the price at time $t+500$ms but is priced in at $t+1000$ms.
This suggests the possibility that cross-sectional arbitrages can survive for more than $500$ms between a strong leading market and a strong lagging market.
Note that this agrees well with our previous identification of Bitmex and Bybit as lagging exchanges, and Binance and Huobi as leading exchanges.

Conversely, when the $500$ms $R^2$ is much larger than its $1000$ms counterpart on a given market, this indicates that this market is prone to the arrival of new information in the time window $t+500$ms to $t+1000$ms which is not yet encapsulated in the feature values at time $t$.
We would predict this to be the case for strongly leading markets, as indeed appears to be the case when we compare the $R^2$ values with our previous findings on the leader-lagger network.

Our next observation pertains to the comparison between in-sample and out-of-sample values.
On average, we find that on the $500$ms time horizon, the out-of-sample $R^2$ is around 91\% of the in-sample one.
The largest discrepancy, both in relative and absolute terms, can be seen for the FTX perpetual where the out-of-sample $R^2$ is 25\% lower than the in-sample $R^2$.
Overall, the small to moderate loss in accuracy as we pass from the in-sample to the out-of-sample values ameliorates concerns of overfitting to an extent.

\begin{figure}[!ht]
\hspace*{-1cm}
\vspace{-4mm}
\centering
\includegraphics[width=0.75\textwidth]{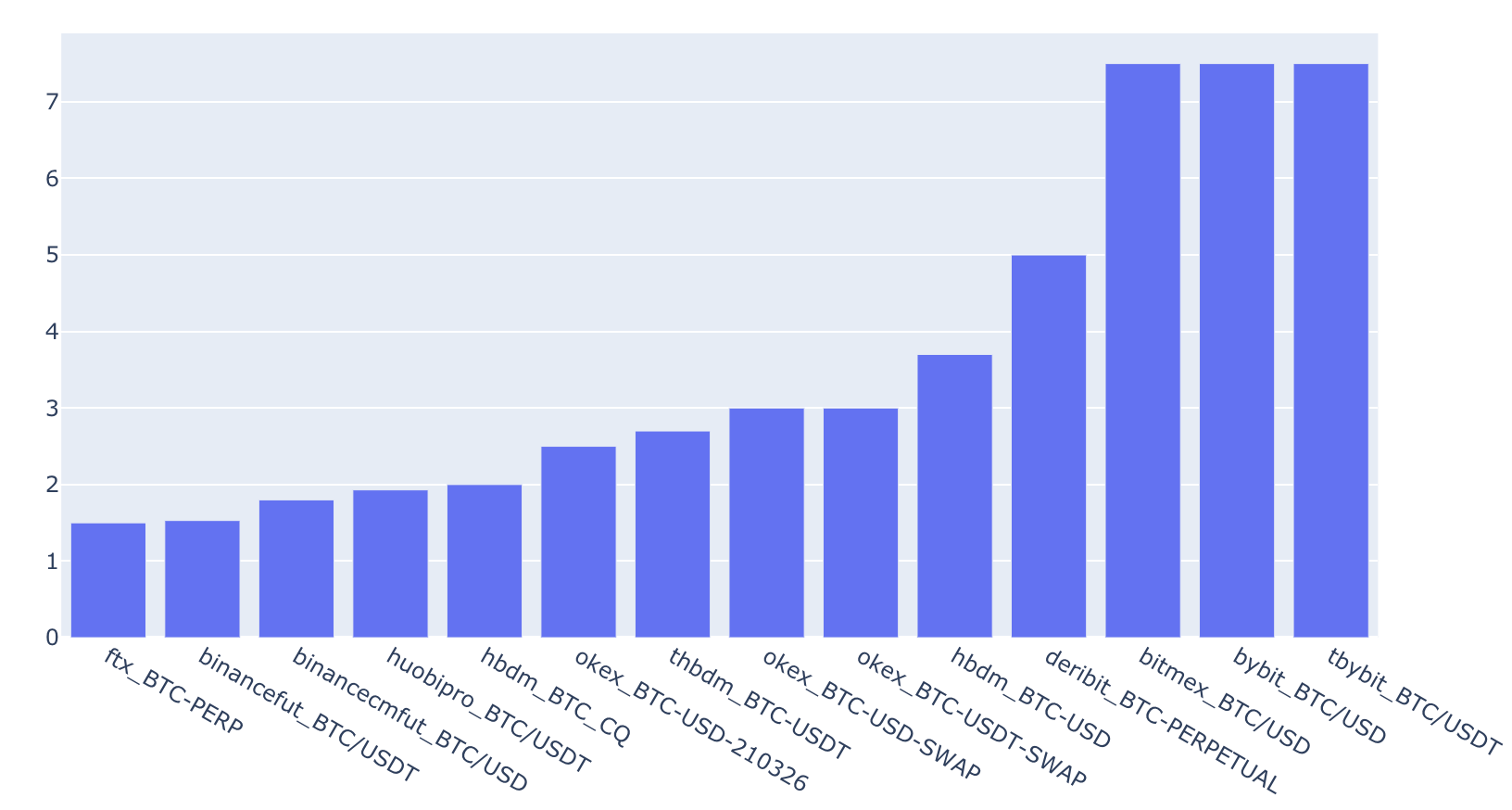}
\captionsetup{width=0.97\linewidth}
\caption{Lowest possible (across tiers) taker fee per market (in basis points).}
\label{fig:lowest_taker_fee}
\hspace{0.0\textwidth} 
\vspace{-3mm}
\end{figure}

Next, we note that by using a market's actual ``leadingness" or ``laggardness", as measured by $R^2$, in conjunction with its fee structure, one  can form a set of expectations about what PnL values could be obtained.  
See Figure~\ref{fig:lowest_taker_fee} for a visual overview of each market's lowest possible taker fee (usually accessible only to highly active traders).
As we reasoned in the previous section, we would a-priori expect a market with a low taker fee to be a leading market, which should express itself in a low $R^2$ value.
If we find that a market which ``should" be a leader based on its fee structure is in fact a lagger, we would predict a high PnL value for this market.
When we compare Figure~\ref{fig:lowest_taker_fee} with Table~\ref{tbl:baseline_models_r2s} we observe that, remarkably, the FTX perpetual has both the highest $R^2$ (indicating that its returns can be anticipated well) and the lowest taker fee.
On this basis, one would predict a large PnL value for those VIP traders with access to this fee.

The Bybit BTC perpetual is among the most lagging markets, although this is in agreement with the fact that it has the highest taker fee, so we would not necessarily foresee a high PnL value for this market, unless its lagging nature is so significant that it outweighs the large fee.

On the other end of the spectrum, the most leading market, the Binance USDT perpetual, also has one of the lowest taker fees, so it is not straightforward to predict what the PnL might be on this market.
Whether a trader subject to the lowest fee is capable of producing a positive PnL boils down to the question of whether the general leading nature of this market can be overcome by its low taker fee.

In Table~\ref{tbl:baseline_models_pnls} we display the PnL values produced by the trading strategies corresponding to the models $M_{i,500\text{ms}}^{\text{baseline}}$ for $i=1,\ldots, 14$.
The values are computed by the same procedure we introduced in Subsection~\ref{sec:trading_realities}.
That is, we map a model to a natural trading strategy and calculate the PnL of this strategy in a synthetic walk-forward over the out-of-sample period.
As before, we compute the PnL in three different ways: (1) ignoring execution fees, (2) adjusting for the default execution fee, (3) adjusting for the lowest possible (``VIP") execution fee.
The PnL formulae are precise analogues of the ones from Equations~\eqref{eqn:pnl1},~\eqref{eqn:pnl2}, and~\eqref{eqn:pnl3}. 
Note that, as before, the PnL therefore tells us what the total number of accumulated basis points over the two day test period is.

In line with the expectations we formed on the basis of predictability and fee regime, we find that FTX gives the largest $\operatorname{PnL}_3$ value.
Somewhat surprisingly, even the most leading market, the Binance USDT perpetual, exhibits a positive $\operatorname{PnL}_3$ value.
This can be interpreted as a suggestion that the extremely low taker fee is sufficient to ``overpower" the market's relative lack in predictability.
The lowest PnL is found on Deribit which has one of the highest taker fee and middling $R^2$ values.
Similarly, Bitmex has slightly larger $R^2$ values but also a larger taker fee, so it is not surprising to see it having the second smallest $\operatorname{PnL}_3$ value. 
\begin{table}
\centering
\caption{PnLs of baseline models}
\label{tbl:baseline_models_pnls}
\begin{tabular}{llll}
\toprule
{} &    $\operatorname{PnL}_1$ &    $\operatorname{PnL}_2$ &    $\operatorname{PnL}_3$ \\
\midrule
ftx\_BTC-PERP          &  19792.4 & -27527.6 &   9652.4 \\
huobipro\_BTC/USDT     &  28844.0 & -28251.0 &   5645.4 \\
binancecmfut\_BTC/USD  &  16519.1 & -22010.9 &   2648.3 \\
binancefut\_BTC/USDT   &  12107.6 & -17724.4 &    696.8 \\
thbdm\_BTC-USDT        &  20784.9 & -10703.1 &   -469.5 \\
hbdm\_BTC-USD          &  17703.1 &  -8186.9 &  -1455.5 \\
hbdm\_BTC\_CQ           &  16432.0 & -23400.0 &  -3484.0 \\
okex\_BTC-USD-SWAP     &  18041.0 & -19319.0 &  -4375.0 \\
okex\_BTC-USD-210326   &  17436.4 & -27273.6 &  -4918.6 \\
okex\_BTC-USDT-SWAP    &  15012.4 & -20097.6 &  -6053.6 \\
bybit\_BTC/USD         &  19811.2 & -11508.8 & -11508.8 \\
tbybit\_BTC/USDT       &  17036.4 & -15828.6 & -15828.6 \\
bitmex\_BTC/USD        &  17804.5 & -16140.5 & -16140.5 \\
deribit\_BTC-PERPETUAL &  10401.1 & -19738.9 & -19738.9 \\
\bottomrule
\end{tabular}
\end{table}

Note that all $\operatorname{PnL}_2$ numbers are negative. This hints to ``the rich get richer" phenomena  in the crypto markets, whereby traders with access to the lowest taker fee have a much greater ability to turn profits than novice traders.
Since a trader's fee regime is determined by trade volume, and in some cases, holdings in the exchange's native cryptocurrency, the privilege of being in the best fee tier clearly correlates highly with the trader's net worth.
On the other hand, a more novice trader wishing to ascend to the best fee tier in order to boost profits might have to ``burn through" some losses in lower fee tiers until they have accumulated enough trading volume to be granted a better fee.

\subsection{LASSO Regression Approach}

The models we trained in the preceding subsection used a total of 70 features from 14 different markets.
Many of these features exhibit high cross-correlations, as evidenced  in Table~\ref{tbl:pret_corrmatrix} where we display the cross-correlations amongst the 500ms past returns features across markets.
\begin{table}
\centering
\caption{Correlation matrix of 500ms past returns. For brevity we used the market acronyms from Table~\ref{tbl:acronym_dict2}. }
\label{tbl:pret_corrmatrix}
\begin{tabular}{lrrrrrrrrrrrrrr}
\toprule
{} &   m1 &  m2 &  m3 &   m4 &   m5 &  m6 &   m7 &  m8 &  m9 &  m10 &  m11 &  m12 &  m13 &  m14 \\
\midrule
m1      &  1.0 &    0.5 &  0.4 &  0.4 &  0.5 &  0.4 &  0.5 &    0.5 &  0.4 &      0.5 &      0.5 &  0.4 &  0.6 &     0.6 \\
m2   &  0.5 &    1.0 &  0.3 &  0.4 &  0.4 &  0.6 &  0.5 &    0.6 &  0.4 &      0.6 &      0.5 &  0.3 &  0.6 &     0.6 \\
m3     &  0.4 &    0.3 &  1.0 &  0.3 &  0.4 &  0.2 &  0.3 &    0.3 &  0.3 &      0.3 &      0.3 &  0.3 &  0.4 &     0.3 \\
m4      &  0.4 &    0.4 &  0.3 &  1.0 &  0.4 &  0.3 &  0.4 &    0.4 &  0.3 &      0.4 &      0.3 &  0.3 &  0.4 &     0.4 \\
m5      &  0.5 &    0.4 &  0.4 &  0.4 &  1.0 &  0.3 &  0.4 &    0.4 &  0.4 &      0.4 &      0.4 &  0.4 &  0.5 &     0.4 \\
m6     &  0.4 &    0.6 &  0.2 &  0.3 &  0.3 &  1.0 &  0.4 &    0.5 &  0.3 &      0.7 &      0.4 &  0.2 &  0.6 &     0.6 \\
m7      &  0.5 &    0.5 &  0.3 &  0.4 &  0.4 &  0.4 &  1.0 &    0.5 &  0.4 &      0.5 &      0.5 &  0.3 &  0.6 &     0.6 \\
m8   &  0.5 &    0.6 &  0.3 &  0.4 &  0.4 &  0.5 &  0.5 &    1.0 &  0.4 &      0.5 &      0.5 &  0.3 &  0.6 &     0.6 \\
m9     &  0.4 &    0.4 &  0.3 &  0.3 &  0.4 &  0.3 &  0.4 &    0.4 &  1.0 &      0.4 &      0.4 &  0.3 &  0.5 &     0.5 \\
m10 &  0.5 &    0.6 &  0.3 &  0.4 &  0.4 &  0.7 &  0.5 &    0.5 &  0.4 &      1.0 &      0.4 &  0.2 &  0.6 &     0.6 \\
m11 &  0.5 &    0.5 &  0.3 &  0.3 &  0.4 &  0.4 &  0.5 &    0.5 &  0.4 &      0.4 &      1.0 &  0.3 &  0.5 &     0.5 \\
m12     &  0.4 &    0.3 &  0.3 &  0.3 &  0.4 &  0.2 &  0.3 &    0.3 &  0.3 &      0.2 &      0.3 &  1.0 &  0.4 &     0.3 \\
m13     &  0.6 &    0.6 &  0.4 &  0.4 &  0.5 &  0.6 &  0.6 &    0.6 &  0.5 &      0.6 &      0.5 &  0.4 &  1.0 &     0.7 \\
m14  &  0.6 &    0.6 &  0.3 &  0.4 &  0.4 &  0.6 &  0.6 &    0.6 &  0.5 &      0.6 &      0.5 &  0.3 &  0.7 &     1.0 \\
\bottomrule
\end{tabular}
\end{table}

In this subsection, we will pursue an alternative methodology for fitting the parameters of Eqn.~\eqref{eqn:baseline_model_i} based on \emph{LASSO regression}.
The difference between an OLS fit and a LASSO fit resides in an additional regularization term introduced in the objective function of the latter.
Specifically, OLS regression minimizes the standard least-squares objective function,  
while LASSO regression minimizes the regularized objective function 
\begin{equation}
\min_{w\in \mathbb R^p} \|{y} - \mathbf{X}w \|^2_2 + \lambda \|w\|_1, 
\end{equation}
where $\mathbf{X} \in \mathbb R^{n\times p}$, $y\in \mathbb R^n$ and $\lambda \in \mathbb R_{>0}$ is the regularization parameter.
The additional $\ell_1$ regularization term has the effect of zeroing a subset of the parameters.
This subset grows in size as the regularization parameter $\lambda$ is increased.
For a full discussion on LASSO regression, the reader is referred to~\cite{hastie01statisticallearning}.

We fit the coefficients in Eqn.~\eqref{eqn:baseline_model_i} using a LASSO regression with regularization parameters from the set $\{ 0.001 \cdot 2^k \Big| k=0,\ldots,8 \}$.
The lower bound $\lambda=0.01$ of this set was chosen since it was found to be the smallest (round) number which, when used as the regularization parameter in the LASSO regression, results in a positive number of zero coefficients for each market's model.
Similarly, the upper bound $\lambda=0.256=0.01\cdot2^8$ was selected since it was the largest doubling of the lower bound $\lambda=0.01$ for which a positive number of nonzero coefficients persist.

For each market and each of the nine regularization parameters, we then examine the number of nonzero coefficients, their in-sample and out-of-sample explanatory power over future returns, and the PnL values of their naturally associated trading strategies. Furthermore, we will also inspect exactly which coefficients tend to survive as $\lambda$ is increased.
Comparisons of the results in this subsection with the results of baseline approach are deferred until we introduce our third alternative methodology for training powerful linear models to explain short-term future returns.

\textbf{Model Selection Results.} First, we inspect the number of  
nonzero 
coefficients of each of the LASSO fits, as the regularization parameter $\lambda$ is increased. The results can be seen in Table~\ref{tbl:lasso_models_surviving_coeffs}. With $\lambda = 0.001$, the majority of the 70 coefficients are found to be nonzero. In line with expectation, as $\lambda$ is increased the number of surviving coefficients decreases. With our largest choice of $\lambda = 0.256$, we typically do not retain more than a handful of features in the model.
\begin{table}
\centering
\caption{Number of surviving coefficients, as we vary the $\lambda$ regularization coefficient.} 
\label{tbl:lasso_models_surviving_coeffs}
\begin{tabular}{llllllllll}
\toprule
{} & 0.001 & 0.002 & 0.004 & 0.008 & 0.016 & 0.032 & 0.064 & 0.128 & 0.256 \\
\midrule
deribit\_BTC-PERPETUAL &    67 &    64 &    61 &    53 &    44 &    36 &    25 &     9 &     4 \\
binancefut\_BTC/USDT   &    67 &    65 &    62 &    54 &    43 &    26 &    16 &     6 &     1 \\
ftx\_BTC-PERP          &    67 &    64 &    57 &    50 &    45 &    40 &    26 &    14 &     6 \\
huobipro\_BTC/USDT     &    67 &    67 &    61 &    51 &    41 &    28 &    21 &     9 &     2 \\
binancecmfut\_BTC/USD  &    67 &    62 &    59 &    49 &    35 &    27 &    17 &     9 &     2 \\
hbdm\_BTC-USD          &    66 &    64 &    56 &    43 &    34 &    27 &    15 &     6 &     6 \\
okex\_BTC-USD-SWAP     &    66 &    62 &    53 &    43 &    38 &    33 &    20 &    10 &     5 \\
okex\_BTC-USDT-SWAP    &    66 &    64 &    58 &    47 &    41 &    35 &    21 &     9 &     4 \\
thbdm\_BTC-USDT        &    66 &    61 &    56 &    51 &    33 &    20 &    14 &    10 &     3 \\
hbdm\_BTC\_CQ           &    65 &    63 &    58 &    52 &    32 &    26 &    22 &     6 &     3 \\
okex\_BTC-USD-210326   &    65 &    61 &    58 &    48 &    37 &    30 &    24 &     9 &     2 \\
bitmex\_BTC/USD        &    62 &    60 &    54 &    48 &    38 &    28 &    22 &    15 &     4 \\
bybit\_BTC/USD         &    62 &    60 &    55 &    33 &    28 &    23 &    17 &     9 &     5 \\
tbybit\_BTC/USDT       &    62 &    61 &    51 &    43 &    39 &    26 &    19 &    10 &     5 \\
\bottomrule
\end{tabular}
\end{table}

Our next set of observations pertains to the explanatory power of each of the LASSO models.
It would be reasonable to hypothesize that perhaps explanatory power (particularly out-of-sample) might increase as we discard superfluous features in a LASSO model, via a suitable regularization parameter. 
However, this turns out to be the case only very rarely, as evidenced in Tables~\ref{tbl:lasso_models_r2s} and~\ref{tbl:lasso_models_r2s_oos}, where we show in-sample and out-of-sample (respectively) $R^2$ values for each model. In fact, $R^2$ seems to generally be monotonically decreasing in the regularization parameter $\lambda$.
That said, the decrease happens quite slowly, especially for $\lambda \leq 0.032$.
For example, the LASSO models with $\lambda = 0.016$ retain on average $97.2$\% of the out-of-sample $R^2$ compared to the LASSO models with $\lambda=0.001$.
When we compare the $\lambda=0.001$ models with the $\lambda=0.032$ ones, we still find an average decrease in out-of-sample $R^2$ of only $5.9$\%.
Since the number of features retained in the latter models ranges between 20 and 40, we can conclude therefore that only around half of the original 70 features are sufficient to retain the vast majority of explanatory power. When we increase the regularization parameter to $\lambda=0.064, 0,128$ and $0.256$, we lose on average $13.3$\%, $25.97$\%, and $39$\% of explanatory power relative to the $\lambda=0.001$ models, respectively.
Both of the aforementioned tables were created using the $500$ms future returns horizon.
For the $1000$ms versions, we refer to Tables~\ref{tbl:lasso_models_r2s_1000ms} and~\ref{tbl:lasso_models_r2s_oos_1000ms} in the Appendix. The results are very similar in both cases. 

\begin{table}
\centering
\captionsetup{width=0.97\linewidth}
\caption{In-sample $R^2$ values for LASSO models using $500$ms future return horizon, as we vary the $\lambda$ regularization coefficient.} 
\label{tbl:lasso_models_r2s}
\begin{tabular}{llllllllll}
\toprule
{} &  0.001 &  0.002 &  0.004 &  0.008 &  0.016 &  0.032 &  0.064 &  0.128 &  0.256 \\
\midrule
ftx\_BTC-PERP          &  0.364 &  0.364 &  0.363 &  0.362 &   0.36 &  0.354 &  0.335 &  0.291 &  0.237 \\
bybit\_BTC/USD         &  0.293 &  0.293 &  0.291 &  0.289 &  0.286 &  0.283 &   0.27 &  0.237 &  0.131 \\
thbdm\_BTC-USDT        &  0.273 &  0.273 &  0.272 &  0.269 &  0.262 &  0.252 &  0.234 &  0.203 &  0.107 \\
hbdm\_BTC-USD          &  0.229 &  0.229 &  0.227 &  0.224 &  0.221 &  0.212 &  0.202 &  0.177 &  0.092 \\
huobipro\_BTC/USDT     &  0.219 &  0.219 &  0.218 &  0.216 &   0.21 &  0.201 &  0.177 &  0.108 &  0.064 \\
bitmex\_BTC/USD        &  0.217 &  0.216 &  0.215 &  0.214 &  0.212 &  0.206 &  0.192 &  0.165 &  0.088 \\
tbybit\_BTC/USDT       &  0.214 &  0.214 &  0.213 &  0.212 &   0.21 &  0.205 &  0.191 &  0.168 &    0.1 \\
binancecmfut\_BTC/USD  &  0.209 &  0.209 &  0.208 &  0.206 &  0.202 &  0.193 &  0.171 &  0.141 &  0.076 \\
deribit\_BTC-PERPETUAL &  0.187 &  0.187 &  0.186 &  0.184 &  0.182 &  0.177 &  0.163 &  0.121 &  0.069 \\
hbdm\_BTC\_CQ           &  0.185 &  0.185 &  0.185 &  0.182 &  0.179 &  0.173 &  0.154 &  0.108 &  0.058 \\
okex\_BTC-USDT-SWAP    &  0.173 &  0.173 &  0.172 &  0.171 &  0.169 &  0.164 &  0.148 &  0.117 &   0.07 \\
okex\_BTC-USD-210326   &   0.17 &   0.17 &   0.17 &  0.168 &  0.165 &  0.161 &  0.144 &    0.1 &  0.059 \\
okex\_BTC-USD-SWAP     &  0.164 &  0.164 &  0.163 &  0.162 &   0.16 &  0.154 &  0.138 &  0.108 &  0.059 \\
binancefut\_BTC/USDT   &  0.106 &  0.106 &  0.105 &  0.103 &  0.096 &  0.083 &  0.058 &  0.036 &  0.002 \\
\bottomrule
\end{tabular}
\end{table}

\begin{table}
\centering
\captionsetup{width=0.97\linewidth}
\caption{Out-of-sample $R^2$ values for LASSO models using $500$ms future return horizon, as we vary the $\lambda$ regularization coefficient.} 
\label{tbl:lasso_models_r2s_oos}
\begin{tabular}{llllllllll}
\toprule
{} &  0.001 &  0.002 &  0.004 &  0.008 &  0.016 &  0.032 &  0.064 &  0.128 &  0.256 \\
\midrule
bybit\_BTC/USD         &  0.289 &  0.289 &  0.287 &  0.283 &  0.281 &  0.279 &  0.272 &  0.263 &  0.236 \\
ftx\_BTC-PERP          &  0.273 &  0.272 &  0.272 &  0.271 &  0.269 &  0.265 &   0.25 &  0.211 &  0.193 \\
thbdm\_BTC-USDT        &   0.25 &   0.25 &  0.249 &  0.247 &  0.242 &  0.234 &  0.223 &  0.211 &  0.149 \\
tbybit\_BTC/USDT       &  0.219 &  0.219 &  0.218 &  0.217 &  0.216 &  0.212 &  0.202 &  0.194 &  0.167 \\
hbdm\_BTC-USD          &  0.213 &  0.213 &  0.212 &   0.21 &  0.206 &  0.198 &  0.193 &  0.188 &  0.176 \\
bitmex\_BTC/USD        &  0.193 &  0.193 &  0.192 &  0.192 &  0.191 &  0.188 &  0.181 &  0.171 &  0.141 \\
huobipro\_BTC/USDT     &  0.191 &   0.19 &   0.19 &  0.188 &  0.183 &  0.176 &  0.159 &  0.095 &  0.073 \\
binancecmfut\_BTC/USD  &  0.187 &  0.187 &  0.186 &  0.184 &  0.181 &  0.173 &  0.153 &  0.136 &  0.101 \\
deribit\_BTC-PERPETUAL &  0.178 &  0.178 &  0.178 &  0.177 &  0.176 &  0.173 &  0.167 &  0.143 &  0.122 \\
hbdm\_BTC\_CQ           &  0.171 &  0.171 &   0.17 &  0.168 &  0.164 &   0.16 &  0.146 &  0.107 &  0.082 \\
okex\_BTC-USD-210326   &  0.158 &  0.158 &  0.158 &  0.156 &  0.154 &  0.151 &   0.14 &  0.103 &  0.083 \\
okex\_BTC-USDT-SWAP    &  0.156 &  0.156 &  0.156 &  0.155 &  0.153 &   0.15 &  0.138 &  0.115 &  0.096 \\
okex\_BTC-USD-SWAP     &  0.151 &  0.151 &   0.15 &  0.149 &  0.148 &  0.144 &  0.131 &  0.111 &   0.09 \\
binancefut\_BTC/USDT   &  0.094 &  0.094 &  0.093 &  0.092 &  0.087 &  0.076 &  0.051 &  0.038 &  0.026 \\
\bottomrule
\end{tabular}
\end{table}

Next, we examine the profits and losses attained by the LASSO models. In Tables~\ref{tbl:lasso_models_pnl1} and~\ref{tbl:lasso_models_pnl2} of the Appendix, we show the $\operatorname{PnL}_1$ and $\operatorname{PnL}_2$ values (respectively) for each model. Most interesting are the $\operatorname{PnL}_3$ values which we display in Table~\ref{tbl:lasso_models_pnl3}. 
We note that in most cases the maximal PnL is achieved with either $\lambda = 0.001$ or $\lambda = 0.002$, although the difference between the maximum PnL and any other PnL with $\lambda \leq 0.032$ is marginal.
This is in agreement with our prior observation that explanatory power is almost identical between all models with $\lambda \leq 0.032$. The only markets where the maximal PnL is not attained with $\lambda \in \{0.001, 0.002\}$ are the Huobi spot market and the Binance USDT perpetual. The former achieves its maximum with $\lambda=0.128$ and nine features while the latter does so with $\lambda=0.032$ and 26 features.

\begin{table}
\centering
\caption{$\operatorname{PnL}_3$ values of LASSO models as regularization parameter varies}
\label{tbl:lasso_models_pnl3}
\begin{tabular}{llllllllll}
\toprule
{} &   0.001 &   0.002 &   0.004 &   0.008 &   0.016 &   0.032 &   0.064 &   0.128 &   0.256 \\
\midrule
ftx\_BTC-PERP          &    9576 &    9439 &    9371 &    9253 &    9444 &    9407 &    9370 &    8847 &    6987 \\
huobipro\_BTC/USDT     &    5738 &    5679 &    5654 &    5637 &    5610 &    6196 &    7264 &    8908 &    3598 \\
binancecmfut\_BTC/USD  &    2576 &    2497 &    2229 &    1802 &    1991 &    1855 &    1511 &    -179 &   -2188 \\
binancefut\_BTC/USDT   &     766 &     824 &     844 &    1274 &    1627 &    2335 &    2095 &    1340 &    1297 \\
thbdm\_BTC-USDT        &    -502 &    -516 &    -732 &   -1013 &   -1246 &   -1756 &   -2527 &   -3722 &   -8110 \\
hbdm\_BTC-USD          &   -1517 &   -1572 &   -1648 &   -1551 &   -1816 &   -2104 &   -2483 &   -3197 &   -4461 \\
hbdm\_BTC\_CQ           &   -3579 &   -3669 &   -3711 &   -4104 &   -4232 &   -4061 &   -4129 &   -5479 &   -9427 \\
okex\_BTC-USD-SWAP     &   -4380 &   -4376 &   -4603 &   -4979 &   -5314 &   -5929 &   -6544 &   -7829 &  -12021 \\
okex\_BTC-USD-210326   &   -4951 &   -5006 &   -5327 &   -5784 &   -5966 &   -6776 &   -7439 &   -8122 &  -11823 \\
okex\_BTC-USDT-SWAP    &   -6050 &   -6105 &   -6418 &   -6770 &   -6951 &   -7605 &   -7752 &   -8334 &  -11518 \\
bybit\_BTC/USD         &  -11631 &  -11440 &  -11504 &  -11808 &  -11895 &  -12510 &  -13771 &  -15709 &  -19746 \\
tbybit\_BTC/USDT       &  -15855 &  -15917 &  -16264 &  -16521 &  -16601 &  -17025 &  -17592 &  -19935 &  -25232 \\
bitmex\_BTC/USD        &  -16209 &  -16091 &  -16341 &  -16386 &  -16311 &  -15813 &  -16328 &  -18577 &  -23969 \\
deribit\_BTC-PERPETUAL &  -19666 &  -19621 &  -19784 &  -20189 &  -20368 &  -20438 &  -20441 &  -20387 &  -25174 \\
\bottomrule
\end{tabular}
\end{table}

\begin{figure}[!ht]
\vspace{-3mm}
\centering
\subcaptionbox[]{Regularization $\lambda=0.032$. Only features with a count of at least seven are included for ease of legibility.
\label{fig:lasso_feature_count_032}
}[ 0.49\textwidth ]
{ \includegraphics[width=0.5\textwidth, trim=0cm 0cm 0cm 0.0cm,clip] {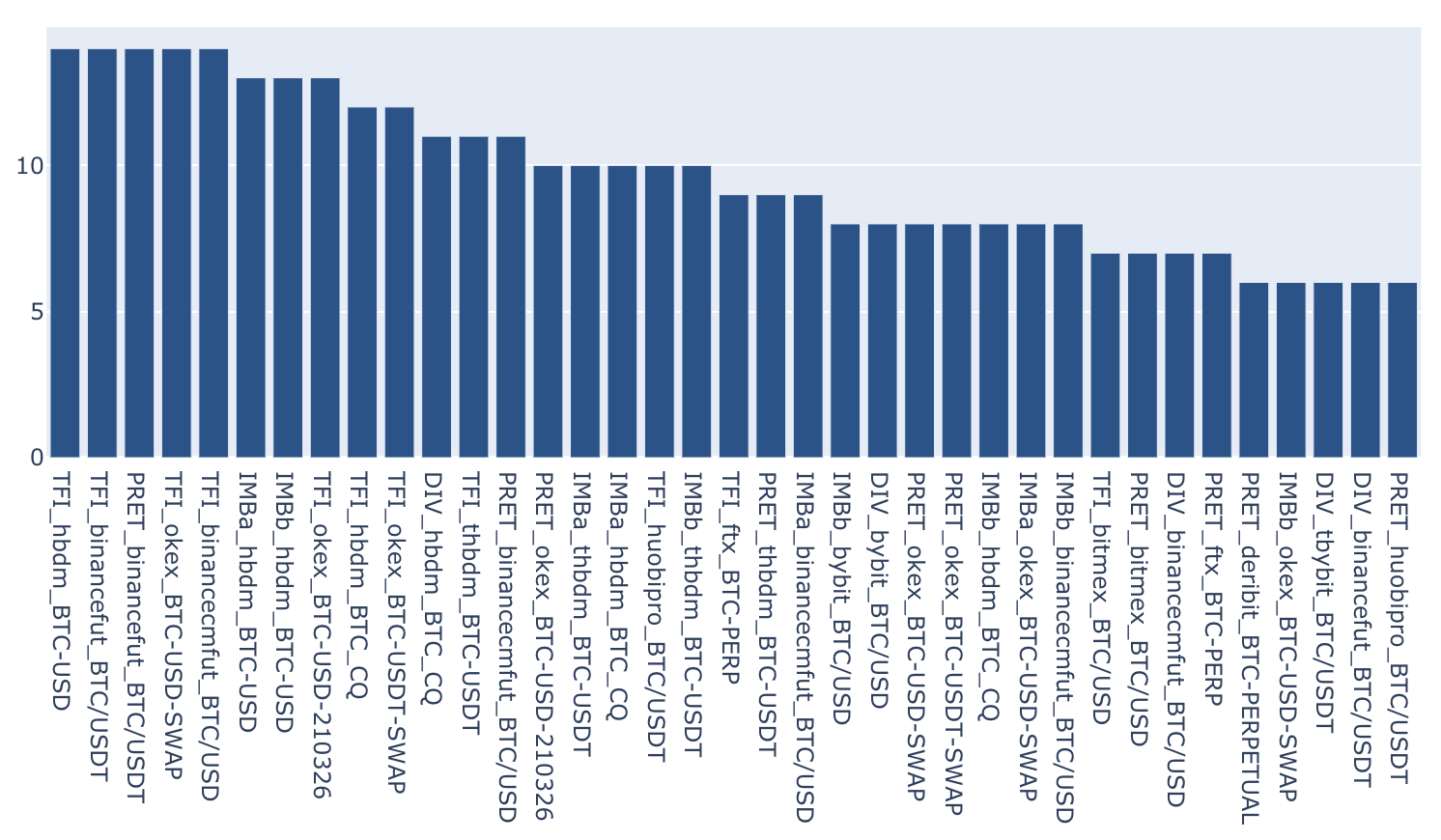} }
\hspace{0.0\textwidth} 
\subcaptionbox[]{Regularization $\lambda=0.256$
\label{fig:lasso_feature_count_256}
}[ 0.49\textwidth ]
{\includegraphics[width=0.5\textwidth, trim=0cm 0cm 0cm 0.0cm,clip]{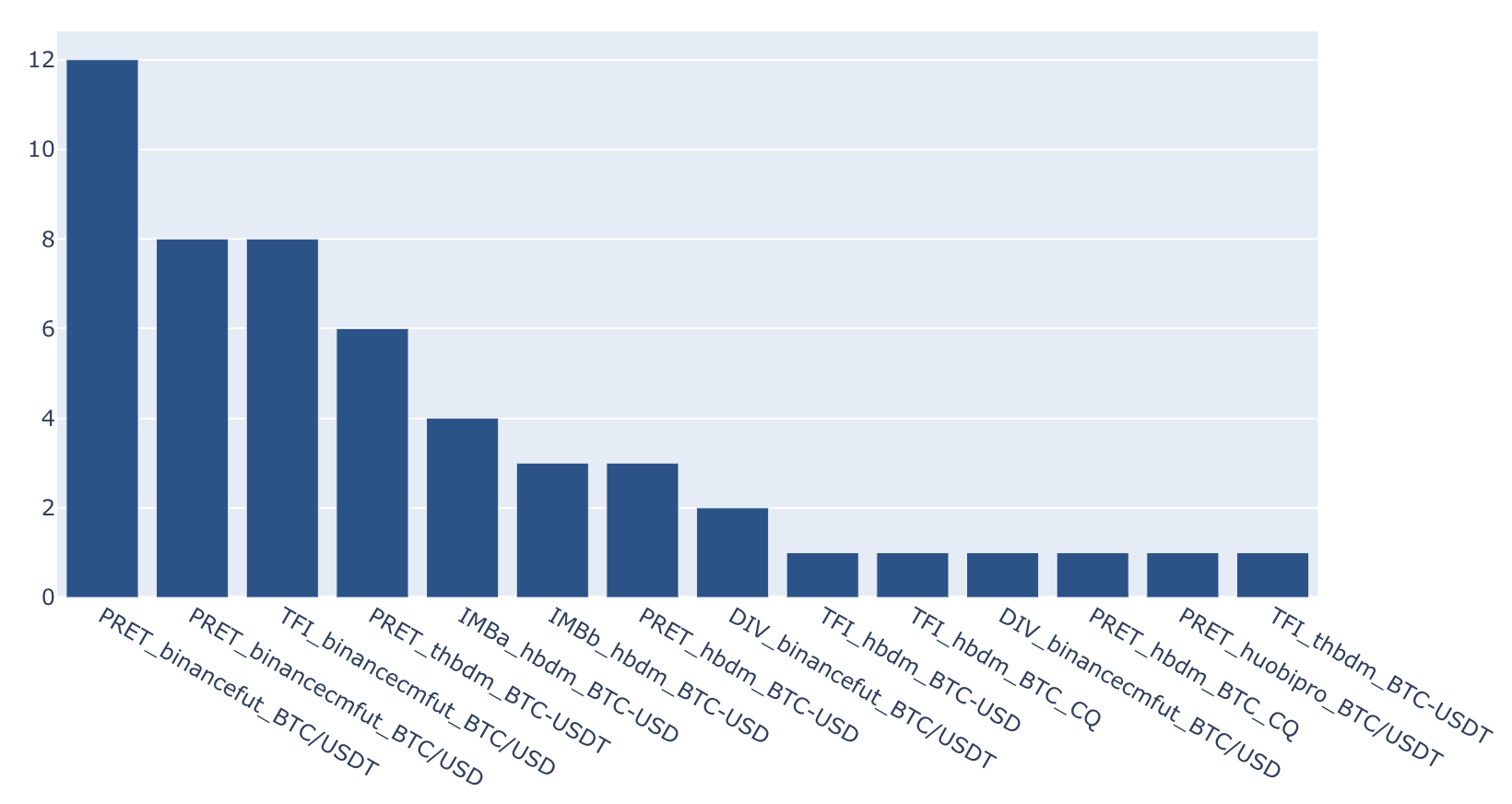} }
\vspace{-1mm}
\captionsetup{width=0.98\linewidth}
\caption[Short Caption]{Count of nonzero features}
\vspace{-3mm}
\end{figure}

We have seen that the PnL and explanatory power of a market's LASSO models is almost unchanged for any regularization parameters $\lambda \leq 0.032$. It is interesting to examine which of the 70 original features are retained in these models. In Figure~\ref{fig:lasso_feature_count_032} we provide an illustration of the number of markets whose $\lambda = 0.032$ model uses the features listed on the $x$-axis. For ease of legibility, we include in this plot only those features which appear in the models of at least six markets.

We are immediately struck in Figure~\ref{fig:lasso_feature_count_256} that the features used by any of the $\lambda=0.256$ models are \emph{exclusively} from a market on Binance or Huobi.
This underlines the outsized role that these two exchanges play in price discovery.
Let us next inspect Figure~\ref{fig:lasso_feature_count_032}.
Here we observe the same dominance of Binance and Huobi with notably large counts also occurring from Okex markets.
We note that the most common type of feature is a (transformed) trade flow imbalance feature.
This suggests signed volume is the most powerful of our indicators.
Of the orderbook imbalance features, the one corresponding to the Huobi BTC perpetual is the most commonly selected.
The most frequent $DIV$ feature comes from the Huobi quarterly futures contract, which indicates that the price difference with this market is particularly useful to predict price action.

\subsection{Meta Features Approach}

In the previous subsection, we found that a much smaller subset retaining approximately 30 of the original 70 features used in the baseline model is sufficient to produce models whose explanatory power and PnL values are only marginally lower than those of the baseline models. In this subsection, we pursue an alternative methodology for drastically reducing dimensionality of our linear models. The procedure involves an additional feature processing step, which cuts down the number of features from 70 to five. The resulting models will turn out to achieve a greater average PnL than the baseline models, even though average explanatory power is diminished.

We observed in Table~\ref{tbl:pret_corrmatrix} that past returns features exhibit large cross-correlations across exchanges. This is of course not surprising, since the markets we are considering are either Bitcoin spot markets themselves or derivative contracts whose underlying is an index composed of Bitcoin spot markets.
Motivated by this consideration, we devise a methodology by which we form a single ``meta" past returns feature from the set of all individual past returns features. That is, we will define for each market $i\in\{1,\ldots, 14\}$ a meta feature $mPRET_{i}$ using
\begin{equation}
mPRET_{t}^i := \sum_{j=1}^{14} \alpha_{ij}^{PRET} PRET_{t}^j, 
\end{equation}
where the coefficient $\alpha_{ij}^{PRET}$ is chosen in proportion to the predictive power that the past returns feature of market $j$ has over the future returns of market $i$. Other meta features (orderbook imbalances, trade flow imbalance, mean divergence) are computed analogously.
More specifically, we define a set of so-called \emph{meta features} for each market $i\in \{1,\ldots, 14\}$ by the equations

\begin{minipage}{0.49\textwidth}
$$ mIMB^{a, i}_{t} := \sum_{j=1}^{14} \alpha_{ij}^{IMB^a} IMB^{a, j}_{jt},$$
\end{minipage}
\begin{minipage}{0.49\textwidth}
$$ mIMB^{b,i}_{t}  := \sum_{j=1}^{14} \alpha_{ij}^{IMB^b} IMB^{b, j}_{t},$$
\end{minipage}
\begin{minipage}{0.49\textwidth}
$$ mTFI^i_{t}      := \sum_{j=1}^{14} \alpha_{ij}^{TFI} TFI^j_{t},     $$
\end{minipage}
\begin{minipage}{0.49\textwidth}
$$ mDIV^i_{t}      := \sum_{j=1}^{14} \alpha_{ij}^{DIV} DIV^{ij}_{t}, $$ 
\end{minipage}
where the coefficients are determined by the procedure described below. 
We will demonstrate this process on the example of the past returns feature.
Let us fix $i\in\{1,\ldots, 14\}$ and $\delta \in \left\{ 500\text{ms}, 1000\text{ms} \right\}$.
The steps are then as follows
\begin{enumerate}
\item For every $j\in \{1,\ldots, 14\}$, we train the univariate models $fret_{t}^{\delta, i} = \mu_{ij} + \beta_{ij} PRET_{t}^j +  \epsilon_{ij, t}$ using an OLS regression. We denote its coefficient of determination by $R^2_{ij}$.
\item We set $\alpha_{ij}^{PRET} := \frac{R^2_{ij}}{\max_j R^2_{ij}}$.
\end{enumerate}
The procedure is completely analogous for other features.
Our normalization procedure involves division by a maximum value.
Note that there are other possible weight normalizations, such as division by the $\sum_j R^2_{ij}$; we leave it for future work to compare these approaches.

Now that we have defined the set of meta features, we can define the following linear models for every market $i\in \{1,\ldots, 14\}$ and the two future returns time horizons $\delta \in \left\{ 500\text{ms}, 1000\text{ms} \right\}$
\begin{equation}
fret_{t}^{\delta, i} = \mu_{i} + {\beta_{i,1} mIMB_{t}^{a, j} + \beta_{i, 2} mIMB_{t}^{b, j}} + \beta_{i, 3} mTFI_{t}^{i} + \beta_{i, 4} mPRET_{t}^{i} + \beta_{i, 5} mDIV_{t}^{i} + \epsilon_{i, t}
    \label{eqn:meta_model_i}
\end{equation}
We fit these models using an OLS regression and analyze their performance along the same axes we have previously considered (explanatory power and PnL).

\textbf{Results.} One main benefit of the meta feature models is their simplicity.
Compared to the baseline models, we have cut down the number of covariates from 70 to five.
In Table~\ref{tbl:meta_models_tstats_pvalues}, we report t-statistics and p-values for each of the coefficients in all of the 14 models we trained. We find strong statistical significance for all meta features in each of the models. The smallest absolute t-statistic corresponds to the meta past returns feature for the model predicting future returns on the Okex quarterly futures contract.
It is perhaps not too surprising that we would see a comparably small t-statistic in this case, since the futures premium can fluctuate somewhat freely in the sense that it is not linked by any strong arbitrage bounds to the price of its underlying. This argument applies more strongly the more distant the expiration of the futures contract is (note that in our case the expiration date lays approximately one month in the future).
\begin{table}
\small 
\centering
\caption{T-statistics and p-values for meta models}
\label{tbl:meta_models_tstats_pvalues}
\begin{tabular}{lllllllllll}
\toprule
{} & \multicolumn{5}{l}{t-stat} & \multicolumn{5}{l}{p-value} \\
{} &   mDIV &  mIMBa &  mIMBb &    mTFI &  mPRET &    mDIV & mIMBa & mIMBb & mTFI & mPRET \\
\midrule
hbdm\_BTC-USD          &  28.85 &  53.29 &  -24.7 &   64.47 &  47.59 &     0.0 &   0.0 &   0.0 &  0.0 &   0.0 \\
hbdm\_BTC\_CQ           &  -9.37 &  44.81 & -12.37 &   98.57 & -16.33 &     0.0 &   0.0 &   0.0 &  0.0 &   0.0 \\
bitmex\_BTC/USD        &  44.19 &  37.24 & -16.03 &   37.36 &  73.89 &     0.0 &   0.0 &   0.0 &  0.0 &   0.0 \\
deribit\_BTC-PERPETUAL &  28.31 &  41.18 & -12.16 &   50.14 &  51.19 &     0.0 &   0.0 &   0.0 &  0.0 &   0.0 \\
bybit\_BTC/USD         &  54.91 &  34.91 & -31.58 &   50.55 &  74.97 &     0.0 &   0.0 &   0.0 &  0.0 &   0.0 \\
binancefut\_BTC/USDT   &  31.76 &  33.36 &  -6.21 &  104.06 & -35.93 &     0.0 &   0.0 &   0.0 &  0.0 &   0.0 \\
okex\_BTC-USD-SWAP     &  -6.72 &   48.1 & -24.71 &   62.38 &  20.09 &     0.0 &   0.0 &   0.0 &  0.0 &   0.0 \\
okex\_BTC-USD-210326   &  -13.5 &  38.55 &  -13.5 &   79.51 &   2.11 &     0.0 &   0.0 &   0.0 &  0.0 &  0.04 \\
ftx\_BTC-PERP          &  53.45 &  40.31 & -10.22 &   77.34 &  95.26 &     0.0 &   0.0 &   0.0 &  0.0 &   0.0 \\
huobipro\_BTC/USDT     &  47.64 &   27.6 &  -5.86 &   110.8 & -30.58 &     0.0 &   0.0 &   0.0 &  0.0 &   0.0 \\
okex\_BTC-USDT-SWAP    & -12.02 &  46.86 & -21.47 &   57.19 &  29.97 &     0.0 &   0.0 &   0.0 &  0.0 &   0.0 \\
tbybit\_BTC/USDT       &  28.17 &  34.46 & -21.09 &   26.64 &  88.36 &     0.0 &   0.0 &   0.0 &  0.0 &   0.0 \\
thbdm\_BTC-USDT        &  29.99 &  57.25 & -20.41 &  108.79 &  18.96 &     0.0 &   0.0 &   0.0 &  0.0 &   0.0 \\
binancecmfut\_BTC/USD  &  19.76 &  60.37 &  -22.4 &   89.62 &  15.52 &     0.0 &   0.0 &   0.0 &  0.0 &   0.0 \\
\bottomrule
\end{tabular}
\end{table}
The coefficients of determination for the meta models defined by Eqn.~\eqref{eqn:meta_model_i} are given in Table~\ref{tbl:meta_models_r2s}, where we show the in-sample and out-of-sample $R^2$ values for the two time horizons under consideration. Here we find that a similar set of observations holds true as in the baseline and LASSO models.

First, the average difference between the $500$ms and $1000$ms $R^2$ values tends to be quite small.
For example, in case of the in-sample $R^2$ values we observe that the $500$ms version is on average 7\% larger than the $1000$ms one, in line with the expectation that shorter time horizons would be more easily  predicted. Some outliers are the lagging markets Bybit USDT perpetual and Bitmex perpetual, where passing from the $500$ms to the $1000$ms future returns horizon yields an increased $R^2$ of $32.6$\% and 37\%, respectively. On the other hand, markets that see a notably high decrease in accuracy in passing from the $500$ms to the $1000$ms horizon are leading markets such as the Binance USDT perpetual or the Huobi spot market, where we see decreases of $39.7$\% and $40.5$\%, respectively. A similar phenomenon was noted for the baseline and LASSO models. 

Secondly, the in-sample $R^2$ is not significantly larger than its out-of-sample counterpart, which provides evidence of the absence of significant overfitting.
We find average decreases of $10.7$\% for the $500$ms horizon and $4.9$\% for the $1000$ms horizon.
\begin{table}
\small 
\centering
\caption{In-sample and out-of-sample $R^2$ values for the meta models.}
\label{tbl:meta_models_r2s}
\begin{tabular}{lllll}
\toprule
{} & \multicolumn{2}{l}{$R^2$} & \multicolumn{2}{l}{$R^2_{oos}$} \\
{} &  $500$ms & $1000$ms &       $500$ms & $1000$ms \\
\midrule
ftx\_BTC-PERP          &  0.306 &  0.288 &       0.213 &  0.212 \\
bybit\_BTC/USD         &  0.262 &    0.3 &       0.261 &  0.294 \\
thbdm\_BTC-USDT        &  0.215 &   0.18 &       0.202 &  0.173 \\
hbdm\_BTC-USD          &  0.201 &  0.229 &        0.19 &  0.217 \\
bitmex\_BTC/USD        &  0.192 &  0.263 &       0.174 &  0.249 \\
tbybit\_BTC/USDT       &   0.19 &  0.252 &       0.198 &   0.26 \\
binancecmfut\_BTC/USD  &  0.156 &  0.139 &       0.137 &   0.13 \\
deribit\_BTC-PERPETUAL &  0.141 &  0.166 &       0.152 &  0.186 \\
okex\_BTC-USDT-SWAP    &   0.13 &  0.144 &       0.117 &  0.139 \\
huobipro\_BTC/USDT     &  0.126 &  0.075 &       0.106 &  0.069 \\
okex\_BTC-USD-SWAP     &  0.122 &  0.131 &       0.112 &  0.124 \\
hbdm\_BTC\_CQ           &  0.116 &  0.091 &       0.104 &   0.09 \\
okex\_BTC-USD-210326   &  0.112 &  0.105 &       0.105 &  0.102 \\
binancefut\_BTC/USDT   &  0.068 &  0.041 &       0.055 &  0.041 \\
\bottomrule
\end{tabular}
\end{table}

In Table~\ref{tbl:meta_models_pnls}, we report PnL values for the meta models~\eqref{eqn:meta_model_i}. 
As before, we compute three separate values: $\operatorname{PnL}_1$ which is the total number of basis points accumulated over the test period by the strategy associated with the respective meta model, $\operatorname{PnL}_2$ where we adjust $\operatorname{PnL}_1$ by the default fee, and $\operatorname{PnL}_3$ where we adjust $\operatorname{PnL}_1$ by the VIP fee.
\begin{table}
\centering
\caption{PnLs of meta models}
\label{tbl:meta_models_pnls}
\begin{tabular}{llll}
\toprule
{} &    $\operatorname{PnL}_1$ &    $\operatorname{PnL}_2$ &    $\operatorname{PnL}_3$ \\
\midrule
huobipro\_BTC/USDT     &  21220.4 &  -6377.1 &  10007.1 \\
ftx\_BTC-PERP          &  17140.6 & -19357.4 &   9319.6 \\
binancefut\_BTC/USDT   &  10790.5 & -12097.5 &   2035.8 \\
binancecmfut\_BTC/USD  &  10085.4 & -13284.6 &   1672.2 \\
thbdm\_BTC-USDT        &  12393.6 &  -6910.4 &   -636.6 \\
hbdm\_BTC\_CQ           &   8051.1 & -10916.9 &  -1432.9 \\
hbdm\_BTC-USD          &  15057.7 &  -8112.3 &  -2088.1 \\
okex\_BTC-USD-210326   &   8739.4 & -15220.6 &  -3240.6 \\
okex\_BTC-USD-SWAP     &  10398.4 & -13091.6 &  -3695.6 \\
okex\_BTC-USDT-SWAP    &   9803.9 & -13946.1 &  -4446.1 \\
bybit\_BTC/USD         &  19613.6 & -14451.4 & -14451.4 \\
deribit\_BTC-PERPETUAL &   9116.7 & -14623.3 & -14623.3 \\
bitmex\_BTC/USD        &  17561.6 & -17433.4 & -17433.4 \\
tbybit\_BTC/USDT       &  17659.5 & -18580.5 & -18580.5 \\
\bottomrule
\end{tabular}
\end{table}
As before with the other models, we find particularly large $\operatorname{PnL}_3$ values for FTX where the fee structure implies that it ``should be" a leader, but its large $R^2$ value points at lagging behaviour.
We also note that the Binance USDT perpetual yields a positive $\operatorname{PnL}_3$ value despite the fact that it has (by quite a margin) the lowest predictability score (as measured by $R^2$).
An interpretation of this result could be that the extremely low VIP fee on this market is sufficiently small to overcome its general leadingness behaviour.

The Huobi spot market has one of the lowest VIP taker fees and its corresponding meta model has middling $R^2$ value. Based on these observations, we would predict a relatively large $\operatorname{PnL}_3$ value although we would not anticipate it having the highest $\operatorname{PnL}_3$, as it turns out to be the case. As mentioned before, this might be partially explained by the fact that this market tends to have extremely low top of the book liquidity, implying that the strategy whose PnL we are computing is not immediately scalable to the deployment of substantial amounts of trading capital.
The additional cost of slippage could become an important consideration at that point.

Surprisingly, the Binance BTC-margined perpetual, which has a similarly low taker fee as the USDT-margined one (see Figure~\ref{fig:lowest_taker_fee}), exhibits a lower $\operatorname{PnL}_3$ when we compare the two (a relative difference of 18\%), despite its considerably higher $R^2$ values.
At the other end of the spectrum, the four markets whose $\operatorname{PnL}_3$ is lowest are the precisely ones with the largest taker fees.
This is a similar finding to what we noted for other models: the greater predictability is not sufficient to overcome the larger fee.

\subsection{Comparisons of Models}

We pursued three alternative methodologies for fitting linear models that anticipate price change on each market. The starting point were the baseline models that used all 70 features. Next, we introduced $\ell_1$ regularization to the baseline model with nine different regularization parameters to obtain more parsimonious models with fewer non-zero coefficients. Finally, we devised an alternative methodology for reducing the number of features down to only five using an additional feature processing step. In this subsection, we set out to draw comparisons between the respective explanatory powers and PnLs of each of the trained models.

\begin{figure}[!ht]
\hspace*{-1cm}
\vspace{-3mm}
\centering
\includegraphics[width=0.8\textwidth]{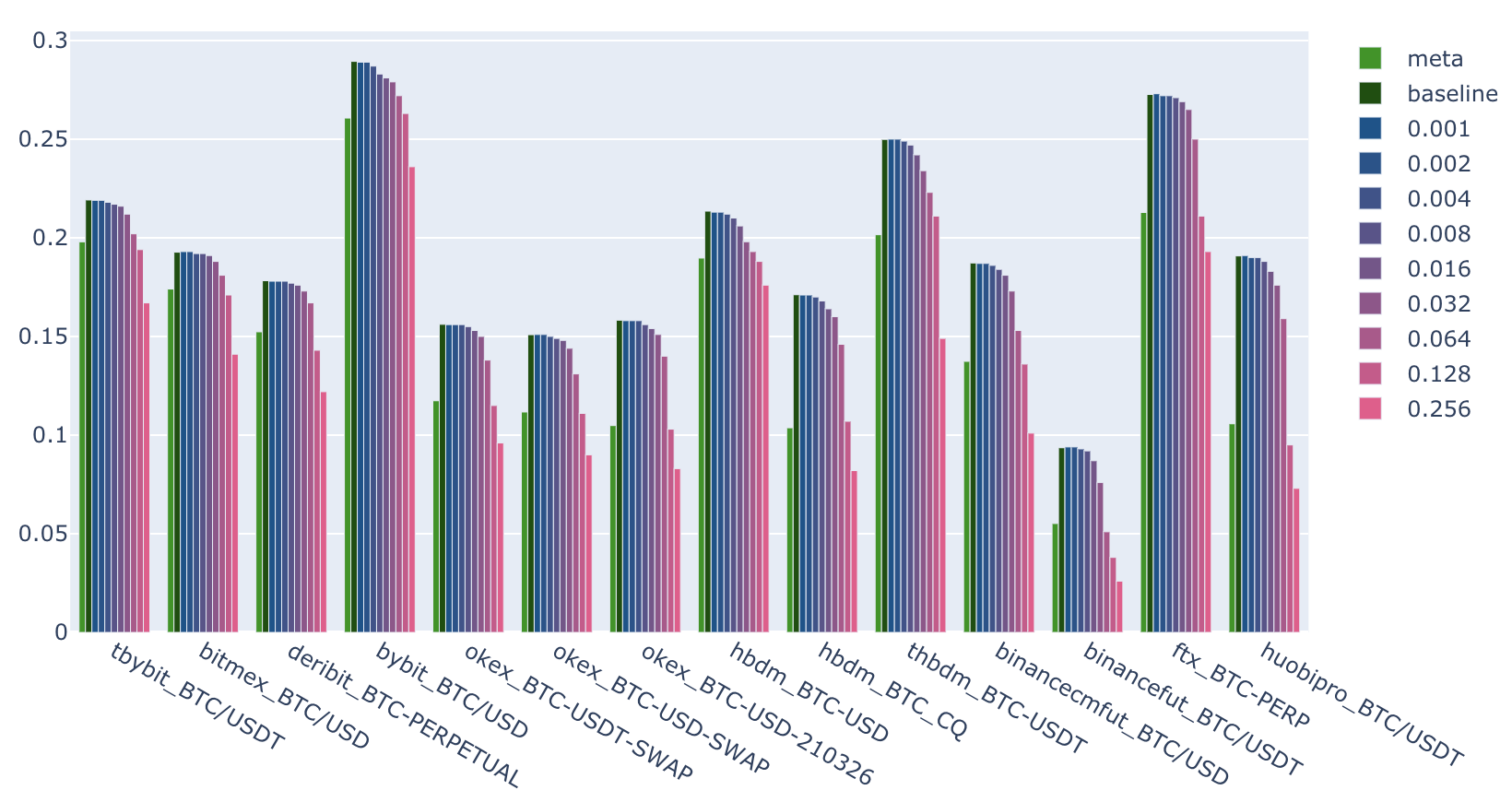}
\caption{500ms out-of-sample $R^2$ values.}
\label{fig:r2_oos_500ms_comparisons}
\hspace{0.0\textwidth} 
\vspace{-3mm}
\end{figure}

\begin{figure}[!ht]
\hspace*{-1cm}
\vspace{-3mm}
\centering
    \includegraphics[scale=0.45]{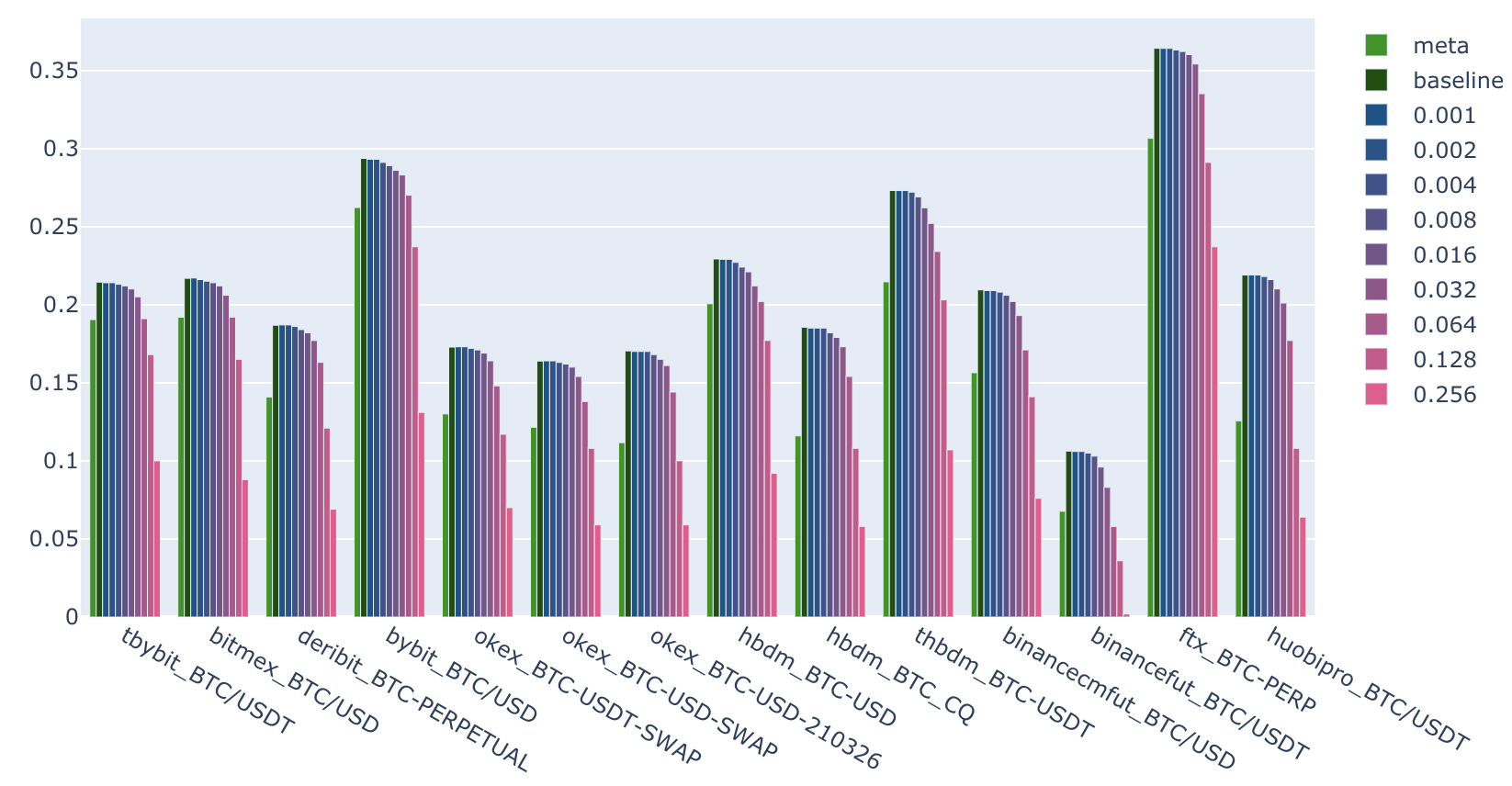}
    \caption{500ms in-sample $R^2$ values}
    \label{fig:r2_500ms_comparisons}
\hspace{0.0\textwidth} 
\vspace{-3mm}
\end{figure}

\begin{figure}[!ht]
\hspace*{-1cm}
\vspace{-3mm}
\centering
    \includegraphics[scale=0.45]{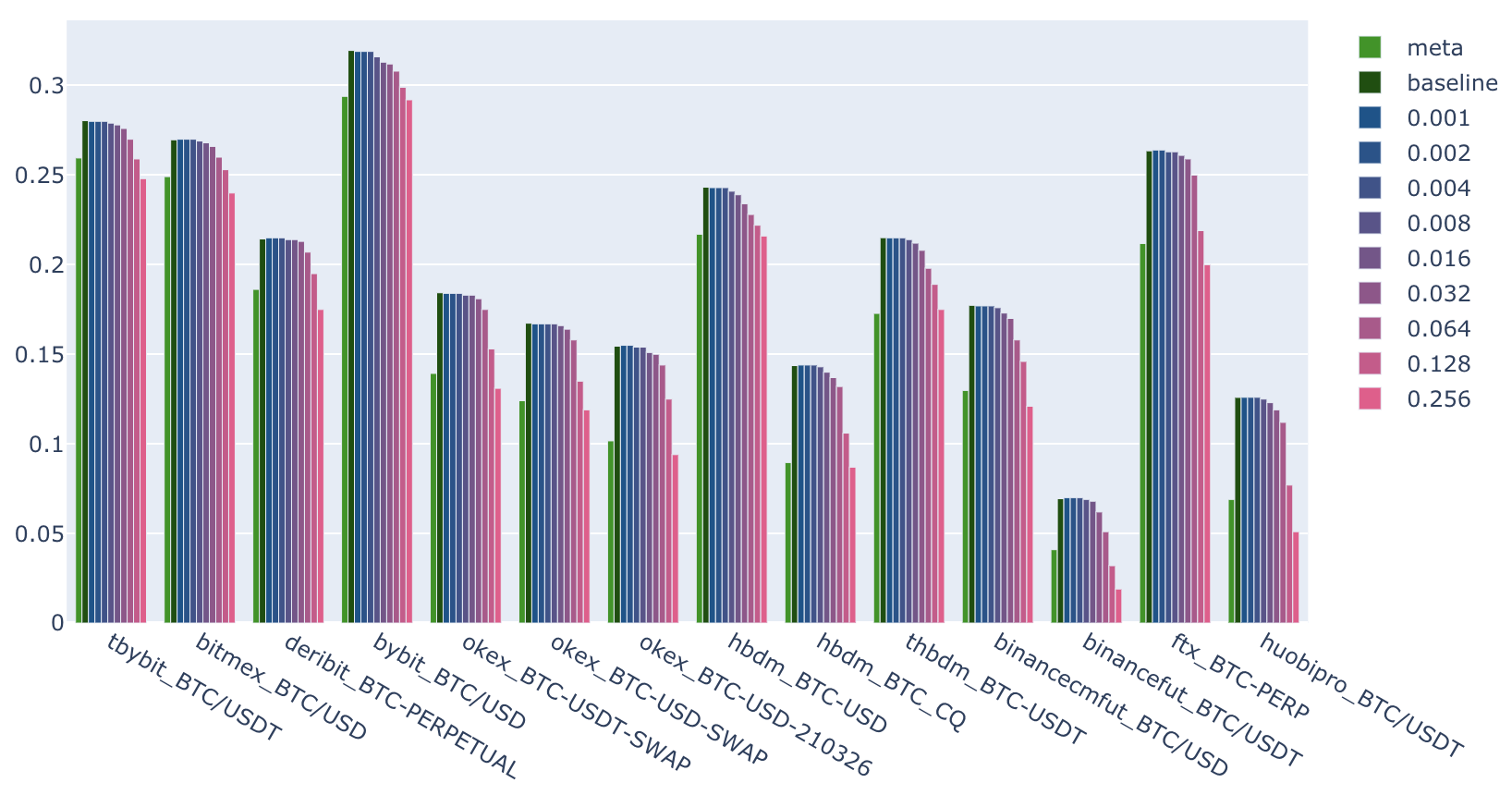}
    \caption{1000ms out-of-sample $R^2$ values.}
    \label{fig:r2_oos_1000ms_comparisons}
\hspace{0.0\textwidth} 
\vspace{-3mm}
\end{figure}

\begin{figure}[!ht]
\hspace*{-1cm}
\vspace{-3mm}
\centering
    \includegraphics[scale=0.45]{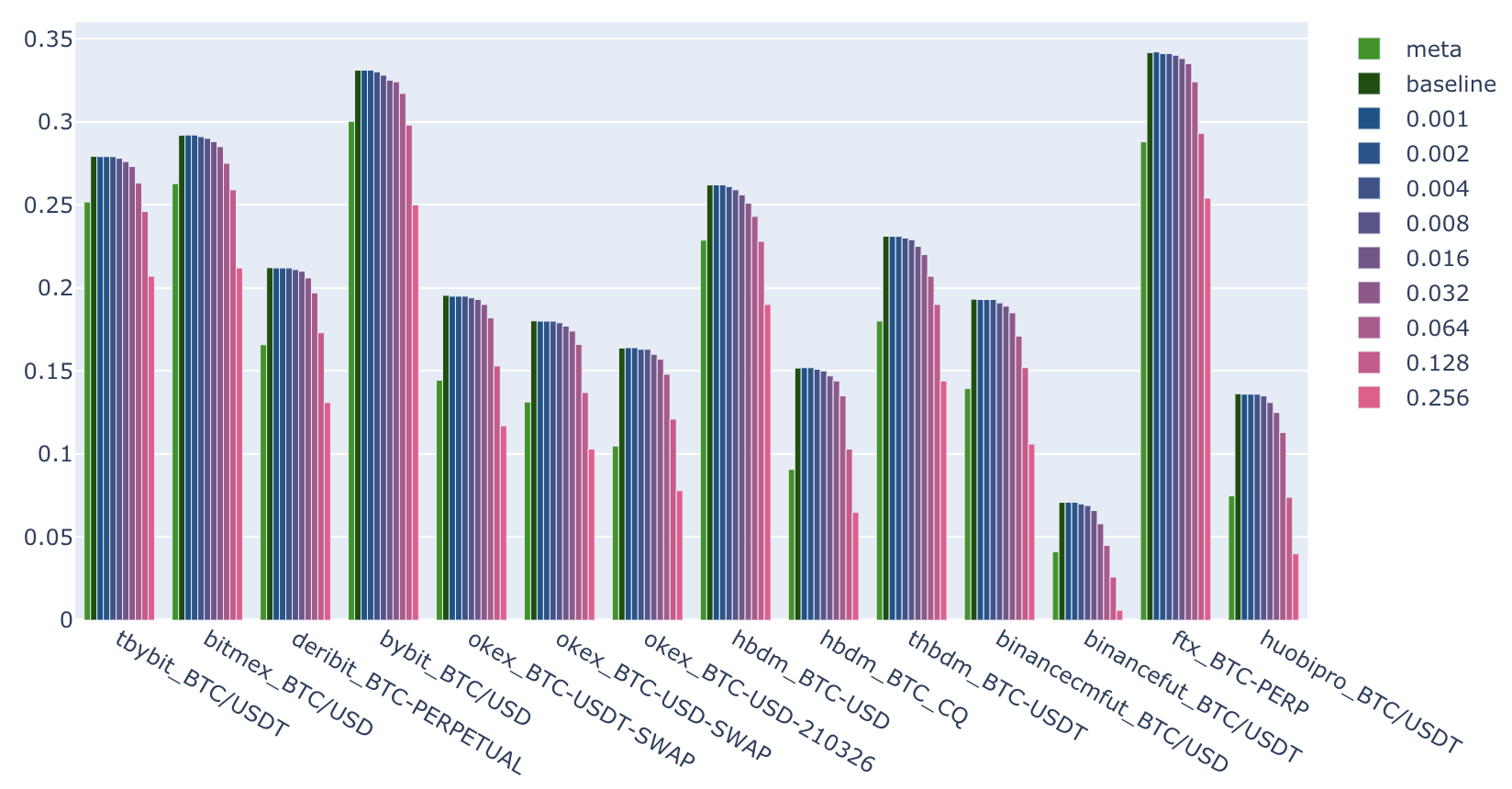}
    \caption{1000ms in-sample $R^2$ values.}
    \label{fig:r2_1000ms_comparisons}
\hspace{0.0\textwidth} 
\vspace{-3mm}
\end{figure}

We begin by investigating how the $R^2$ values of our models compare to one another.
Figure~\ref{fig:r2_oos_500ms_comparisons} shows the out-of-sample $R^2$ values of each model.
The in-sample and $1000$ms versions of the same illustration were deferred to the Appendix since all of the $R^2$-based plots strongly resemble one another.
The interested reader can find these in Figures~\ref{fig:r2_oos_1000ms_comparisons},~\ref{fig:r2_500ms_comparisons}, and~\ref{fig:r2_1000ms_comparisons}.
In examining Figure~\ref{fig:r2_oos_500ms_comparisons}, we first note that the maximal $R^2$ is attained by the baseline model in each case. LASSO models with small regularization parameter $\lambda$ perform similarly. For example, the LASSO models with $\lambda=0.016$ experience an average drop of only $2.8$\% relative to the baseline model and the $\lambda=0.032$ models experience an average decrease of $5.9$\%.
The average number of nonzero coefficients retained is 38 for $\lambda=0.016$, and 29 for $\lambda=0.032$.
As noted earlier, this suggests that of the 70 original features used by the baseline model, around 30-40 contribute very little to the model's performance. When $\lambda$ is increased beyond $0.032$, and as more and more coefficients are thus set to zero, we find progressively larger drops in explanatory power.

The meta model uses only five features and is by far the most parsimonious in terms of explanatory power per feature. To illustrate this point, we can compare the results of the meta model with those of the $\lambda=0.256$ LASSO models which have a similar number of features (namely an average of four).
Here we find a lower average $R^2$ of $21.2$\% for the LASSO models. The $R^2$ values of the meta models are actually quite similar to those of the $\lambda=0.128$ models which retain an average of nine features.
The average difference in $R^2$ values between the two is $3.5$\%. When we compare the meta models to the baseline models, we note a substantial reduction in explanatory power. The average reduction is $23.7$\%.    

\begin{figure}[!ht]
\hspace*{-1cm}
\vspace{-3mm}
\centering
\includegraphics[width=0.8\textwidth]{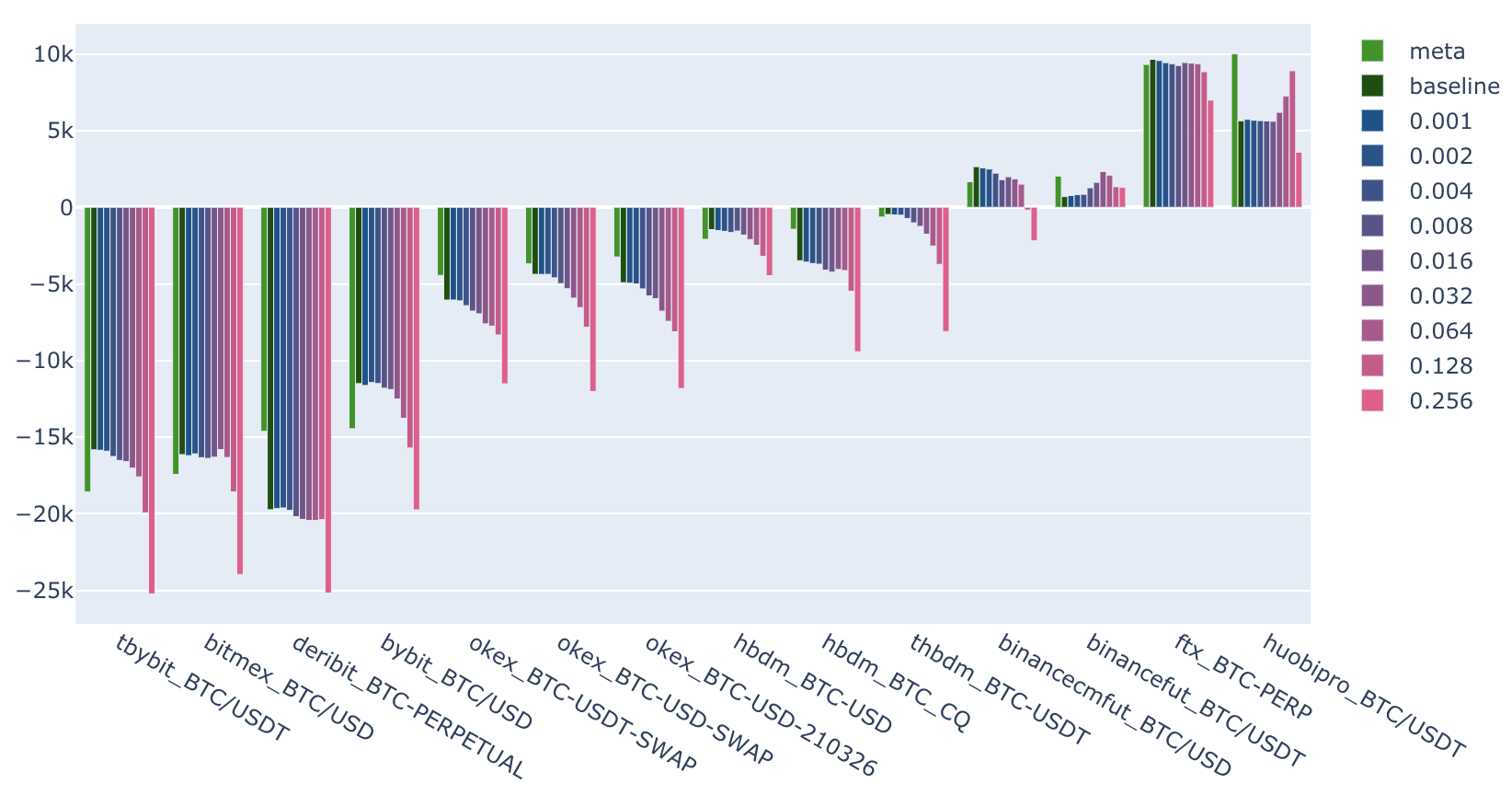}
\captionsetup{width=0.97\linewidth} 
\caption{$\operatorname{PnL}_3$ values for each model.}
\label{fig:pnl3_comparisons}
\hspace{0.0\textwidth} 
\vspace{-3mm}
\end{figure}

\begin{figure}[!ht]
\hspace*{-1cm}
\vspace{-3mm}
\centering
\includegraphics[width=0.7\textwidth]{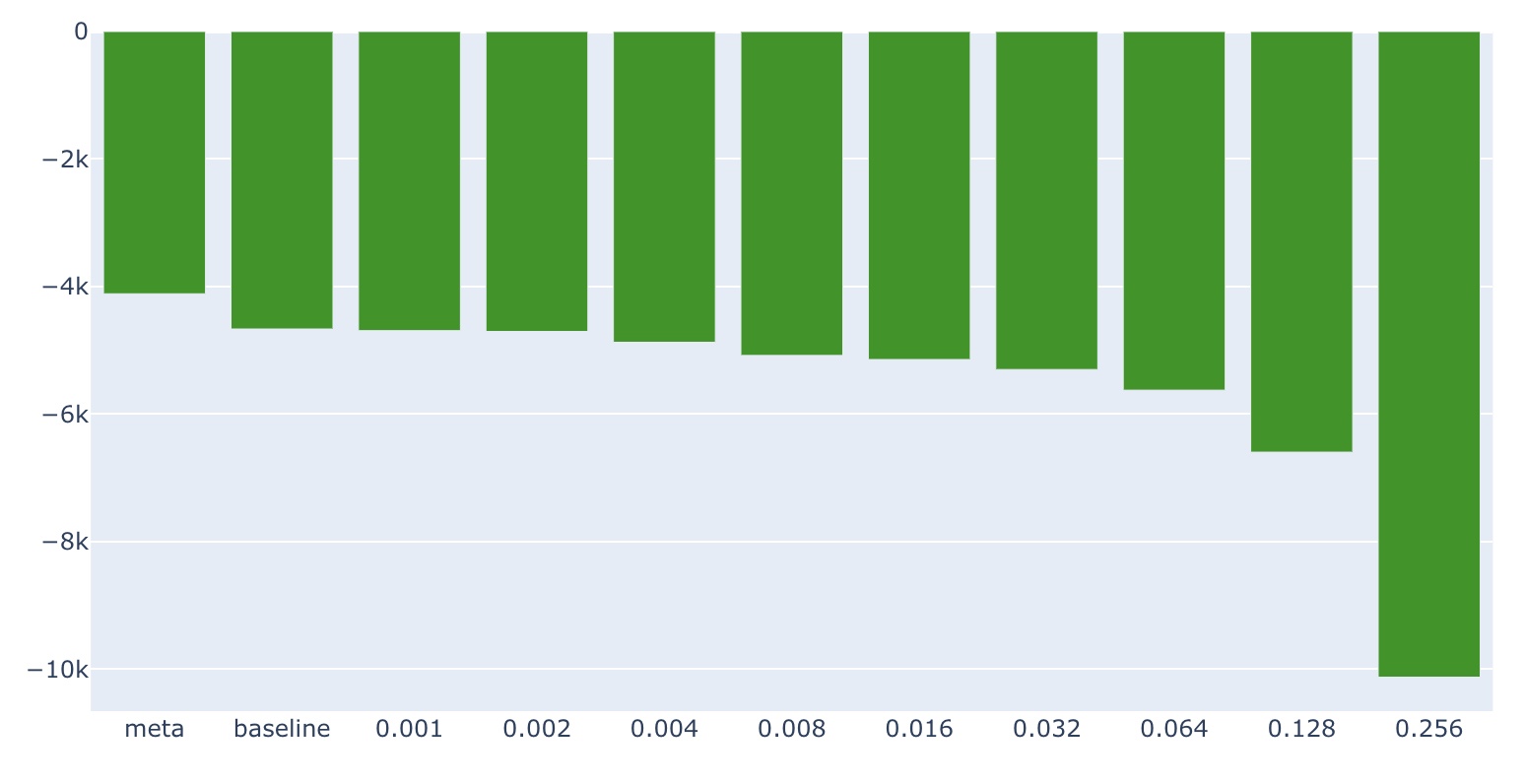}
\captionsetup{width=0.97\linewidth} 
\caption{$\operatorname{PnL}_3$ values averaged across markets for each type of model. The numeric values denote the LASSO regularization parameters.}   
    \label{fig:pnl3_averages}      
\hspace{0.0\textwidth} 
\vspace{-3mm}
\end{figure}

Let us now compare the $\operatorname{PnL}_3$ values of each of the models shown in Figure~\ref{fig:pnl3_comparisons}.
Remarkably, the meta models exhibit the largest average $\operatorname{PnL}_3$ of $-4114$ bpts compared to $-4666$ bpts for the baseline models, which have the second largest average.
These findings are unexpected in the sense that they do not agree with the earlier findings on explanatory power, which potentially hints at the power of the meta models and their simplicity for actual trading applications.
With the exception of the meta models, the average $\operatorname{PnL}_3$ is in fact monotonically decreasing in the number of coefficients.
That is, as the regularization parameter is increased, the average $\operatorname{PnL}_3$ decreases.
See Figure~\ref{fig:pnl3_averages} for a visual overview of these average values across markets.

\begin{figure}[!ht]
\hspace*{-1cm}
\vspace{-3mm}
\centering
\includegraphics[width=0.8\textwidth]{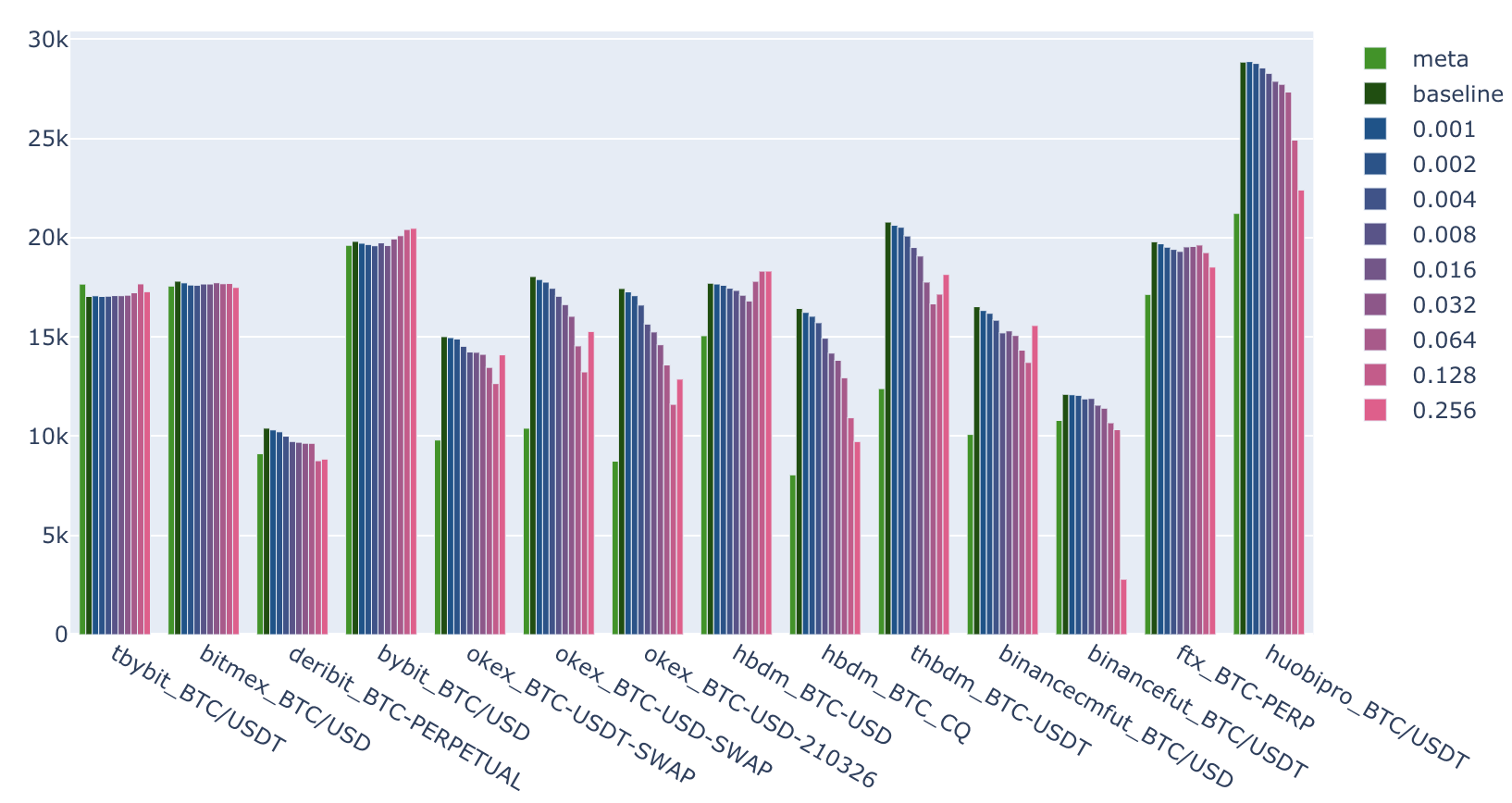}
\caption{$\operatorname{PnL}_1$ values for each model.}
\label{fig:pnl1_comparisons}
\hspace{0.0\textwidth} 
\vspace{-3mm}
\end{figure}

It is interesting to contrast these findings with the analogous results for the the $\operatorname{PnL}_1$ values of all models where execution cost is ignored, as shown in Figure~\ref{fig:pnl1_comparisons}. 
One is immediately struck by the fact that the meta models perform considerably worse than, say, the baseline model according to $\operatorname{PnL}_1$.
As a matter of fact, the meta models now have the \emph{smallest} average value of 13402 whereas the baseline model has an average of 17695.
For the LASSO models, the average $\operatorname{PnL}_1$ decreases monotonically in the regularization parameter $\lambda$, starting at 17602 for $\lambda=0.001$ and ending with 15128 for $\lambda=0.256$.

How can we reconcile the fact that the meta models yielded the best results in terms of $\operatorname{PnL}_3$, but yet delivered the worst results if we consider $\operatorname{PnL}_1$ as the metric? Since $\operatorname{PnL}_3^i = \operatorname{PnL}_1^i - \text{fee}_i\cdot n_i$ where $\text{fee}_i$ is the taker fee on market $i$ and $n_i$ is the number of trades on market $i$, we must have a smaller number of trades for the meta models compared with the baseline models. That is, the strategy associated with the baseline model trades more often than that associated with the meta model, hence accumulating basis points in profit when fees are ignored, but accumulating losses when fees are accounted for. 

\begin{figure}[!ht]
\hspace*{-1cm}
\vspace{-3mm}
\centering
\includegraphics[width=0.8\textwidth]{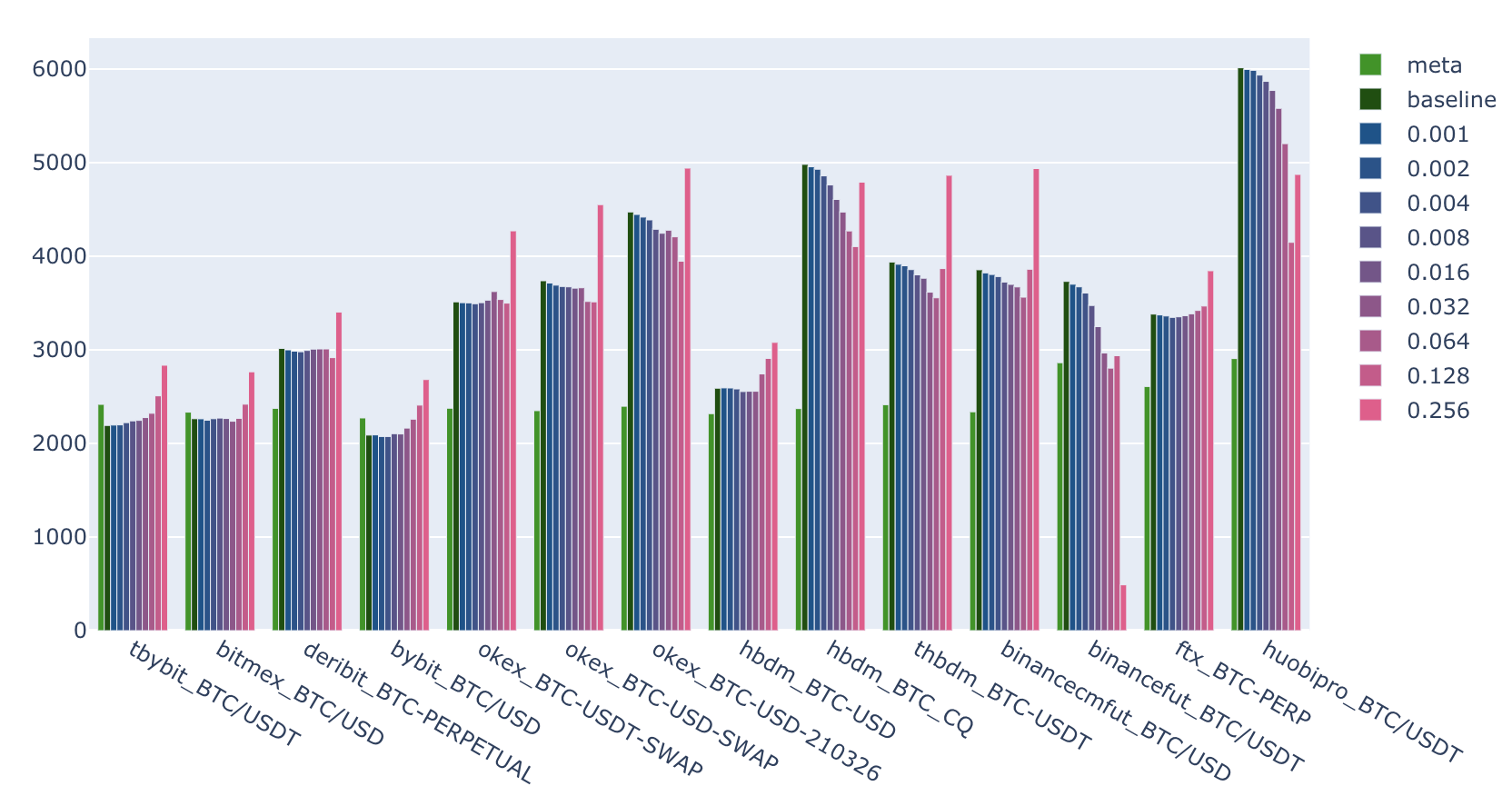}
\caption{Number of trades for each model}
\label{fig:trade_count_comparison}
\hspace{0.0\textwidth} 
\vspace{-3mm}
\end{figure}

The difference in the number of trades per model is illustrated in Figure~\ref{fig:trade_count_comparison}.
The strategy we associate to the meta models trade an average of $25.3$\% less often relative to its baseline model counterpart. The discrepancy is especially large on the Huobi spot market and quarterly futures contracts, where the number of trades is less than half that of the baseline models.

\section{Market Making Experiments}
\label{sec:6}

The strategies we associated to each of the previously trained linear models have been \emph{taker} strategies.
That is, they rely on limit orders that would lead to immediate execution: buy orders use the top ask as their limit price, while sell orders use the top bid. We found positive PnL values even after taking execution cost into account, in a number of cases when the taker fee was particularly low. Some markets (such as the Bybit BTC perpetual) which have among the highest taker fees gave low PnL values, despite their strong tendency of being laggards.

However, as we noted before, a large taker fee usually also implies a large maker rebate.
This is because the net fee (the difference between taker fee and maker rebate) is largely uniform across markets, as one would expect in a competitive exchange environment, since a market with comparatively very large net fees would naturally find it difficult to attract traders. On the other hand, a market which undercuts competing exchanges with exceptionally low net fees would likely struggle to be profitable. 
Let us consider more closely those markets which are laggards, have a high taker fee and consequently also a high maker rebate. The maker rebate on the Bybit BTC perpetual, for instance, is $2.5$ bpts, which is the maximum among our set of $14$ markets. 
It is natural to wonder whether we can leverage the large maker rebate on Bybit in conjunction with its lagging nature to instead devise a \emph{maker} strategy which achieves positive PnL values. 

In contrast with taker strategies, maker strategies employ \emph{passive} limit orders (also called maker orders). That is, for a sell order, the limit price is strictly greater than the top bid price, while for a buy order, the limit price is strictly smaller than the top ask price. The order thus submitted rests ``passively" in the orderbook, can be observed by others, and consumed by others with a taker order (an exception is the iceberg order type supported by some exchanges which allows one to submit passive orders that are not visible to other market participants). Execution of the maker order hinges on the arrival of a taker order which consumes it. It is therefore uncertain at what point in time a passive order will be executed, if at all. The compensation for this uncertainty is that a maker order submitted at time $t$ necessarily achieves a better price (if it is executed) than a taker order submitted at time $t$ would.
In addition, maker orders incur less (in some cases negative) execution fees relative to taker orders. The trade-off between passive and aggressive orders can be concisely summarized as one of immediacy on one hand, versus ``goodness" of execution price on the other hand.  

What is required of a successful maker strategy? Since the execution of a maker order relies on the arrival of a taker order in the opposite direction, the maker order is naturally subject to a certain adverse price move. If we use the notion of microprice defined below, then it is true by definition that at the precise time of the fill, there is an adverse price move against the maker order. 
We define the microprice as 
$$p_m := \frac{p\cdot a + p'\cdot a'}{a + a'}$$
where $p$ and $p'$ are the top bid and ask prices, respectively, while $a$ and $a'$ are the top bid and ask amounts, respectively.
The goal of a maker strategy therefore must be to minimize the magnitude of the adverse price move following a fill event.
In order to achieve this goal, the trader must constantly monitor market conditions for the potential future arrival of a significant adverse price move.
When such a price move is imminent, the trader should cancel their passive order (and possibly resubmit at a different price level).

In this section, we set out to devise a maker strategy that builds on the previously trained meta models. These strategies will then be rigorously tested which, due to the difficulties in backtesting maker strategies, will involve a series of real world trading experiments on Bybit where the maker rebate is largest. Finally, we draw comparisons between the results of our maker strategy with those of a naive benchmark strategy, in order to showcase the efficacy of the trading alpha provided by the meta models.   

\subsection{Strategy Specification}

This subsection is devoted to constructing a maker strategy that builds on the meta models.
The reason we use the meta models as the foundation is two-fold.
First, the meta models are far simpler than for instance the baseline models if we gauge model complexity by the number of covariates.
Second, while the meta models had lower explanatory power than some of the other models, their efficiency in the sense of $R^2$ per feature was by far the largest, and they achieved the largest PnL values when fees were taken into account.

What is the simplest, most natural maker strategy one can define?
Let us first consider what data is required in order to fully specify a maker strategy.
For the sake of concreteness, let us consider in the following the case of a single passive sell order.
It is straightforward to dualize this discussion to the case of a passive buy order.
The specification of a strategy requires
\begin{enumerate}
    \item a criterion for when to post a new order,
    \item a limit price $p > \text{top bid}$ and an amount $a\in \mathbb N$ for the posted order, and 
    \item a criterion for when to cancel the ask order. 
\end{enumerate}
In order to reduce the problem to its most plain form, we shall utilize the following simplifications
\begin{itemize}
    \item $p := \text{top bid} + \text{tick size}$,
    \item $a := 2000 \text{ USD}$, and
    \item Post a new order $:\iff$ cancel criterion returns ``\textit{do not cancel}" and we currently do not have an ask order posted.
\end{itemize}
Here the \textit{tick size} refers to the minimum price movement of the market in question. 
That is, we use a fixed price, namely the lowest possible ask level and a fixed (somewhat  arbitrarily chosen) small order amount for the sake of testing. It is left for future work to investigate the scalability to larger order amounts. With these 
simplifications in mind, we need only to specify a condition for when to cancel an order posted at the top ask. From this,  we will then obtain a full maker strategy. The cancel criterion for a passive order at the top ask level should be capable of anticipating when an adverse price move in excess of the maker rebate is imminent, and in that case return a \textit{``cancel"} decision in time to successfully pull the order. As mentioned earlier, adverse selection is unavoidable - the question is just how much of it our fills are subject to it. If this is less than the maker rebate received per fill, we achieve profitability.

How can we leverage the meta models to define a natural cancel criterion?
Let us fix a market $i\in\{1,\ldots,14\}$ on which to trade.
In fact, as previously noted, we will conduct our experiments on the Bybit BTC perpetual.
The meta model for market $i$ is the fitted linear model
\begin{equation}
fret_{t}^{\delta, i} = \mu_{i} + {\beta_{i,1} mIMB_{t}^{a, j} + \beta_{i, 2} mIMB_{t}^{b, j}} + \beta_{i, 3} mTFI_{t}^{i} + \beta_{i, 4} mPRET_{t}^{i} + \beta_{i, 5} mDIV_{t}^{i} + \epsilon_{i, t}, 
\label{eqn:fitted_meta_model} 
\end{equation} 
and we shall fix a future returns horizon $\delta = 1000\text{ ms}$ for the remainder of this section. Let
\begin{equation}
F_i:\mathbb R^5 \to \mathbb R, 
\end{equation}
denote the market-specific function that maps an observation of meta features to the prediction of the fitted meta model according to Eqn.~\eqref{eqn:fitted_meta_model}. 
Suppose we observe a new sample $x_t\in \mathbb R^5$ at a time $t$, where we have a sell order posted on the top ask level.
We can then define the following cancel criterion parametrized by a constant $T_0$
\begin{equation}
\text{cancel top ask order} :\iff F_i(x_t) > T_0. 
\end{equation}
That is, when the meta model predicts a rise price of a certain magnitude (parametrized by $T$) based on the most recent feature observation, we shall cancel an order posted at the top ask if indeed we have one posted there at the time.

We are now left with the choice of the threshold $T$. If we select a value which is too large, we will end up cancelling very rarely and hence experiencing a lot of adverse selection.
In the limit case $T = \infty$, we in fact never cancel (which will serve as our benchmark for comparison, as we will describe later). On the other hand, if we select a value $T$ which is too close to zero, we will cancel extremely often and thus receive very little fills (and consequently collect very few maker rebates). As a matter of fact, if we were to choose $T = -\infty$ the cancel condition would always return ``cancel" and we would receive no fills whatsoever, since we would never post a new order to begin with. Given the trade-offs highlighted  by the limiting cases, how can we arrive at a sensible parameter value where we maximize the number of fills we receive, while simultaneously minimizing the amount of adverse selection our fills are subject to? 

Our approach is to come up with an approximate answer to this question by learning from cancellation behaviour of other market participants. That is, we will attempt to characterize, by means of meta model predictions, times at which a significant amount of cancellations from the top ask level is imminent. More specifically, we search for meta feature observations in the union of training and test set (spanning Feburary 22 -- March 1) where, at times $t$ the following conditions holds
\begin{itemize}
    \item $a_t > M$, 
    \item $p_{t+\delta} = p_t$ and $a_{t+\delta} < m$ for some $\delta\in (0, 500\text{ms})$.
\end{itemize}
Here $a_t$ and $p_t$ denote the top ask liquidity and price at time $t$, respectively.
The constant $M > 0$ represents an at least moderately large amount of liquidity, while the scalar $m > 0$ represents a small amount of liquidity. Experimentation has revealed definitions of $m$ and $M$ as the first and second quartile (respectively) to be sensible ones. The quartiles are computed over all observations of the top ask liquidity in the training and test sets combined. Let us denote the subset of meta feature samples defined by the above two condition by $\mathcal F$.

\begin{figure}[!ht]
\vspace{-3mm}
\centering
\subcaptionbox[]{Random subset of size $20,000$ from the set $\mathcal F$
\label{fig:distr_of_preds_on_F}
}[ 0.49\textwidth ]
{ \includegraphics[width=0.49\textwidth, trim=0cm 0cm 0cm 0.0cm,clip] {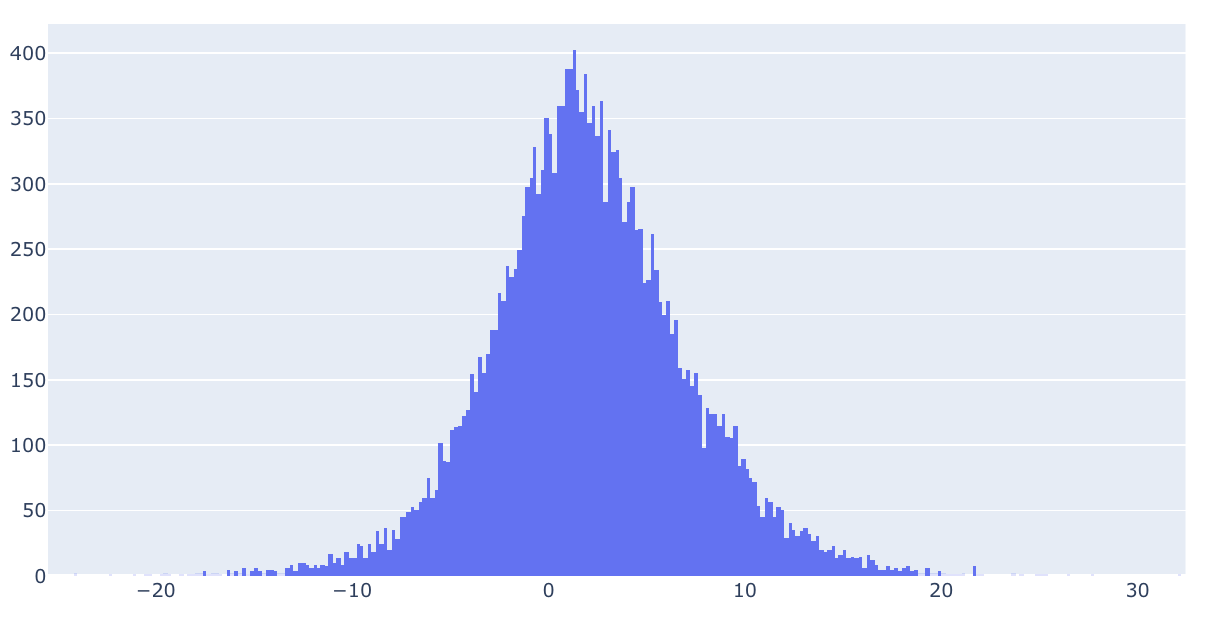} }
\hspace{0.0\textwidth} 
\subcaptionbox[]{\emph{Any} random set of $20,000$ meta feature samples
\label{fig:distr_of_preds_on_random_set}
}[ 0.49\textwidth ]
{\includegraphics[width=0.49\textwidth, trim=0cm 0cm 0cm 0.0cm,clip]{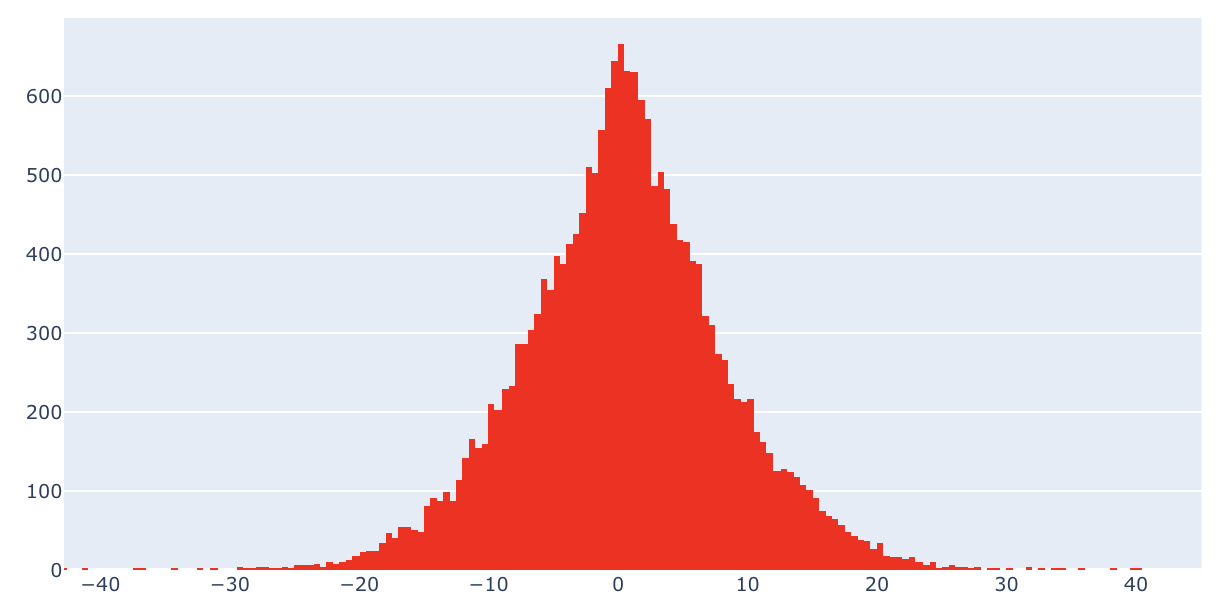} }
\vspace{-2mm}
\captionsetup{width=0.98\linewidth}
\caption[Short Caption]{Distribution of meta model predictions.}
\vspace{-3mm}
\end{figure}

The intuition behind our approach is to scan our data set for examples where we pass from a moderate-high liquidity regime (on the top ask level) to a low liquidity regime.
These are cases representing the cutoff point at which the market (i.e.\ most market participants) agree that it is no longer profitable to post an order at the top ask level.
We want to understand such cases by means of our meta models.
To this end, consider Figure~\ref{fig:distr_of_preds_on_F} where we display a histogram of values from the set
\begin{equation}
\mathcal P   :=  \{ F_i(x) : x \in \mathcal F\}. 
\label{eqn:preds_on_F}
\end{equation} 
That is, we show the distribution of predictions made by the meta model on samples from the set $\mathcal F$ defined by the two conditions above.
More precisely, we uniformly select a random subset of size $20,000$ from these predictions in order to avoid many successive samples and to keep the histogram more legible. 
The average and median predictions are $2.04$ and $1.83$ respectively, while the standard deviation is $5.14$. We compare this with the histogram shown in Figure~\ref{fig:distr_of_preds_on_random_set}, where we display a uniformly random size $20,000$ subset of $\{ F_i(x) : x \text{ is any training or test sample}\}$.
Here we find an average prediction value of $0.47$, a median of $0.44$ and a standard deviation of $7.88$.

From the preceding analysis, we might extract $T:= 1.83$, the median prediction on the set $\mathcal F$, as a reasonable candidate for the definition of our cancel criterion and hence the full specification of the maker strategy. In the following, we will examine this choice from another perspective and draw comparisons with other possible values for $T$.

\begin{figure}[!ht]
\hspace*{-1cm}
\vspace{-3mm}
\centering
\includegraphics[scale=0.5]{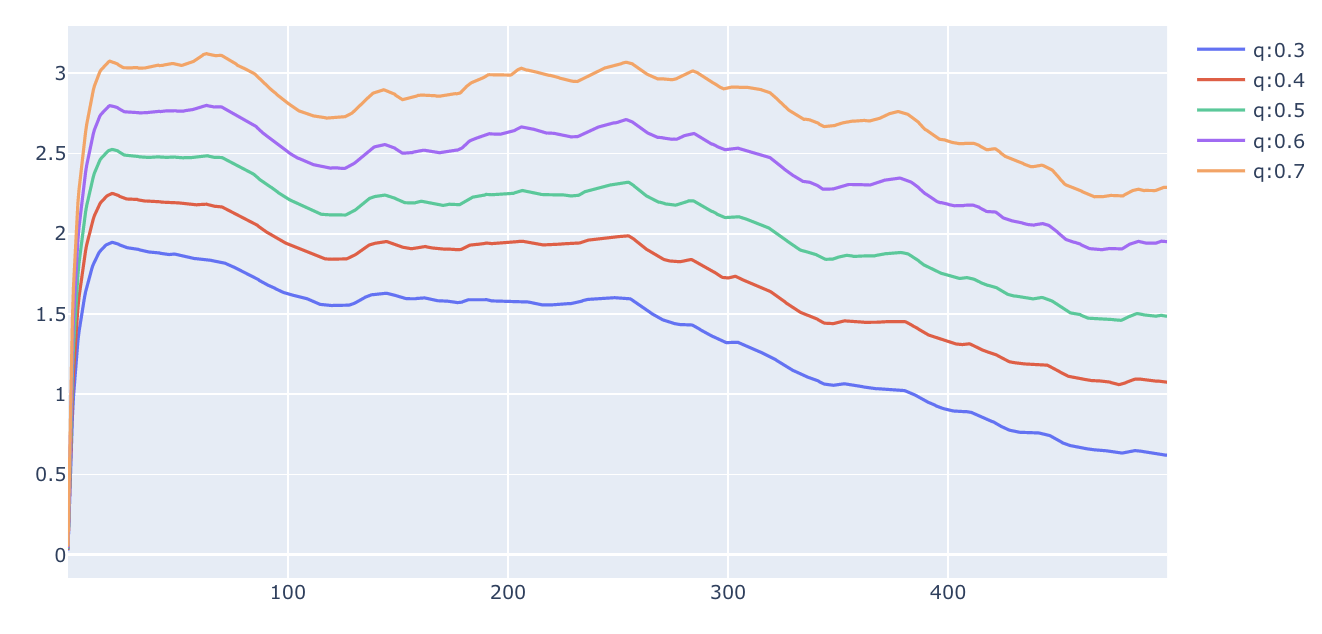}
\captionsetup{width=0.97\linewidth} 
\caption{Average top ask price returns after $\delta$ second conditioned on meta model prediction greater than $T_q$}
\label{fig:adverse_price_move_by_quantile}
\hspace{0.0\textwidth} 
\vspace{-3mm}
\end{figure}

Next, we would like to measure the expected adverse selected when the prediction exceeds $T$. 
To this end, let us consider Figure~\ref{fig:adverse_price_move_by_quantile}
where 
we computed the average price (top ask) increase (in basis points) after $\delta$ seconds, for $\delta \leq 500\text{s}$, conditioned on a meta model prediction greater than $T_q$ for $q\in \{0.3, 0.4, 0.5, 0.6, 0.7 \}$. The quantity $T_q$ is defined as the $q$-th quantile of the set $\mathcal P$ from Equation~\ref{eqn:preds_on_F}. Note that $T_{0.5} = 1.83$ is our candidate threshold.
We observe in Figure~\ref{fig:adverse_price_move_by_quantile} that our candidate threshold (corresponding to $q=0.5$) exactly attains the $2.5$ bpts mark after a few seconds, which is precisely the breakeven point of tolerable adverse selection, since the maker rebate is $2.5$ bpts. This provides us with further evidence of the suitability of the $T=1.83$ threshold.

\textbf{Impossibility of backtesting.} At this point, we have a candidate maker strategy (or a set of candidate strategies) that we would like to test, in order to determine whether it is capable of achieving positive PnL. But how exactly can we test performance of a maker strategy? It turns out that there are a number of difficulties associated with this, making it essentially impossible to do so with any reasonable degree of precision. Let us contrast this with the case of backtesting a taker strategy on historical data.

The situation is straightforward in the taker case due to the immediacy of execution for taker orders: when a candidate strategy provides a buy signal, we assume execution at the top ask price $p_a$ or, in more elaborate versions, at a price $p > p_a$ that takes (i.e. \textit{sweeps}) liquidity from a number of ask levels into account. Similarly, when a sell signal occurs, we assume execution at a price less or equal to the top bid price.
As we scan over a historical test set in this manner (computing buy and sell signals at each point), we obtain a sequence of hypothetical buy and sell orders, which can be used as the  foundation for subsequent PnL calculations. While this analysis neglects some market  micro-structural effects such as market impact and latency concerns, it nevertheless provides a good approximation of the actual PnL of the strategy.
Moreover, there are also fairly natural ways of mitigating some of the aforementioned concerns.
For instance, latency concerns largely relate to uncertainty of execution, and can be addressed by assuming that hypothetical orders fail with some nonzero probability, which could, for example, be calibrated from historical execution data.
Market impact can be accounted for by using an impact model from the large body of literature on this generally well understood topic. 

How does this compare to the case when we would like to backtest a maker strategy?
The underlying difficulty is the fact that the execution of a posted maker order is uncertain.
It is not known when it will occur, or if it will occur at all.
Execution hinges on the arrival of a taker order in the opposite direction of the posted maker order.
This means that if a candidate strategy which is being tested posts a hypothetical maker order at some orderbook level, we must henceforth keep track of the queue position of our hypothetical maker order. While this hypothetical order remains posted in the orderbook (i.e.\ it has not been cancelled in the meantime), we must monitor all opposite-side taker orders and evaluate, based on the size of the taker order and our inferred queue position, whether it would have taken (i.e filled) our maker order. There are a number of factors that present challenges to doing this accurately, whose severity ranges from mild to critical. On the mild side of the spectrum, we are facing a similar pair of problems as in the taker case: 
\begin{itemize}
    \item The submission of a passive order has an impact on the market. One would expect that the presence of a large maker order discourages aggressive orders that take from the orderbook level at which the maker order is posted. Hence, when we keep track of historical taker orders and attempt to infer whether they filled a hypothetical maker order of ours, we must account for this by assuming a nonzero probability that the taker order in fact does not arrive or that its size would have been less than the one observed historically. This sort of market impact of passive orders is far less well-understood and understudied relative to the impact of taker orders. Nevertheless, we rate this concern as only mild in severity.
    \item There are latency issues similar to those that one must contend with when backtesting a taker strategy. For instance, at time $t$ our maker strategy outputs a decision to cancel a posted order and at time $t'\approx t$ a large taker order arrives that would certainly have consumed our passive order. It can be difficult to know with certainty whether the cancellation would have been processed first by the exchange's matching engine, or whether the taker order would in fact have consumed our passive order. To address this concern, we must associate a probability of failure to our cancellation that we could, for instance, calibrate from historical cancellation data and which depends on the time difference $t' - t$. We rate this concern as mild.
\end{itemize}

\begin{figure}[h]
    \fbox{\includegraphics[width=0.29\textwidth, trim=0cm 0cm 0cm 0.0cm,clip]{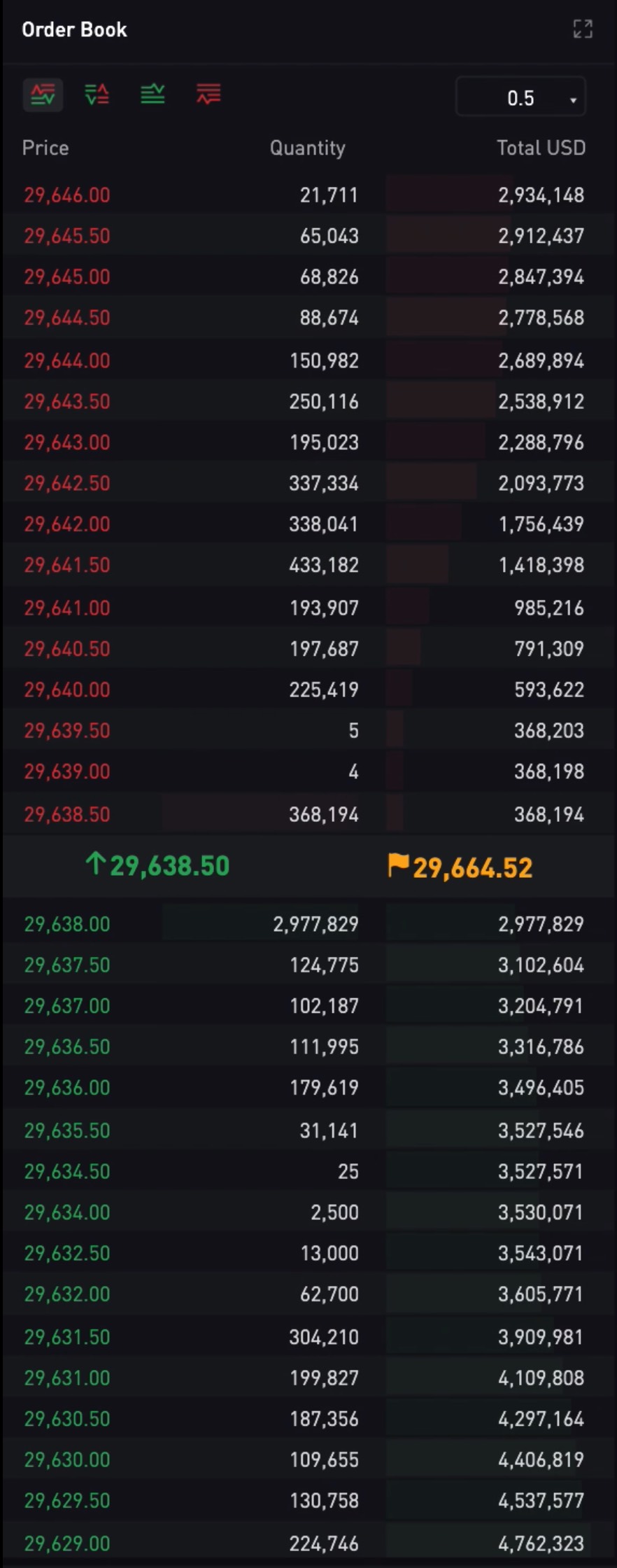}}
    \hspace{30px}
    \fbox{\includegraphics[width=0.29\textwidth, trim=0cm 0cm 0cm 0.0cm,clip]{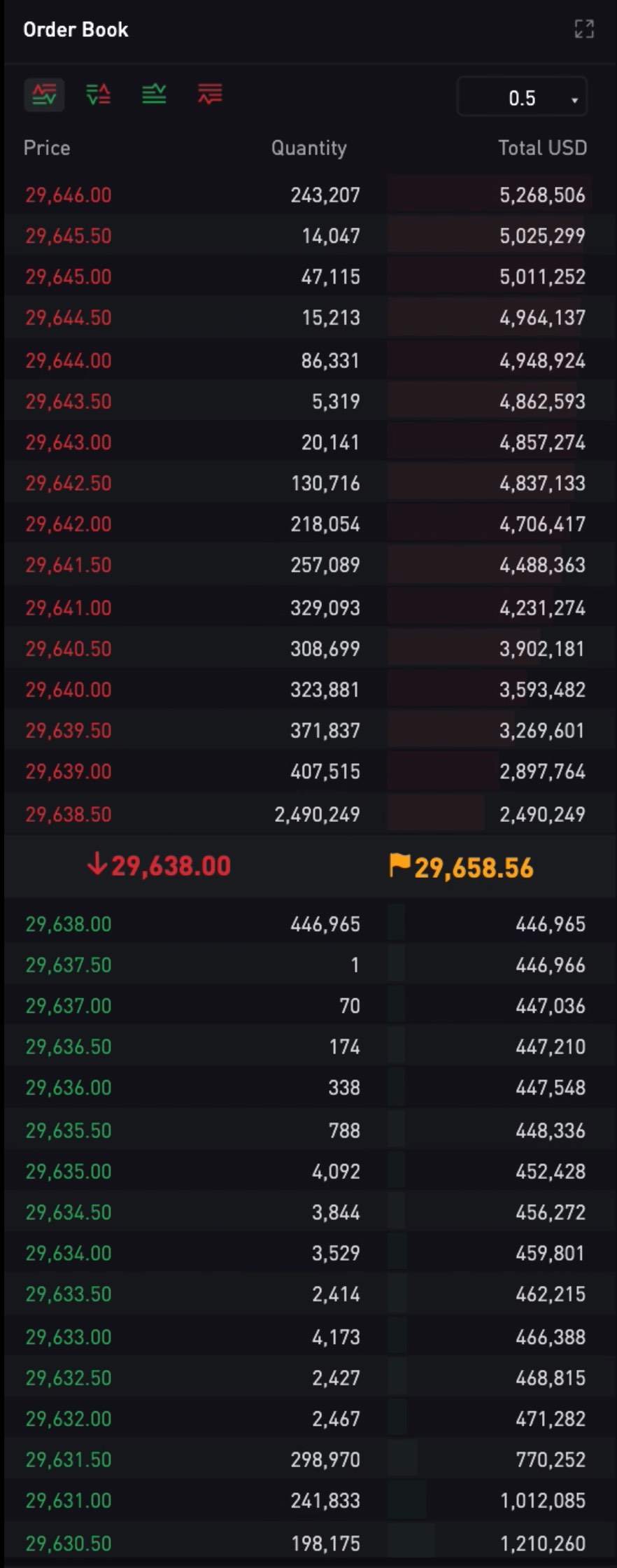}}
    \hspace{30px}
    \fbox{\includegraphics[width=0.29\textwidth, trim=0cm 0cm 0cm 0.0cm,clip]{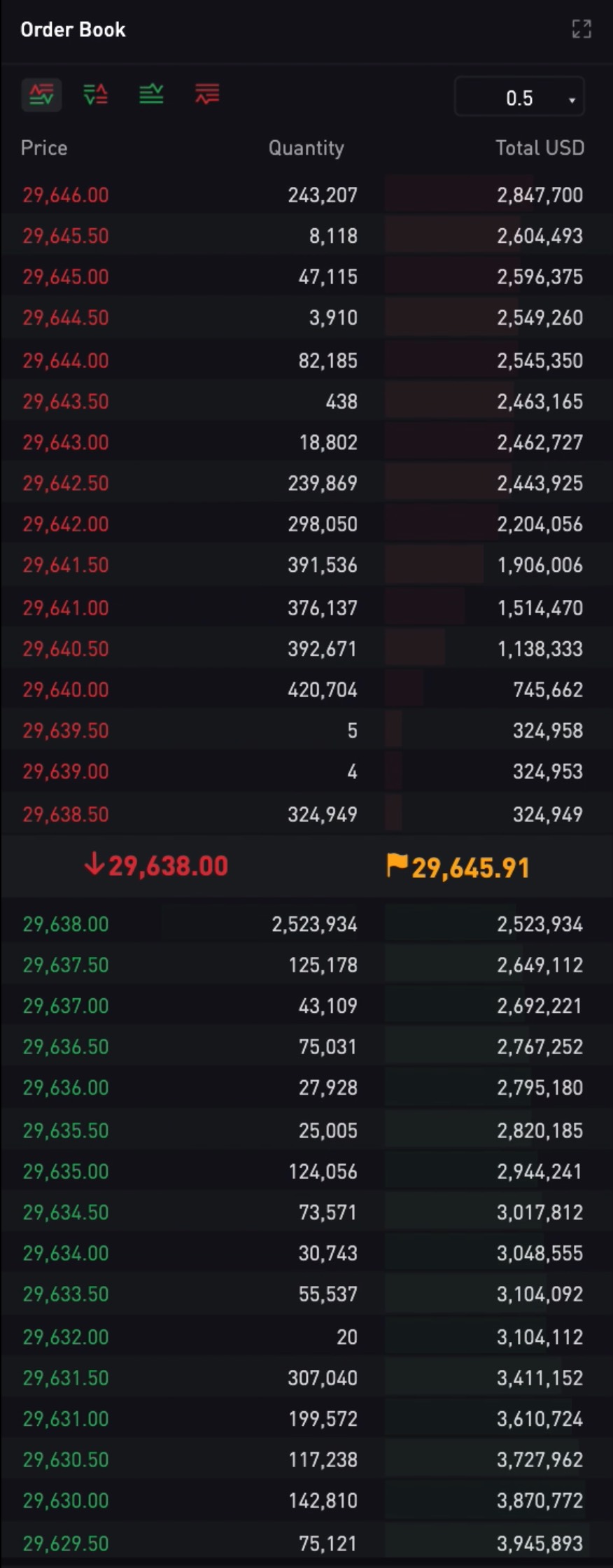}}
\captionsetup{width=0.97\linewidth} 
\caption{Consecutive orderbook snapshots, highlighting a scenario when it is virtually impossible to backtest a maker strategy, mainly to issues arising from the lack of knowledge of the queue position.}
    \label{fig:ob_snapshots}
\end{figure}

A much greater challenge, which is essentially detrimental to the 
ability to backtest maker strategies, is related to the fact that we must keep track of our \textit{queue position} in the orderbook, in order to gauge whether any particular opposite-side taker order consumed our passive order. This turns out to be virtually impossible with the data we have at our disposal in this study, and more broadly, in the crypto space, given the lack of Lever 3 (order-by-order) data in this ecosystem. This is best understood by an example.   
Consider Figure~\ref{fig:ob_snapshots} and suppose that the orderbooks displayed there are subsequent snapshots received from the exchange. Let us denote the arrival times of the respective orderbook snapshots by $t_1, t_2$ and $t_3$. Consider an example where our maker strategy behaves in the following manner. 
\begin{enumerate}
    \item At time $t_1$, the maker strategy posts a new hypothetical ask order at the top ask. Hence there is $386194 \approx 368\text{k}$ USD in liquidity in front of this order, in the queue on this level.
    \item At time $t_2$, the maker strategy keeps its order on the top ask level. In light of the large amount of new submissions to the top ask, there is now presumably still $\sim 368\text{k}$ USD of liquidity in front of our hypothetical maker order and additionally $\sim 2122\text{k}$ USD behind our order in the top ask queue.
    \item Suppose at time $t_3$ our maker strategy decides to remain posted on the top ask level. There was a significant amount of cancellations from the top ask, with roughly $2165\text{k}$ USD removed. We do not know whether these cancellations correspond to the orders that were originally \emph{in front} of our hypothetical order or \emph{behind} it in the top ask queue. Hence, at this point our queue position is completely unknown. Our hypothetical maker order could now well be the first liquidity on the top ask level or the last.
    \item Suppose at a time $t_3 + \epsilon$ the exchange informs us about the arrival of a taker buy order of size $100\text{k}$ USD, while the most recent orderbook we received remains the one from time $t_3$. We now have no way of assessing whether the taker order consumed our posted ask order or not. This depends entirely on our queue position which, as noted above, is not known. Modeling the queue position is a very interesting problem in itself, and a high impact topic of independent interest, which merits its own study. While the topic is relatively unexplored in the literature, the interested reader can find more information on this subject in, for instance,~\cite{Moallemi2016AMF}. 
\end{enumerate}
This example sheds light on the difficulty in keeping track of queue position of a hypothetical maker order. Similar examples can be easily constructed. They are the norm, not the exception.
Generally speaking, the longer a hypothetical order remains posted, the more difficult it becomes to know its queue position exactly.
The core issue which makes it impossible to precisely track queue position is related to the nature of the data we receive from the exchange, namely orderbook snapshot data as opposed to, for instance, order-by-order data.
If we instead received order-by-order data that included unique identifiers for each order, we would know exactly which orders were cancelled and what their queue positions were. From this data, we could then infer our exact queue position and hence whether or not a hypothetical passive order of ours would have filled. Alas, this data is not available in crypto markets, so we must accept the impossibility of precise backtesting of candidate maker strategies.
We will circumvent this issue by conducting a series of real world trading experiments, to test performance and compute PnL of maker strategies.

\subsection{Methodology and Experimental Setup}

While backtesting of maker strategies with any reasonable degree of precision is not possible in crypto markets according to our remarks in the previous subsection, it is relatively simple for individuals to trade in an automated fashion on crypto markets due to the ease of creating exchange accounts, and accessing market data and trade infrastructure through public API endpoints.
This is in contrast to traditional financial markets with a high barrier to entry,  where one must usually interface with brokers, undergo a number of elaborate verification procedures, set up a prohibitively elaborate trading infrastructure to interface with relevant markets, and so on.
We leverage the comparatively easy access to crypto markets in order to test, in the real world, the performance of our previously defined candidate maker strategy. In this subsection, we describe the framework and scope of our trading experiments, alog with details of how they are conducted.

\textbf{Some practical matters.} Let us begin by describing a number of practical circumstances involving the real-time data collection and feature calculation process. The data feeds we use here are the same ones used to gather historical data, and thus the descriptions given in Section~\ref{sec:2} apply equally here. That is, we use websocket API feeds for each of the 14 markets whose data we require. This is generally the quickest way of obtaining real-time data in crypto markets, since the exchange automatically pushes orderbook or trades updates to the subscriber without them having to explicitly prompt the exchange for the most recent market data (which is the case for the alternative type of data feed, the so-called REST API requests). 
Orderbook and trades data are received via different websocket feeds.

We are mining this real-time data in a server cloud-colocated to the server of the exchange on which we are trading, which is Bybit. To explain the term ``cloud-colocated", we first remark that Bybit, as most other crypto exchanges, are hosted in a cloud computing facility. In the case of Bybit this is AWS, and specifically the Singapore region. Colocation can then be easily achieved by renting server space in the same AWS region where the exchange is located. In doing so, we achieve a sub-1ms latency with Bybit.

The market data we receive is processed in real time to first compute values of the features (or transformed features, when applicable), then the meta features, and finally the output of the meta model and the corresponding output decision. This entire computation pipeline takes between zero and two milliseconds. The main bottlenecks are the computations of the average prices $p_{a, t}^i(N_i)$ and $p_{b, t}^i(N_i)$ that appear in the definition of the orderbook imbalance~\eqref{eqn:imb}. This is due to the fact that their computation can involve iterations over many orderbook levels. We trigger a new feature calculation (on the latest market data) and decision output roughly every 5-10ms, depending on market volatility.

\textbf{Experimental setup.} We now describe the setup of our experiment, whose goal is to evaluate the PnL of the strategy devised in the previous subsection. 
To this end, let us first recall the specification of the strategy. 
For a sell order, it is defined by the following set of rules:
\begin{enumerate}[(a)]
    \item The sell post price at time $t$ is $p_{b, t} + \Delta$, where $p_{b, t}$ is the top bid price on the Bybit BTC perpetual at time $t$ and $\Delta$ represents the tick size.
    \item The order amount is fixed at $2000$ USD.
    \item Cancel a top ask order $:\iff$ $F_i(x_t) > T := 1.83$ where $F_i(x_t)$ is the prediction of Bybit's meta model on the latest meta feature observation $x_t$
    \item Post a new top ask order $:\iff$ $F_i(x_t) \leq T$. 
\end{enumerate}

However, due to rate limit constraints imposed by the exchange, we have found it necessary in practice to further restrict condition (d).
Specifically, the exchange allows only for a limited number of API requests (order placements or cancellations) per minute.
When the limit is exceeded, any subsequent requests fail for some period of time.
Luckily, the exchange provides information about the number of remaining API requests.
We incorporate this information by additionally requiring that there be at least two remaining API requests.
That is, if we let $\operatorname{rrl}_t$ denote the remaining rate limit at time $t$, we restrict (d) by additionally imposing the condition $\operatorname{rrl}_t > 1$ (we need at least two requests so we can cancel after posting an order).

Even with this additional restriction, however, we ran into some issues.
Whenever $F_i(x_t) \approx T$, there are often many up- and down-crossings of $T$ in a short time window $[t, t+\delta]$. This results in rapid depletion of the number of permitted API requests. 
The consequence is that we must subsequently pause for roughly one minute until a new order can be submitted. The effect is a significant portion of inactivity periods. We have addressed this problem by further restricting condition (d). Specifically, we additionally require certain trade flow imbalance in favor of the potential sell order. More explicitly, this condition takes the form 
\begin{equation*}
    mTFI_t^i < T', 
\end{equation*}
for some threshold $T'$.
This type of condition was chosen not for any strong theoretical reason, but because it works well in practice. The threshold $T'$ was selected on the basis of intuition, but also trial and error. We leave further calibration and a more rigorous approach to future work. In the meantime, the net result of our restriction measures is that we have reduced the number of new submissions and mitigated extended inactivity periods due to quickly depleted rate limits. The updated condition (d) now takes the form 
\begin{equation*}
    \text{Post a new top ask order } :\iff F_i(x_t) \leq T \quad \wedge \quad mTFI_t^i < T'\quad \wedge \quad \operatorname{rrl}_t > 1.
\end{equation*}

Certainly, we want to not only post sell orders but also buy orders. We have decided to avoid subsequent same-side fills to simplify post-analysis of the results. To elaborate on why this simplifies matters, consider a case when we do not impose this restriction and we receive many subsequent sell fills with no buy fills in the intervening period. The result will be a significant short inventory. If there then happens to be a drastic price jump up, the net results will amount to an extremely high PnL loss. On the other hand, a significant price decrease will result in potentially large profits that are not indicative of long term performance.
This is assuming that we use the comparison between average buy and average sell price over the sample of all fills as an evaluation metric, which indeed we do, as we shall discuss further below when we examine results.

This issue is not difficult to address, however.
We do so by additionally tracking a variable $P_t$ which tells us the net position (or inventory) at time $t$.
If the inventory is negative (that is, we are short), we rule out submissions of new sell orders.
By the same token, if we have long inventory, we do not permit the posting of a new buy order.
With this new restriction, the final form of condition (d) from above is as follows
\begin{equation*}
    \text{Post a new top ask order } :\iff F_i(x_t) \leq T \quad \wedge \quad mTFI_t^i < T'\quad \wedge \quad \operatorname{rrl}_t > 1 \quad \wedge \quad P_t > 0.
\end{equation*}

We define a maker \emph{buy} strategy completely analogously by the following set of rules
\begin{enumerate}[(a)]
    \item The buy post price at time $t$ is $p_{a, t} - \Delta$ where $p_{a, t}$ is the top ask price on the Bybit BTC perpetual at time $t$ and $\Delta$ represents the tick size.
    \item The order amount is fixed at $2000$ USD.
    \item Cancel a top bid order $:\iff$ $F_i(x_t) < -T$
    \item Post a new top bid order $:\iff F_i(x_t) \leq T \quad \wedge \quad mTFI_t^i < T'\quad \wedge \quad \operatorname{rrl}_t > 1 \quad \wedge \quad P_t > 0.$
\end{enumerate}

Now we have fully specified our maker buy and sell strategies in their final form.
For each of these, we launch an independent ``bot" that operates according to the above specified rules. That is, we launch a \emph{sell bot} with access to real time data (including the remaining rate limit $\operatorname{rrl}_t$ and net position $P_t$) which acts according to the set of rules for the sell strategy. Likewise, we launch a \emph{buy bot} which uses the same data feeds, trades on the same exchange, and acts according to the rules of the buy strategy. Note that the data feeds $\operatorname{rrl}_t$ and $P_t$ are shared. This means, for instance, that if the buy bot is barred from posting a new buy order due to $P_t > 0$ (see condition (d) above) and the sell bot receives a fill at time $t+\delta$ resulting in $P_{t+\delta} < 0$, then the buy bot will receive the updated value $P_{t+\delta}$ and become active again. Both bots operate in perpetuity according to these guidelines. We initialize the position by $P_0 := 1$.

In order to be able to better interpret results and draw comparisons, we launch another pair of buy and sell bots operating the same way as we outlined above, with the exception that they use $T=\infty$ and fixed order amount $a:=100$ (we reduced the amount since this strategy is expected to be extremely unprofitable).
The modification of the threshold $T$ has large ramifications: the resulting pair of bots never cancel after posting an order, hence have no capability of avoiding adverse selection.
We shall call this the \textit{naive benchmark} or just \textit{benchmark} strategy.
It will also be useful to have a name for the other pair of bots (the non-benchmark ones), which we will refer to as the \textit{metaMM} bots or strategy.

\subsection{Results and Comparison with Benchmark}

In the preceding two subsections we defined a maker strategy associated to the meta model on the Bybit BTC perpetual, and we outlined our methodology for conducting a real-world market making experiment on the aforementioned market.
This subsection is devoted to the examination of the results.

The total filled amount for the metaMM strategy is 1.537 million USD.
The bots were active from \texttt{2021-07-08 13:44:02} until \texttt{2021-07-13 23:54:02+00:00}.
This implies an hourly fill volume of 11.810k USD, which corresponds to approximately six ``units" since the order size is fixed at 2,000 USD. 
Over the entire sample, the difference between the average sell and buy price is $-4.4749$ bpts.
More specifically, let us define the average sell price by
\begin{equation*}
p_s := \frac{1}{A} \sum_{(p,a)\in \mathcal G^{(s)}} p\cdot a, 
\end{equation*}
where $G^{(s)}$ is the set of all sell fills over the sample period, and $A := \sum_{(p,a)\in \mathcal G^{(s)}} a$.
In the same manner, we define the average buy price
\begin{equation*}
p_b := \frac{1}{A} \sum_{(p,a)\in \mathcal G^{(b)}} p\cdot a.
\end{equation*}
Then it holds that
\begin{equation*}
\operatorname{bpts}_{\operatorname{metaMM}} := \left( \frac{p_s}{p_b} - 1 \right) \cdot 10000 = -4.4749, 
\end{equation*}
over the set $G^{(b)}$ of buy fills.
In other words, a roundtrip trade loses on average $-4.4749$ bpts.
However, recall that each leg (buy or sell) of the trade receives a $2.5$ bpts maker rebate so that a roundtrip trade nets a total of $5$ bpts in rebates.
This implies an average profit per roundtrip trade of $5 - 4.4749 = 0.5251$ bpts.
The median interarrival time between opposite-side fills is 328 seconds.

Note that the trading period of the metaMM strategies is more than four months after the training and test periods that were used to calibrate the models underlying metaMM.
We expect that better results could be achieved if we repeat the experiment using more recent data for the model calibration.
Indeed, we performed similar real-world trading experiments at the end of February and March which exhibited slightly better performance than the one shown in this paper.
These experiments were conducted over a longer period of time and used larger order sizes (up to 20k USD).
In total, the notional USD turnover in this initial live trading experiment was more than 100 million USD. The methodological approach was slightly different, however, and does not match well with the narrative of the present work, thus we decided to redesign the experiment to a more fitting approach.

For the benchmark strategy we employed the 20-fold smaller order amount $a=100$.
The total filled amount was $153.26k$ USD representing 1532 units.
The activity period of the benchmark bots was from \texttt{2021-07-17 12:11:27} until \texttt{2021-07-18 06:59:26}, hence considerably shorter than for the metaMM strategy
This amounts to 8.152k USD in fill volume per hour.
The benchmark strategy achieved the average value (defined similarly to the metaMM case)
\begin{equation*}
\operatorname{bpts}_{\operatorname{benchmark}} = -6.8184, 
\end{equation*}
on a roundtrip trade which, in line with our expectation, did not yield a profit despite the 5 bpts maker rebate per roundtrip trade. However, the median interarrival time between opposite-side fills is far lower for the benchmark bots, namely only 27 seconds, which is less than 10\% of the corresponding metaMM value. This is sensible, since the metaMM strategy cancels much more often after it posts, and therefore accumulates fills much more slowly.
 
\begin{figure}[!ht]
\vspace{-3mm}
\centering
\subcaptionbox[]{metaMM strategy
\label{fig:results_bps_diff_distr}
}[ 0.49\textwidth ]
{ \includegraphics[width=0.49\textwidth, trim=0cm 0cm 0cm 0.0cm,clip] {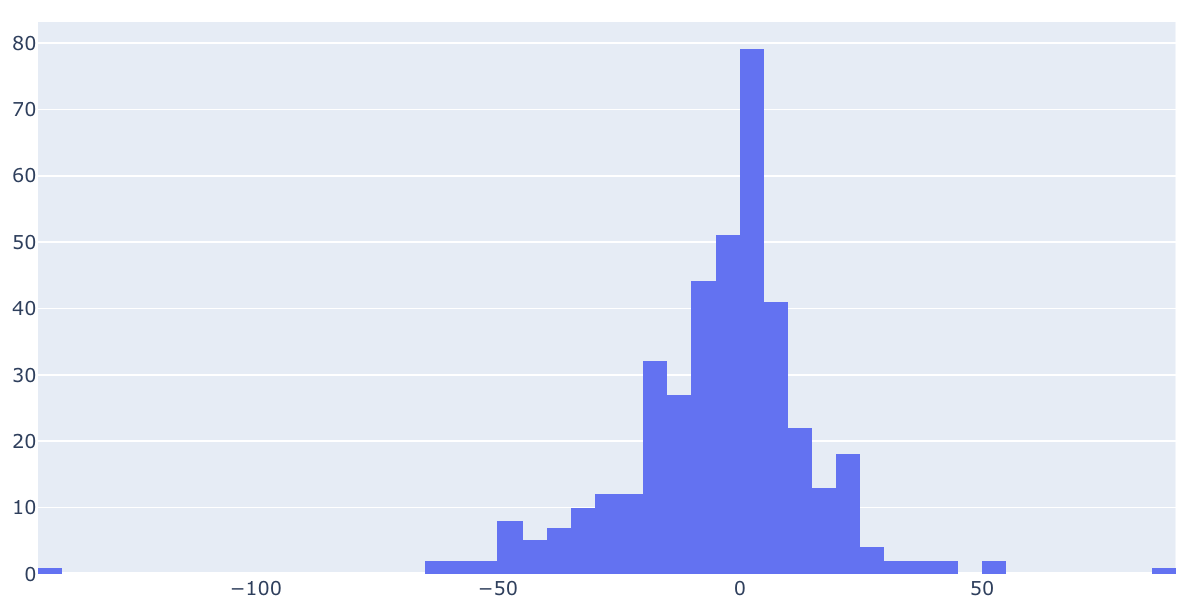} }
\hspace{0.0\textwidth} 
\subcaptionbox[]{naive benchmark
\label{fig:results_bps_diff_distr_benchmark}
}[ 0.49\textwidth ]
{\includegraphics[width=0.49\textwidth, trim=0cm 0cm 0cm 0.0cm,clip]{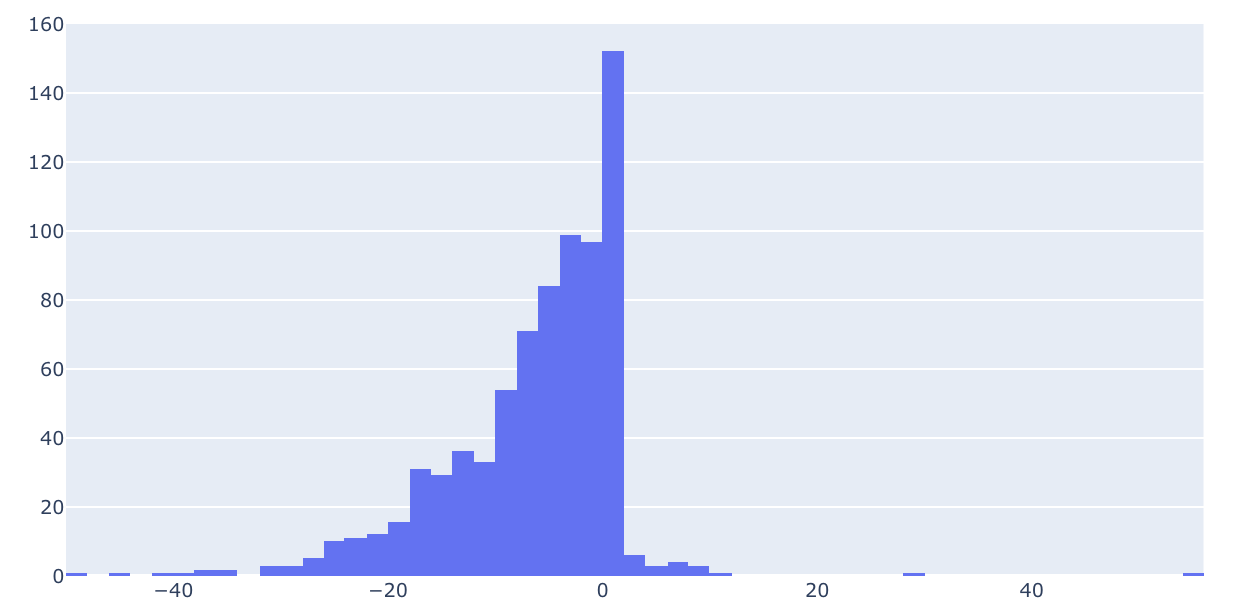} }
\vspace{-2mm}
\captionsetup{width=0.98\linewidth}
\caption[Short Caption]{Distribution of the basis point difference between buy and subsequent sell fill.}
\vspace{-3mm}
\end{figure}

In Figures~\ref{fig:results_bps_diff_distr} and~\ref{fig:results_bps_diff_distr_benchmark} we show histograms, one for the metaMM strategy and one for the benchmark bots, of the differences in basis points between pairs of sell and subsequent buy orders.
It is interesting to note here that the benchmark strategy appears to have a much smaller proportion of roundtrip trades with positive basis point difference.

\begin{figure}[!ht]
\vspace{-3mm}
\centering
\subcaptionbox[]{MetaMM strategy. 
\label{fig:results_pnl_by_time}
}[ 0.49\textwidth ]
{ \includegraphics[width=0.49\textwidth, trim=0cm 0cm 0cm 0.0cm,clip] {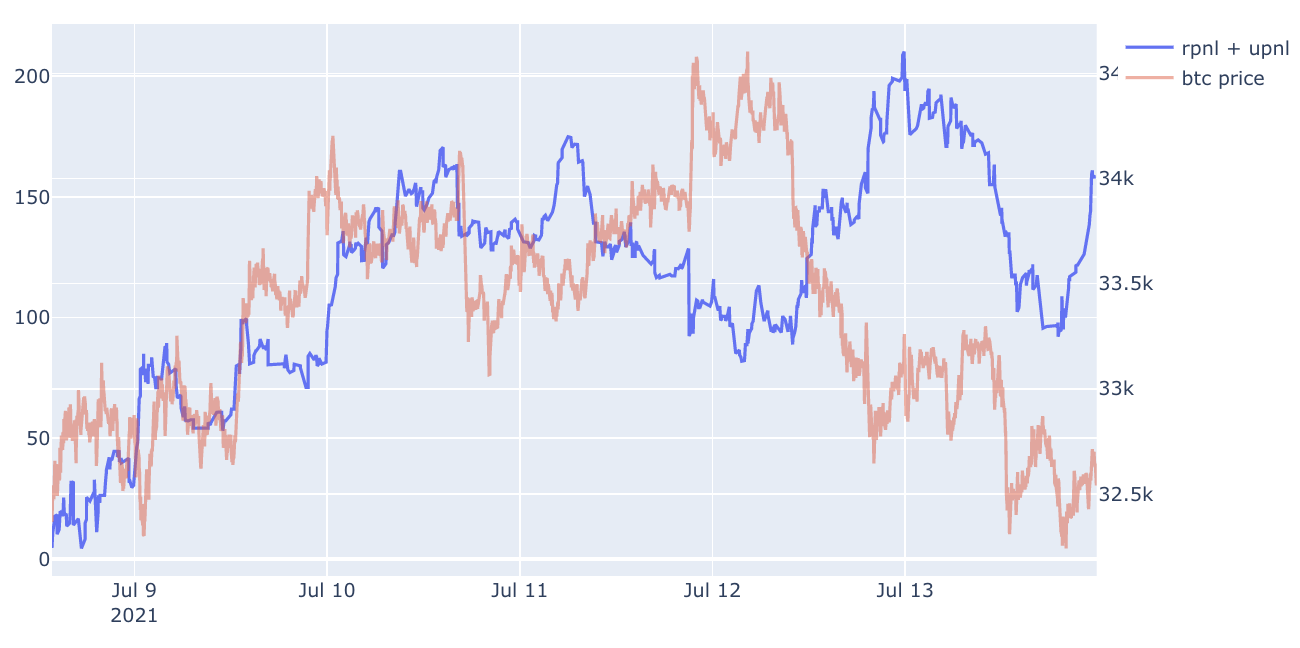} }
\hspace{0.0\textwidth} 
\subcaptionbox[]{Naive benchmark strategy.
\label{fig:results_pnl_by_time_benchmark}
}[ 0.49\textwidth ]
{\includegraphics[width=0.49\textwidth, trim=0cm 0cm 0cm 0.0cm,clip]{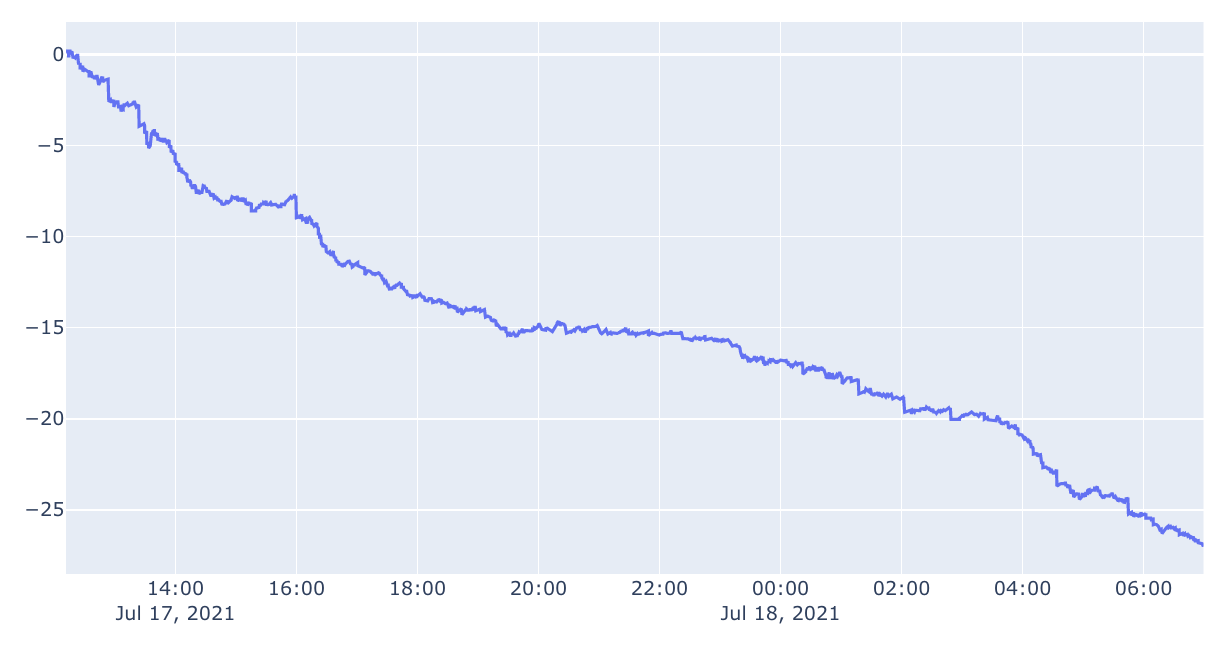} }
\vspace{-2mm}
\captionsetup{width=0.98\linewidth}
\caption[Short Caption]{Cumulative PnL in USD, across time.} 
\vspace{-3mm}
\end{figure}

In Figure~\ref{fig:results_pnl_by_time}, we show PnL over time for the metaMM bots, as well as an overlayed Bitcoin price chart, which we included to give an indication of price volatility over the sample period.
We suspect that the strategy's expected performance is worse during periods of high volatility - a potentially interesting subject of investigation for future work.
Furthermore, we hypothesize that the variation in the strategy's PnL is far greater during highly volatile times. This is made plausible by the observation that high volatility implies more frequent price jumps and hence larger PnL jumps due to inventory held during a large price rise or fall (even if the inventory is small).
The plot~\ref{fig:results_pnl_by_time} includes both realized and unrealized PnL.
To elaborate on what exactly this means, let us introduce some notation.
If at time $t$ the set of (buy and sell) fills is $\mathcal G_t$, the inventory is $x\neq 0$ and the last fill price is $p$, we define the set $\mathcal G_t' := \mathcal G_t \cup \{ (-x, p) \}$.
The realized and unrealized PnL computed is then the PnL computed over the set $\mathcal G_t'$,  while the PnL computed over the set $\mathcal G_t$ is just the realized PnL.
If we denote the output of this calculation by $\operatorname{PnL}_t$, then the pair $t, \operatorname{PnL}_t)$ represents a point on the ``rpnl + upnl" graph shown in Figure~\ref{fig:results_pnl_by_time}.
That is, the quantity $\operatorname{PnL}_t$ takes into account the ``PnL potential" stemming from the fact that the inventory can be nonzero.
 Note that by design of our experiment, we have $|x| \leq 2000$ so the ``unrealized" component is generally quite small (it never exceeds 30 USD).

For comparison, we show a similar plot for the benchmark strategy in Figure~\ref{fig:results_pnl_by_time_benchmark}.
It is striking how steadily the PnL decreases in this case.
The contrast between the two PnL graphs provides clear visual evidence of the additional ``alpha" that the meta model provides for this maker strategy.

For our last performance evaluation metric, we sought a measure for the adverse selection that our fills were subject to.
Intuitively, this means that we want to compare the fill price with another price some time into the future.
In order to rigorously define the notion of ``adverse selection", one needs to define a time horizon and a notion of price.
For our comparisons, we used the last traded price on the Bybit BTC perpetual as the reference price and employed a number of different time horizons, namely $\delta \in \left\{ 0.5\text{s}, 5\text{s}, 10\text{s}, 30\text{s}, 60\text{s}, 150\text{s}, 300\text{s}, 600\text{s}, 1200\text{s}, 2400\text{s}, 4800\text{s} \right\}$.
For each of these time windows, we measure how much the price moved against our fill price on average, what the median adverse price move is, the 25th and 75th percentle, and the minimum and maximum. We additionally show the standard deviation in adverse price move.
The results are shown in Table~\ref{tbl:results_adverse_price_move_by_time} for the metaMM bots,  and in Table~\ref{tbl:results_adverse_price_move_by_time_benchmark} for the naive benchmark.

\begin{table}
\centering
\caption{Adverse selection over different time horizons for metaMM strategy}
\label{tbl:results_adverse_price_move_by_time}
\begin{tabular}{lrrrrrrrrrrr}
\toprule
{} &   0.5s &     5s &    10s &    30s &    60s &   150s &   300s &    600s &   1200s &   2400s &   4800s \\
\midrule
avg &  -0.51 &  -0.97 &  -1.24 &  -1.69 &  -1.99 &  -2.27 &  -2.44 &   -2.19 &   -2.88 &   -2.91 &   -1.73 \\
std  &   1.61 &   2.74 &   3.57 &   5.41 &   7.54 &  12.52 &  16.21 &   23.12 &   30.04 &   40.94 &   57.06 \\
min  & -12.66 & -24.18 & -31.75 & -23.75 & -37.99 & -53.18 & -69.18 & -139.03 & -129.19 & -215.28 & -217.02 \\
25\%  &  -0.15 &  -1.07 &  -1.94 &  -3.74 &  -5.71 &  -9.48 & -12.45 &  -15.26 &  -20.32 &  -25.09 &  -34.52 \\
50\%  &   0.00 &   0.00 &  -0.15 &  -0.74 &  -1.22 &  -2.32 &  -2.56 &   -3.14 &   -3.69 &   -1.64 &   -1.80 \\
75\%  &   0.00 &   0.00 &   0.00 &   0.15 &   0.30 &   4.73 &   6.82 &   11.52 &   14.24 &   18.27 &   33.22 \\
max  &   3.87 &  13.72 &  17.41 &  28.14 &  30.46 &  42.77 &  61.28 &  110.43 &  108.80 &  192.34 &  199.02 \\
\bottomrule
\end{tabular}
\end{table}
\begin{table}
\centering
\caption{Adverse selection over different time horizons for naive benchmark strategy.}
\label{tbl:results_adverse_price_move_by_time_benchmark}
\begin{tabular}{lrrrrrrrrrrr}
\toprule
{} &   0.5s &     5s &    10s &    30s &    60s &   150s &    300s &    600s &   1200s &   2400s &   4800s \\
\midrule
avg &  -0.49 &  -1.65 &  -2.28 &  -2.78 &  -2.94 &  -3.32 &   -3.26 &   -3.00 &   -3.16 &   -3.24 &   -3.06 \\
std  &   1.25 &   2.96 &   4.12 &   7.10 &  10.40 &  16.33 &   23.63 &   30.26 &   38.49 &   51.81 &   89.72 \\
min  & -12.05 & -19.97 & -26.90 & -38.16 & -61.74 & -91.39 & -122.95 & -164.41 & -164.73 & -184.83 & -368.74 \\
25\%  &  -0.34 &  -2.78 &  -4.15 &  -6.31 &  -8.58 & -13.23 &  -17.64 &  -22.91 &  -28.00 &  -36.19 &  -45.80 \\
50\%  &  -0.00 &  -0.43 &  -1.19 &  -2.16 &  -2.41 &  -3.47 &   -3.08 &   -2.99 &   -3.33 &   -3.31 &   -3.64 \\
75\%  &   0.08 &   0.00 &   0.00 &   0.09 &   2.04 &   6.51 &   11.70 &   16.21 &   21.17 &   30.10 &   40.48 \\
max  &   3.53 &  16.77 &  31.85 &  54.40 &  50.37 &  73.54 &  119.77 &  132.37 &  159.60 &  182.54 &  365.70 \\
\bottomrule
\end{tabular}
\end{table}
When comparing the adverse selection that the benchmark fills are subject to with that of the metaMM fills, we note that after a ``settling period" or around 10s, the difference between the two stabilizes at around one basis point. That is, the benchmark fills experience an average adverse price move which is one basis point worse than that experienced by the metaMM fills. While a mere basis point may not sound particularly substantial, we see in the PnL graphs what ramifications this has. The $500$ms time horizon is anomalous in the sense that the benchmark sees less adverse selection over this time period. The standard deviation in adverse selection is smaller for the metaMM strategy, with exception of the $500$ms time window. The difference becomes more pronounced the larger the time horizon.

\section{Conclusion}
\label{sec:7}

This study used a granular data set comprised of market data from the most liquid Bitcoin markets,  in order to closely examine the Bitcoin price formation process on a micro scale. To this end, we defined a set of features that encapsulate microstructural information on each market, which we subsequently leveraged to generate a leader-lagger network illuminating the cross-impact relationships inherent in our set of markets. Perpetual swaps and quarterly futures on Binance and Huobi were found to be particularly strong leading markets, while the Bybit and FTX perpetuals were found to be strong laggards. 

The generated features were subsequently utilized further for the development of linear price prediction models. We compared three methodologies of training such linear models. The first was based on a simple OLS regression using the full set of $70$ features, the second employed a LASSO regression instead with a number of different regularization parameters, and the third inserted an additional dimensionality reduction step before applying an OLS regression.
The largest explanatory powers over future returns were achieved with the first approach, though LASSO models with only around half of the features surviving (that is, having nonzero coefficient) yielded similar $R^2$ values. On the most lagging markets, we were able to achieve $R^2$ values as high as 36\% for the task of predicting $500$ms future returns. Even the most leading market still yielded an $R^2$ value larger than 10\%.

In a next step, we defined a natural trading strategy associated to each model, in order to assess  how well predictability (as measured by $R^2$) translates to trading profit. These strategies employ taker orders, in contrast with another strategy we define later based on maker orders. 
One would expect large $R^2$ to correspond to high PnL values, but this turned out not to be the case. We noted that a market's 
propensity 
to either lead or lag is intimately tied to its fee structure. A high fee regime consisting of a large taker fee and a large maker rebate predispose a market to lagging behaviour. Conversely, a low fee regime, where both taker fee and maker rebate are small, implies a predisposition to leading behaviour. Only when a market's actual laggingness exceeded the one implied by its fee structure, did our strategy yield large (and positive) PnL values therein. A notable example was the FTX perpetual, which was among the most predictable markets despite having an extremely low taker fee. 

Finally, we constructed a simple maker strategy associated to our linear models in order to address the question of whether the large maker rebates can be converted into profit on exchanges where we found highly negative PnLs with the previously tested taker strategy. Due to the difficulties involved in testing maker strategies on historical data, we devised a live trading experiment on Bybit, where we implemented and deployed real capital to the strategy in the real world. It was found to yield approximately one basis point of ``alpha" compared to a naive benchmark strategy, which was sufficient to produce positive PnL over a sample period of roughly five days.

\textbf{Future work.} Our framework opens up a variety of avenues for future research. 
For one, a natural extension would be to incorporate alternative cryptocurrencies, such as Ethereum, into our analysis. Nowadays, daily Ethereum trade volumes routinely represent around half of the Bitcoin ones (and occasionally even exceeding them). Given the high correlation between these two assets, we would certainly expect significant cross-impact to take place, even on small time scales and potentially as a useful predictive signal. Similarly, one could also explore the interplay between equities and Bitcoin markets, particularly around times of market open and close where equities trade volumes tend to be largest. 

A second avenue for potential future work is to consider extensions of the taker and maker strategies we presented in this paper. One could, for instance, allow for greater inventory risk in the maker strategy, and possibly vary the post (i.e. submission) price depending on the size of the inventory, thus trading off the strength of the alpha with the inventory risk. For the taker strategy, one could explore how profitability is affected by different execution styles, which go beyond the simple top-of-the-book execution that we assume.   

Further related to the market making strategy, it would be interesting to investigate its scalability to larger order amounts, along with the interplay of its profitability with volatility conditions. Building models for estimating queue position and incorporating this into the trading behaviour would likely lead to more realistic backtesting, and potential additional profitability when deployed in a live market experiment.

Lastly, at a more foundational level, our feature set can certainly be refined and extended in a number of ways, or simply processed differently. For instance, instead of applying our nonlinear feature transform to the base features, one might experiment with utilizing nonlinear methods, such as tree ensembles or neural networks, applied directly on the set of base features.

\bibliography{citations}{}
\bibliographystyle{plain}

\clearpage
\section*{Appendix}

In this appendix, we will occasionally abbreviate the names of markets according to the acronyms in Table~\ref{tbl:acronym_dict2}, when it is necessary to save space.
\begin{table}[H]
\centering
\caption{Market acronyms}
\label{tbl:acronym_dict2}
\begin{tabular}{ll}
\toprule
{} & Acronym \\
\midrule
hbdm\_BTC-USD          &      m1 \\
hbdm\_BTC\_CQ           &      m2 \\
bitmex\_BTC/USD        &      m3 \\
deribit\_BTC-PERPETUAL &      m4 \\
bybit\_BTC/USD         &      m5 \\
binancefut\_BTC/USDT   &      m6 \\
okex\_BTC-USD-SWAP     &      m7 \\
okex\_BTC-USD-210326   &      m8 \\
ftx\_BTC-PERP          &      m9 \\
huobipro\_BTC/USDT     &     m10 \\
okex\_BTC-USDT-SWAP    &     m11 \\
tbybit\_BTC/USDT       &     m12 \\
thbdm\_BTC-USDT        &     m13 \\
binancecmfut\_BTC/USD  &     m14 \\
\bottomrule
\end{tabular}
\end{table}

\begin{longtable}{lrrrrrrrrrr}
\caption{Summary statistics for all models $M_{ij}$ from Eqn.~\eqref{eqn:model_ij}.}
\label{tbl:model_ij_summary_stats}\\
\toprule
{} &   IMB\_a &   IMB\_b &     TFI &    PRET &     DIV &   IMB\_a &   IMB\_b &     TFI &    PRET &     DIV \\
{} &  t-stat &  t-stat &  t-stat &  t-stat &  t-stat & p-value & p-value & p-value & p-value & p-value \\
\midrule
\endfirsthead
\caption[]{Summary statistics for all models $M_{ij}$} \\
\toprule
{} &   IMB\_a &   IMB\_b &     TFI &    PRET &     DIV &   IMB\_a &   IMB\_b &     TFI &    PRET &     DIV \\
{} &  t-stat &  t-stat &  t-stat &  t-stat &  t-stat & p-value & p-value & p-value & p-value & p-value \\
\midrule
\endhead
\midrule
\multicolumn{11}{r}{{Continued on next page}} \\
\midrule
\endfoot

\bottomrule
\endlastfoot
i=m1, j=m1           &   88.26 &  -79.91 &   82.37 &   14.41 &    0.00 &    0.00 &    0.00 &     0.0 &    0.00 &    0.00 \\
i=m1, j=m2        &   52.70 &  -41.83 &   42.66 &   85.17 &   75.42 &    0.00 &    0.00 &     0.0 &    0.00 &    0.00 \\
i=m1, j=m3          &   56.80 &  -56.04 &   51.03 &   59.07 &    2.78 &    0.00 &    0.00 &     0.0 &    0.00 &    0.01 \\
i=m1, j=m4           &   39.28 &  -24.55 &   60.34 &   64.63 &   48.47 &    0.00 &    0.00 &     0.0 &    0.00 &    0.00 \\
i=m1, j=m5           &   73.35 &  -74.25 &   18.61 &   80.13 &   -8.36 &    0.00 &    0.00 &     0.0 &    0.00 &    0.00 \\
i=m1, j=m6          &    7.95 &   -0.96 &   36.20 &  100.94 &   79.07 &    0.00 &    0.33 &     0.0 &    0.00 &    0.00 \\
i=m1, j=m7           &   48.24 &  -40.49 &   57.14 &   86.57 &   81.40 &    0.00 &    0.00 &     0.0 &    0.00 &    0.00 \\
i=m1, j=m8        &   21.00 &  -16.93 &   51.09 &   91.29 &   87.88 &    0.00 &    0.00 &     0.0 &    0.00 &    0.00 \\
\end{longtable}

\setlength\LTleft{-1cm}
\begin{longtable}{llllllllllllllll}
\caption{T-statistics for all baseline models}
\label{tbl:baseline_models_tstats}\\
\toprule
    &     &    m1 &    m2 &    m3 &    m4 &    m5 &    m6 &    m7 &    m8 &     m9 &    m10 &   m11 &   m12 &   m13 &   m14 \\
\midrule
\endfirsthead
\caption[]{T-statistics for all baseline models} \\
\toprule
    &     &    m1 &    m2 &    m3 &    m4 &    m5 &    m6 &    m7 &    m8 &     m9 &    m10 &   m11 &   m12 &   m13 &   m14 \\
\midrule
\endhead
\midrule
\multicolumn{16}{r}{{Continued on next page}} \\
\midrule
\endfoot

\bottomrule
\endlastfoot
IMBa & m1 &  32.3 &  10.7 &   7.7 &  15.2 &   8.1 &  13.1 &  15.2 &  12.2 &   14.1 &   11.0 &  14.2 &  13.7 &  15.9 &  15.2 \\
    & m2 &   3.3 &  32.8 &   2.8 &   4.5 &   5.0 &  17.9 &   9.9 &  17.2 &   10.4 &   24.3 &   9.4 &   2.1 &  10.1 &   9.4 \\
    & m3 &  -3.6 &  -6.5 &  21.0 &  -1.2 &  -8.1 & -11.5 &  -1.5 &  -4.2 &  -10.7 &   -8.1 &  -2.6 &  -3.7 &  -9.3 &  -7.1 \\
    & m4 &  -4.8 &  -5.2 &  -7.5 &   3.8 &  -4.6 &  -1.6 &  -3.3 &  -3.5 &   -5.5 &   -5.3 &  -3.4 &  -0.7 &  -6.3 &  -3.1 \\
    & m5 &   9.7 &   0.2 &   6.8 &   6.2 &  49.8 &   2.4 &   1.3 &   2.6 &    3.6 &    2.7 &   4.5 &   8.5 &   2.9 &   4.9 \\
    & m6 &   1.4 &   3.8 &   5.6 &   7.0 &   2.6 &  12.0 &   4.8 &   4.8 &    8.4 &    7.9 &   3.1 &   3.8 &   3.5 &   4.0 \\
    & m7 &   1.9 &   7.7 &   2.7 &   1.3 &   4.3 &  11.4 &   9.8 &   8.1 &    3.1 &   10.8 &   8.4 &   1.2 &   1.6 &   7.3 \\
    & m8 &   0.2 &   0.8 &  -0.0 &   2.2 &  -0.9 &   6.5 &  -1.1 &   5.0 &    0.2 &    4.9 &  -0.7 &   0.0 &  -1.6 &   0.8 \\
\end{longtable}

\setlength\LTleft{-1cm}
\begin{longtable}{llllllllllllllll}
\caption{p-values for all baseline models}
\label{tbl:baseline_models_pvalues}\\
\toprule
    &     &    m1 &    m2 &    m3 &    m4 &    m5 &    m6 &    m7 &    m8 &    m9 &   m10 &   m11 &   m12 &   m13 &   m14 \\
\midrule
\endfirsthead
\caption[]{p-values for all baseline models} \\
\toprule
    &     &    m1 &    m2 &    m3 &    m4 &    m5 &    m6 &    m7 &    m8 &    m9 &   m10 &   m11 &   m12 &   m13 &   m14 \\
\midrule
\endhead
\midrule
\multicolumn{16}{r}{{Continued on next page}} \\
\midrule
\endfoot

\bottomrule
\endlastfoot
IMBa & m1 &   0.0 &   0.0 &   0.0 &   0.0 &   0.0 &   0.0 &   0.0 &   0.0 &   0.0 &   0.0 &   0.0 &   0.0 &   0.0 &   0.0 \\
    & m2 &   0.0 &   0.0 &   0.0 &   0.0 &   0.0 &   0.0 &   0.0 &   0.0 &   0.0 &   0.0 &   0.0 &  0.04 &   0.0 &   0.0 \\
    & m3 &   0.0 &   0.0 &   0.0 &  0.23 &   0.0 &   0.0 &  0.12 &   0.0 &   0.0 &   0.0 &  0.01 &   0.0 &   0.0 &   0.0 \\
    & m4 &   0.0 &   0.0 &   0.0 &   0.0 &   0.0 &  0.12 &   0.0 &   0.0 &   0.0 &   0.0 &   0.0 &  0.49 &   0.0 &   0.0 \\
    & m5 &   0.0 &  0.87 &   0.0 &   0.0 &   0.0 &  0.02 &  0.19 &  0.01 &   0.0 &  0.01 &   0.0 &   0.0 &   0.0 &   0.0 \\
    & m6 &  0.15 &   0.0 &   0.0 &   0.0 &  0.01 &   0.0 &   0.0 &   0.0 &   0.0 &   0.0 &   0.0 &   0.0 &   0.0 &   0.0 \\
    & m7 &  0.06 &   0.0 &  0.01 &   0.2 &   0.0 &   0.0 &   0.0 &   0.0 &   0.0 &   0.0 &   0.0 &  0.23 &   0.1 &   0.0 \\
    & m8 &  0.86 &  0.42 &  0.99 &  0.03 &  0.36 &   0.0 &  0.26 &   0.0 &  0.84 &   0.0 &   0.5 &  0.99 &  0.12 &  0.42 \\
\end{longtable}

\begin{table}
\centering
\captionsetup{width=0.85\linewidth}
\caption{In-sample $R^2$ values for LASSO models, for the $1000$ms future return horizon, as the $\lambda$ regularization parameter varies.}
\label{tbl:lasso_models_r2s_1000ms}
\begin{tabular}{llllllllll}
\toprule
{} &  0.001 &  0.002 &  0.004 &  0.008 &  0.016 &  0.032 &  0.064 &  0.128 &  0.256 \\
\midrule
ftx\_BTC-PERP          &  0.342 &  0.341 &  0.341 &   0.34 &  0.338 &  0.335 &  0.324 &  0.293 &  0.254 \\
bybit\_BTC/USD         &  0.331 &  0.331 &   0.33 &  0.328 &  0.325 &  0.324 &  0.317 &  0.298 &   0.25 \\
bitmex\_BTC/USD        &  0.292 &  0.292 &  0.291 &   0.29 &  0.288 &  0.285 &  0.275 &  0.259 &  0.212 \\
tbybit\_BTC/USDT       &  0.279 &  0.279 &  0.279 &  0.278 &  0.276 &  0.273 &  0.263 &  0.246 &  0.207 \\
hbdm\_BTC-USD          &  0.262 &  0.262 &  0.261 &  0.259 &  0.256 &  0.251 &  0.243 &  0.228 &   0.19 \\
thbdm\_BTC-USDT        &  0.231 &  0.231 &   0.23 &  0.229 &  0.225 &   0.22 &  0.207 &   0.19 &  0.144 \\
deribit\_BTC-PERPETUAL &  0.212 &  0.212 &  0.212 &  0.211 &   0.21 &  0.206 &  0.197 &  0.173 &  0.131 \\
okex\_BTC-USDT-SWAP    &  0.195 &  0.195 &  0.195 &  0.194 &  0.193 &   0.19 &  0.182 &  0.153 &  0.117 \\
binancecmfut\_BTC/USD  &  0.193 &  0.193 &  0.193 &  0.191 &  0.189 &  0.185 &  0.171 &  0.152 &  0.106 \\
okex\_BTC-USD-SWAP     &   0.18 &   0.18 &   0.18 &  0.179 &  0.177 &  0.174 &  0.166 &  0.137 &  0.103 \\
okex\_BTC-USD-210326   &  0.164 &  0.164 &  0.163 &  0.163 &   0.16 &  0.157 &  0.148 &  0.121 &  0.078 \\
hbdm\_BTC\_CQ           &  0.152 &  0.152 &  0.151 &   0.15 &  0.147 &  0.144 &  0.135 &  0.103 &  0.065 \\
huobipro\_BTC/USDT     &  0.136 &  0.136 &  0.136 &  0.135 &  0.131 &  0.125 &  0.113 &  0.074 &   0.04 \\
binancefut\_BTC/USDT   &  0.071 &  0.071 &   0.07 &  0.069 &  0.066 &  0.058 &  0.045 &  0.026 &  0.006 \\
\bottomrule
\end{tabular}
\end{table}

\begin{table}
\centering
\captionsetup{width=0.85\linewidth}
\caption{Out-of-sample $R^2$ values for LASSO models, for the $1000$ms future return horizon,  as the $\lambda$ regularization parameter varies.}
\label{tbl:lasso_models_r2s_oos_1000ms}
\begin{tabular}{llllllllll}
\toprule
{ }  &  0.001 &  0.002 &  0.004 &  0.008 &  0.016 &  0.032 &  0.064 &  0.128 &  0.256 \\ 
\midrule
bybit\_BTC/USD         &  0.319 &  0.319 &  0.319 &  0.316 &  0.313 &  0.312 &  0.308 &  0.299 &  0.292 \\
tbybit\_BTC/USDT       &   0.28 &   0.28 &   0.28 &  0.279 &  0.278 &  0.276 &   0.27 &  0.259 &  0.248 \\
bitmex\_BTC/USD        &   0.27 &   0.27 &   0.27 &  0.269 &  0.268 &  0.266 &   0.26 &  0.253 &   0.24 \\
ftx\_BTC-PERP          &  0.264 &  0.264 &  0.263 &  0.263 &  0.261 &  0.259 &   0.25 &  0.219 &    0.2 \\
hbdm\_BTC-USD          &  0.243 &  0.243 &  0.243 &  0.241 &  0.239 &  0.234 &  0.228 &  0.222 &  0.216 \\
deribit\_BTC-PERPETUAL &  0.215 &  0.215 &  0.215 &  0.214 &  0.214 &  0.213 &  0.207 &  0.195 &  0.175 \\
thbdm\_BTC-USDT        &  0.215 &  0.215 &  0.215 &  0.214 &  0.212 &  0.208 &  0.198 &  0.189 &  0.175 \\
okex\_BTC-USDT-SWAP    &  0.184 &  0.184 &  0.184 &  0.183 &  0.183 &  0.181 &  0.175 &  0.153 &  0.131 \\
binancecmfut\_BTC/USD  &  0.177 &  0.177 &  0.177 &  0.176 &  0.173 &   0.17 &  0.158 &  0.146 &  0.121 \\
okex\_BTC-USD-SWAP     &  0.167 &  0.167 &  0.167 &  0.167 &  0.166 &  0.164 &  0.158 &  0.135 &  0.119 \\
okex\_BTC-USD-210326   &  0.155 &  0.155 &  0.154 &  0.154 &  0.151 &   0.15 &  0.144 &  0.125 &  0.094 \\
hbdm\_BTC\_CQ           &  0.144 &  0.144 &  0.144 &  0.143 &   0.14 &  0.137 &  0.132 &  0.106 &  0.087 \\
huobipro\_BTC/USDT     &  0.126 &  0.126 &  0.126 &  0.125 &  0.123 &  0.119 &  0.112 &  0.077 &  0.051 \\
binancefut\_BTC/USDT   &   0.07 &   0.07 &   0.07 &  0.069 &  0.068 &  0.062 &  0.051 &  0.032 &  0.019 \\
\bottomrule
\end{tabular}
\end{table}

\begin{table}
\centering
\captionsetup{width=0.85\linewidth}
\caption{$\operatorname{PnL}_1$ values of LASSO models, as the $\lambda$ regularization parameter varies.}
\label{tbl:lasso_models_pnl1}
\begin{tabular}{llllllllll}
\toprule
{} &  0.001 &  0.002 &  0.004 &  0.008 &  0.016 &  0.032 &  0.064 &  0.128 &  0.256 \\
\midrule
huobipro\_BTC/USDT     &  28867 &  28774 &  28551 &  28276 &  27874 &  27723 &  27336 &  24912 &  22400 \\
thbdm\_BTC-USDT        &  20621 &  20522 &  20084 &  19501 &  19073 &  17759 &  16663 &  17153 &  18144 \\
bybit\_BTC/USD         &  19718 &  19654 &  19590 &  19736 &  19604 &  19934 &  20098 &  20410 &  20468 \\
ftx\_BTC-PERP          &  19689 &  19519 &  19406 &  19306 &  19533 &  19556 &  19627 &  19245 &  18516 \\
okex\_BTC-USD-SWAP     &  17891 &  17757 &  17446 &  17046 &  16627 &  16036 &  14545 &  13236 &  15272 \\
bitmex\_BTC/USD        &  17720 &  17613 &  17603 &  17663 &  17663 &  17726 &  17676 &  17692 &  17490 \\
hbdm\_BTC-USD          &  17663 &  17601 &  17451 &  17340 &  17098 &  16810 &  17800 &  18314 &  18316 \\
okex\_BTC-USD-210326   &  17268 &  17073 &  16602 &  15645 &  15248 &  14603 &  13585 &  11602 &  12871 \\
tbybit\_BTC/USDT       &  17069 &  17037 &  17050 &  17078 &  17073 &  17099 &  17222 &  17669 &  17277 \\
binancecmfut\_BTC/USD  &  16328 &  16188 &  15841 &  15201 &  15301 &  15067 &  14331 &  13712 &  15573 \\
hbdm\_BTC\_CQ           &  16232 &  16038 &  15712 &  14931 &  14183 &  13818 &  12942 &  10924 &   9728 \\
okex\_BTC-USDT-SWAP    &  14961 &  14894 &  14527 &  14241 &  14222 &  14120 &  13463 &  12647 &  14095 \\
binancefut\_BTC/USDT   &  12085 &  12057 &  11876 &  11898 &  11559 &  11408 &  10672 &  10325 &   2793 \\
deribit\_BTC-PERPETUAL &  10313 &  10218 &   9995 &   9730 &   9691 &   9641 &   9638 &   8762 &   8845 \\
\bottomrule
\end{tabular}
\end{table}

\begin{table}
\centering
\captionsetup{width=0.85\linewidth}
\caption{$\operatorname{PnL}_2$ values of LASSO models, as the  $\lambda$ regularization parameter varies.}
\label{tbl:lasso_models_pnl2}
\begin{tabular}{llllllllll}
\toprule
{} &   0.001 &   0.002 &   0.004 &   0.008 &   0.016 &   0.032 &   0.064 &   0.128 &   0.256 \\
\midrule
hbdm\_BTC-USD          &   -8256 &   -8308 &   -8358 &   -8189 &   -8461 &   -8749 &   -9609 &  -10755 &  -12463 \\
thbdm\_BTC-USDT        &  -10674 &  -10645 &  -10755 &  -10890 &  -11030 &  -11152 &  -11768 &  -13774 &  -20751 \\
bybit\_BTC/USD         &  -11631 &  -11440 &  -11504 &  -11808 &  -11895 &  -12510 &  -13771 &  -15709 &  -19746 \\
tbybit\_BTC/USDT       &  -15855 &  -15917 &  -16264 &  -16521 &  -16601 &  -17025 &  -17592 &  -19935 &  -25232 \\
bitmex\_BTC/USD        &  -16209 &  -16091 &  -16341 &  -16386 &  -16311 &  -15813 &  -16328 &  -18577 &  -23969 \\
binancefut\_BTC/USDT   &  -17506 &  -17310 &  -16963 &  -15877 &  -14408 &  -12311 &  -11751 &  -13162 &   -1118 \\
okex\_BTC-USD-SWAP     &  -19228 &  -19132 &  -19303 &  -19663 &  -19942 &  -20573 &  -20604 &  -21873 &  -30217 \\
deribit\_BTC-PERPETUAL &  -19666 &  -19621 &  -19784 &  -20189 &  -20368 &  -20438 &  -20441 &  -20387 &  -25174 \\
okex\_BTC-USDT-SWAP    &  -20058 &  -20105 &  -20382 &  -20778 &  -21067 &  -22089 &  -21896 &  -22322 &  -28594 \\
binancecmfut\_BTC/USD  &  -21871 &  -21841 &  -21968 &  -22018 &  -21668 &  -21632 &  -21278 &  -24877 &  -33766 \\
hbdm\_BTC\_CQ           &  -23391 &  -23377 &  -23135 &  -23140 &  -22648 &  -21941 &  -21201 &  -21883 &  -28583 \\
okex\_BTC-USD-210326   &  -27171 &  -27086 &  -27257 &  -27214 &  -27181 &  -28156 &  -28464 &  -27847 &  -36518 \\
ftx\_BTC-PERP          &  -27504 &  -27520 &  -27423 &  -27607 &  -27548 &  -27805 &  -28238 &  -29278 &  -35285 \\
huobipro\_BTC/USDT     &  -28056 &  -28064 &  -27802 &  -27441 &  -26921 &  -25258 &  -22063 &  -14474 &  -23873 \\
\bottomrule
\end{tabular}
\end{table}

\begin{figure}[!ht]
\hspace*{-1cm}
\vspace{-3mm}
\centering
    \includegraphics[scale=0.45]{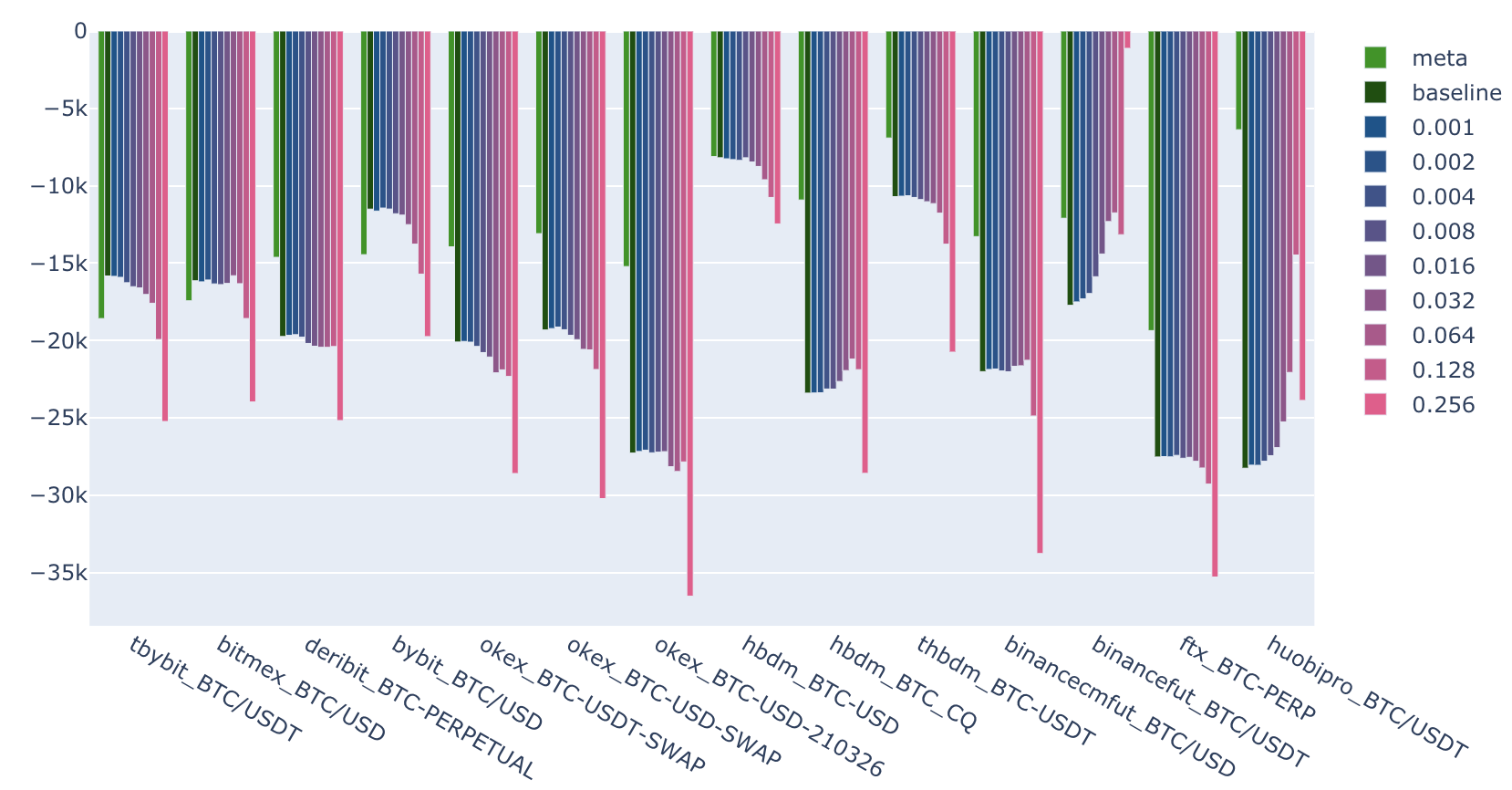}
    \caption{$\operatorname{PnL}_2$  values for each model.}
    \label{fig:pnl2_comparisons}
\hspace{0.0\textwidth} 
\vspace{-3mm}
\end{figure}

\end{document}